\documentclass{lmcs}
\pdfoutput=1

\usepackage{lastpage}
\lmcsdoi{18}{2}{14}
\lmcsheading{}{\pageref{LastPage}}{}{}%
{Oct.~20,~2018}{Jun.~02,~2022}{}

\usepackage[utf8]{inputenc}
\usepackage{microtype}
\usepackage{standalone}
\usepackage[hidelinks,breaklinks]{hyperref}
\usepackage{tikz}
\usepackage{tikz}
\usetikzlibrary{calc}
\usetikzlibrary{positioning}

\definecolor{brown}{RGB}{146,73,0}

\tikzset{->,>=latex,every node/.style={shape=rectangle,draw,},
scope/.style={draw=none},
lnode/.style={draw=none},
trigger/.style={draw=blue,line width=1.7pt},
newnode/.style={red,line width=1.7pt},
guardok/.style={blue,dashed},
guardnok/.style={black,line width=1.0pt},
pics/drawv/.style args={#1/#2/#3/#4}{
	code = {
	\draw[-,dashed] let \p1=(#1.east),\p2=(#2.west),\p3=(#3.north),\p4=(#4.south) in
	({(\x1+\x2)/2},\y3) -- ({(\x1+\x2)/2},\y4);
	}
},
pics/drawvr/.style args={#1/#2}{
	code = {
	\pic{drawv=#1/#2/#2/#2};
	}
},
pics/drawh/.style args={#1/#2/#3/#4}{
	code = {
	\draw[-,dashed] let \p1=(#1.south),\p2=(#2.north),\p3=(#3.west),\p4=(#4.east) in
	(\x3,{(\y1+\y2)/2}) -- (\x4,{(\y1+\y2)/2});
	}
}
}

\usepackage{breakurl}
\usepackage{url}

\title[When Can We Answer Queries Using Result-Bounded Data Interfaces?]{When Can We Answer Queries \texorpdfstring{\\}{}Using Result-Bounded Data Interfaces?}
\author[A.~Amarilli]{Antoine Amarilli\rsuper{a}}
\address{LTCI, T\'el\'ecom Paris, Institut polytechnique de Paris, FR} 
\email{antoine.amarilli@telecom-paris.fr}
\author[M.~Benedikt]{Michael Benedikt\rsuper{b}}
\address{Department of Computer Science, Oxford University, Parks Rd, Oxford OX1 3QD, UK}
\email{michael.benedikt@cs.ox.ac.uk}

\usepackage{colonequals}
\usepackage{multirow}
\usepackage{amsmath}
\usepackage{amsthm}
\usepackage{xspace}
\usepackage{amsfonts}
\usepackage{tabularx}
\usepackage{booktabs}
\usepackage{xcolor}

\newcommand{\coloneqq}{:=}

\newcommand{\smpr}{\kw{SMPR}}

\newcommand{\lift}{\kw{Lin}}

\newcommand{\factsof}{\kw{FactsOf}}
\newcommand{\birthfact}{\kw{BirthFact}}
\newcommand{\gammasep}{\Gamma^{\kw{Sep}}}

\newcommand{\amd}{\kw{AMonDet}}
\newcommand{\canondb}{\kw{CanonDB}}
\newcommand{\ptime}{\kw{PTIME}}
\newcommand{\exptime}{\kw{EXPTIME}}

\newcommand{\arity}{\kw{Arity}}
\newcommand{\adom}{\kw{Adom}}

\newcommand{\elimnd}{\kw{AxiomRB}}

\newcommand{\checkview}{R}

\newcommand{\detby}{\kw{DetBy}}
\newcommand{\relaxs}{\kw{ElimUB}}
\newcommand{\twoexp}{\kw{2EXPTIME}}
\newcommand{\twoexptime}{\twoexp}
\newcommand{\return}{\kw{Return}}

\newcommand{\fds}{\Sigma_{\mathrm{FD}}}
\newcommand{\ids}{\Sigma_{\mathrm{ID}}}
\newcommand{\incd}{\text{ID}}
\newcommand{\uincd}{\text{UID}}
\newcommand{\gtgd}{\kw{GTGD}}

\newcommand{\wrt}{w.r.t.}

\newcommand{\shortcut}{\kw{SC}}

\newcommand{\determines}{\rightarrow}

\newcommand{\aschema}{\kw{Sch}}
\newcommand{\inmap}{\kw{InMap}}
\newcommand{\outmap}{\kw{OutMap}}
\newcommand{\accbind}{\kw{AccBind}}
\newcommand{\abind}{\accbind}

\newcommand{\accpart}{\kw{AccPart}}

\newcommand{\kw}[1]{{\mathsf{#1}}\xspace}
\newcommand{\profinfo}{ \kw{Prof}}

\newcommand{\univdirect}{\udirectory}
\newcommand{\udirectory}{\kw{Udirectory}}

\newcommand{\accessible}{\kw{accessible}}
\newcommand{\acc}[1]{\kw{\scriptscriptstyle Accessed} #1}

\newcommand{\mt}{\kw{mt}}
\newcommand{\aplan}{\kw{PL}}

\newcommand{\card}[1]{\left|#1\right|}

\newcommand{\NN}{\mathbb{N}}

\newcommand{\np}{\kw{NP}}

\newcommand{\sign}{{\mathcal{S}}}

\newcommand{\aselect}{\sigma}
\newcommand{\dep}{\delta} %
\newcommand{\trig}{\tau} %

\newcommand{\bounded}{\kw{Bounded}}%
\newcommand{\acyclic}{\kw{Acyclic}}

\newcommand{\praccess}{\mathsf{pr}}
\newcommand{\udaccess}{\mathsf{ud}}

\newcommand{\attrfmt}[1]{\mathit{#1}}

\begin{document}
\begin{abstract}
We consider answering queries on data 
available through \emph{access methods}, 
that provide lookup access to the tuples matching a given binding.
Such interfaces are common on the Web; further, they often have \emph{bounds} on
how many results they can return, e.g., because of pagination or rate limits.
We thus study \emph{result-bounded methods}, which 
may return only a limited number of tuples.
We study how to decide if a query is \emph{answerable} using result-bounded
  methods, i.e., how to compute a \emph{plan} that returns all answers to
  the query using the methods, assuming that the underlying data satisfies some
   integrity constraints.
  We first show how to reduce answerability to a query containment problem with
  constraints. Second, we show 
  ``schema simplification'' theorems describing when and how result-bounded services can be
  used. Finally, we use these theorems to give decidability and complexity results about  answerability for
  common constraint classes.
\end{abstract}

\maketitle
\section{Introduction}

Web services 
expose programmatic interfaces to data.
Many of these services can be modeled as an
\emph{access method}:
given a set of arguments for some attributes of a relation,
the method returns all matching
tuples for the relation.

\begin{exa} \label{ex:simple}
  Consider a Web service that exposes university employee information.
  The schema has a relation $\profinfo(\attrfmt{id}, \attrfmt{name}, \attrfmt{salary})$
  and an access method $\praccess$ on this relation: the input to $\praccess$ is
  the $\attrfmt{id}$
  of a professor, and an access to this method
  returns the $\attrfmt{name}$ and $\attrfmt{salary}$ of
  the professor. The schema also has a relation $\univdirect(\attrfmt{id},
  \attrfmt{address}, \allowbreak \attrfmt{phone})$, and an access method $\udaccess$: it
  has no input and returns the $\attrfmt{id}$, $\attrfmt{address}$, and
  $\attrfmt{phone}$  %
of all university employees.
\end{exa}

Our goal is to answer queries using such services.
In the setting of Example~\ref{ex:simple}, the user queries are posed
on the relations $\profinfo$ and $\univdirect$, and we wish to answer
them using the methods $\praccess$ and $\udaccess$. 
To do so, we can exploit \emph{integrity constraints} that the data is known to satisfy:
for instance, the referential constraint $\tau$ that says that the
$\attrfmt{id}$ of every tuple in~$\profinfo$ is also in~$\univdirect$.

\begin{exa}
  \label{exa:plani}
  Consider $Q_1(n): \exists i ~ \profinfo(i, n, 10000)$, the query
  that asks for the names of professors with salary~$10000$. 
  If we assume the integrity constraint~$\tau$,
  we can implement
  $Q_1$ as the following \emph{plan}: first access $\udaccess$ to get the set of all ids, and then
  access $\praccess$ with each id
to obtain the salary, filtering the results to return only the names 
with salary~$10000$.
This plan \emph{reformulates} $Q_1$ over the access methods:
it is equivalent to~$Q_1$ on all instances satisfying~$\tau$,
  and it only uses $\praccess$ and $\udaccess$ to access
  $\profinfo$ and $\univdirect$.
\end{exa}

Prior work  (e.g.,~\cite{dln,ustods}) has 
formalized this reformulation task as an \emph{answerability} problem: given a schema
with access methods and integrity constraints, and given a query, determine if
we can answer the query using the methods. The query has to be answered in a
\emph{complete} way, i.e., without missing any results.
This prior work has led to implementations (e.g.,~\cite{usvldb14,usvldb15,bioint})
that can
determine how to evaluate a conjunctive query using
a collection of Web services, by generating a plan that makes calls to the services.

However, all these works assume that when we access a Web service, we 
always obtain \emph{all} tuples that match the access.
This is not realistic: to avoid wasting
resources and bandwidth, virtually all Web services impose a \emph{limit} on how
many results they will return.
For instance, the ChEBI service (chemical entities of
biological interest, see~\cite{bioint}) limits the output of lookup methods to 5000 entries,
while IMDb's web interfaces impose a limit of~10000~\cite{imdb}. 
With some services, we can request more
results beyond the limit, e.g., using pagination or continuation tokens,
but there is often a
\emph{rate limitation} on how many requests can be
made~\cite{fbapi,githubapi,twitterapi}, which also limits the total number of
obtainable results.
Thus, for many Web services, beyond a certain number
of results, we cannot assume that all matching tuples are returned.
In this work, we introduce \emph{result-bounded} methods to reason on these
services.

\begin{exa}
  \label{exa:rbound}
The~$\udaccess$ method in 
Example~\ref{ex:simple} may have a result bound, e.g., it may return
  at most 100 entries. If this is the case, then the plan of
  Example~\ref{exa:plani} is not equivalent to~$Q_1$ as it may miss some result
  tuples.
\end{exa}

Result-bounded methods make it very challenging to reformulate queries. Indeed,
they are \emph{nondeterministic}: if the number of results is more than
the result bound, then the Web service only returns a subset of results, usually according to
unknown criteria. 
For this reason, it is not even clear whether result-bounded methods can be
useful at all to answer queries in a complete way.
However, this may be the case:

\begin{exa} \label{ex:existencecheck} Consider the schema of Example~\ref{ex:simple}
  and assume that $\udaccess$ has a result bound of~$100$ as in
  Example~\ref{exa:rbound}.
  Consider the query $Q_2: \exists i\, a\, p ~ \univdirect(i, a, p)$ asking if
there  is \emph{some} university employee. We can answer $Q_2$ with a plan  that
accesses
the  $\udaccess$ method and returns true if the output is non-empty.
It is not a problem that $\udaccess$ may omit
some result tuples, because we only want to know if it returns something.
This gives a first intuition: 
  result-bounded methods are useful to check for the existence of
  matching tuples.
\end{exa}

Further, 
result-bounded methods can also help under integrity constraints such as keys or
functional dependencies:

\begin{exa} \label{ex:fd}
  Consider the schema
 of Example~\ref{ex:simple} and the access method 
  $\udaccess_2$  on $\univdirect$
  that takes an $\attrfmt{id}$ as input and returns the 
  $\attrfmt{address}$ and phone number of tuples with this $\attrfmt{id}$.
  Assume that $\udaccess_2$ has a result bound of~$1$, i.e., returns at most one
  answer when given an $\attrfmt{id}$. Further assume 
   the functional dependency $\phi$: 
  each employee id 
  has exactly one $\attrfmt{address}$ (but possibly
  many phone numbers).
Consider the query~$Q_3$ asking for the address of 
  the employee with id~12345.
We can answer~$Q_3$ by calling $\udaccess_2$ with 12345 and projecting onto the
  $\attrfmt{address}$ field.
  Thanks to~$\phi$, we know that 
  the result will contain the 
  employee's address,
  even though 
  only one of the phone numbers will be returned.
This gives a second intuition: 
  result-bounded methods are useful when there is a functional dependency that guarantees that
  some projection of the output is complete.
\end{exa}

In this paper, we study how and when we can use result-bounded methods to
reformulate queries and obtain complete answers, formalizing in particular the
intuition of Examples~\ref{ex:existencecheck} and~\ref{ex:fd}.
We then show decidability and complexity results for the answerability problem.
We focus on two common classes of integrity constraints on
databases: \emph{inclusion dependencies} ($\incd$s), as in
Example~\ref{ex:existencecheck}, and \emph{functional dependencies} (FDs), as in
Example~\ref{ex:fd}. But we also show results for more expressive constraints:
see Table~\ref{tab:results} for a summary.

The first step of our study (Section~\ref{sec:reduce}) is to reduce the answerability 
problem to \emph{query containment under constraints}. Such a reduction is well-known
in the context of reformulating queries over views \cite{NSV}, and of answering queries
with access methods without result bounds~\cite{thebook}. But the
nondeterminism of result-bounded methods means that we cannot apply these
results directly. We nevertheless show that 
this reduction technique can still be applied in the presence of result
bounds. This reduction does not suffice to solve the problem, because
the resulting query containment problem involves complex cardinality constraints, 
so it does not immediately lead to decidability results.

Our second step (Section~\ref{sec:simplify})
is to show \emph{schema simplification results}, which explain why some
of the result bounds can be ignored for the answerability problem.
These results characterize how result-bounded methods are
useful: they capture and generalize the examples above.
For instance, 
we show that for constraints consisting of $\incd$s,
result-bounded methods are only useful
as an \emph{existence check} as in Example~\ref{ex:existencecheck}.
We also show that, for FD constraints,
result-bounded methods are only useful to access 
the \emph{functionally-determined part of the
output}, as in
Example~\ref{ex:fd}.
The proofs
utilize  a technique of \emph{blowing up models}, i.e., we enlarge them to increase
the number of outputs of an access, without violating constraints or changing query answers.
The simplest version of this technique is to show limitations on which
queries can be answered with result bounds in the presence of constraints in first-order logic without equality
(Theorem~\ref{thm:simplifychoice}).
This result has some broad similarity to classical finite model theory results
on limitations of first-order logic.
We will show that the blowing-up method can yield similar limitations even in the presence
of equality.

Third, in Section~\ref{sec:complexity}, we use the simplification results to
deduce that answerability is decidable for these constraint classes, and
show tight complexity bounds: we show $\np$-completeness for constraints
consisting of FDs,
and $\exptime$-completeness for $\incd$s. We refine the latter result to show that
answerability is $\np$-complete for \emph{bounded-width} $\incd$s,
which export only a constant number of variables. 
We prove this using
ideas of Johnson and Klug~\cite{johnsonklug}, along with a
\emph{linearization} technique, extending ideas introduced in~\cite{gmp}: we show how
the constraints used to reason about answerability can be ``simulated'' with
 restricted  inclusion dependencies, and that analyzing this simulation
gives finer complexity bounds.

In Section~\ref{sec:simplifychoice}, we study more expressive constraint
classes, beyond $\incd$s and FDs. We do so using a weaker form of simplification, called
\emph{choice simplification}, which replaces all result bounds by~$1$:
this intuitively
implies that the number of results does not matter.  
We show that it suffices to consider the choice
simplification
for a huge class of  constraints, including all
TGDs, and also constraints combining FDs and
$\incd$s. 
In Section~\ref{sec:complexitychoice}, we use this 
technique to show that decidability of  answerability holds much more broadly:
in particular it holds for a wide range of classes where query containment is decidable.
We conclude the paper by giving some limits to schema simplification and decidability of answerability
(Section~\ref{sec:general}), followed by conclusions (Section~\ref{sec:conc}). 
In particular, we explain in the conclusions how our results on answerability
extend to the practically relevant problem of \emph{extracting a plan} in the
case where one exists.

This article is based on the conference paper~\cite{confpaper}. 
In addition to providing full proofs for the major results of \cite{confpaper}, in the appendix 
to this work we give a number of supplementary results, showing the generality of the methods.
We do not include all results claimed in the conference paper. In particular, the conference
paper claims results also for the finite variant of the answerability problem with result bounds.
While we believe these results hold, the proofs in the submission are flawed, and thus we make
no such claims in this work, dealing only with the unrestricted variant.

{\footnotesize
\begin{table}
  {
\caption{Simplifiability and complexity results for monotone answerability}
\label{tab:results}
  \begin{tabular}{lll}
\toprule
{\bfseries Fragment} &
{\bfseries Simplification} & 
    {\bfseries Complexity } \\
\midrule
  IDs & Existence-check  (Thm~\ref{thm:simplifyidsexistence}) & $\exptime$-complete  (Thm~\ref{thm:decidids})\\
  Bounded-width IDs & Existence-check (see above) & $\np$-complete  (Thm~\ref{thm:npidsbounds}) \\
  FDs & FD  (Thm~\ref{thm:fdsimplify})
    & $\np$-complete  (Thm~\ref{thm:decidfd}) \\
  FDs and UIDs & Choice  (Thm~\ref{thm:simplifychoiceuidfd})
    &  $\np$-hard (see above) and in~$\twoexp$  (Thm~\ref{thm:deciduidfd}) \\
  Equality-free FO & Choice  (Thm~\ref{thm:simplifychoice})
    & Undecidable  (Proposition~\ref{prp:undec}) \\
  Frontier-guarded TGDs & Choice  (see above)
    & $\twoexp$-complete  (Thm~\ref{thm:decidegf}) \\
\bottomrule
\end{tabular}
  }
\end{table}
}

\section{Related Work}
\label{sec:related}

Our paper relates
to a line of work about finding plans to answer
queries using access methods. The initial line of work considered finding
equivalent ``executable rewritings''~--- conjunctive queries
where the atoms are ordered in a way compatible with the access patterns. This was studied
first without integrity
constraints~\cite{access1,access2},
and then
with disjunctive TGD constraints \cite{dln}.
Later \cite{ustods,thebook} formulated the problem of finding a \emph{plan}
that answers the query over the access patterns, distinguishing two notions of plans with access methods:
one with arbitrary relational operators in middleware and another without the difference
operator. They
studied the problem of getting plans of both types in the presence of integrity
constraints:
following~\cite{dln},
they reduced the search for executable rewritings to 
query containment under constraints.
Further, \cite{ustods,thebook} also related the reduction to a semantic notion of 
determinacy, originating
from the work of Nash, Segoufin, and Vianu~\cite{NSV} in the context of views.
Our paper extends the reduction to query containment 
in the presence of result bounds,
relying heavily on the techniques of \cite{dln,NSV,ustods,thebook}.

Non-determinism in query
 languages has been studied in other contexts \cite{av,asv}. 
However, the topic of this work, namely, using non-deterministic Web services to
implement deterministic queries, has not been studied.
Result bounds are reminiscent of \emph{cardinality constraints}, for
which
the answerability problem has been studied~\cite{wenfeifloris1}. However, the
two are different: whereas cardinality constraints restrict the 
\emph{underlying data},
result bounds concern the \emph{access methods} to the data, and makes them
\emph{non-deterministic}:
this has not been studied in the past.
In fact, surprisingly, 
our schema simplification results (in Sections~\ref{sec:simplify}
and~\ref{sec:simplifychoice}) imply that answerability with result bounds can be
decided \emph{without} reasoning about cardinality constraints at all.

To study our new setting with result-bounded methods, we introduce several 
specific techniques
to reduce to a decidable query containment problem, e.g., 
determinacy notions for non-deterministic services and the technique of
``blowing up models''.
The additional technical  tools needed to bound the complexity of our problems revolve around analysis of the chase.
While many components of this analysis are specific to the constraints produced by our problem,
our work includes 
a \emph{linearization} method, which we believe is of interest
in  other settings.
 Linearization is a technique
from~\cite{gmp}, which shows that certain entailment problems can
be reduced to entailment of queries from  linear TGDs. We refine this
to show that in certain cases we can reduce to entailments involving a restricted class of
linear TGDs, where more specialized bounds \cite{johnsonklug} can be applied.

\section{Preliminaries} \label{sec:prelims}

\paragraph{Data and queries.}
We consider a \emph{relational signature} $\sign$ that consists of
a set of \emph{relations} with an associated
\emph{arity} (a positive integer) and of a finite
set of \emph{constants}.
The \emph{positions} of a relation~$R$ of~$\sign$ are $1 \ldots n$ where
$n$ is the arity of~$R$.
An \emph{instance} of~$R$ is a
set of~$n$-tuples (finite or infinite), and an \emph{instance} $I$ of~$\sign$
consists of instances
for each relation of~$\sign$,
along with a mapping from the constants of the signature to the \emph{active
domain} $\adom(I)$ of~$I$, i.e., the set of all
the values that occur in facts of~$I$. Note that this means that two signature
constants can be interpreted by the same element.
 We can equivalently see~$I$ as a set
of \emph{facts} $R(a_1 \ldots a_n)$ for each tuple 
$(a_1 \ldots a_n)$ in the instance of each relation~$R$, along with the mapping
of constants.
A \emph{subinstance} $I'$ of~$I$ is an instance that contains a subset of the
facts of~$I$, and $I$ is then a \emph{superinstance} of $I'$.

We will study \emph{conjunctive queries} (CQs), which are logical expressions of the form
$\exists x_1 \ldots x_k  ~ (A_1 \wedge \cdots \wedge A_m)$, where
the~$A_i$ are \emph{relational atoms} of the form
$R(t_1 \ldots t_n)$, with~$R$ being a relation of arity~$n$ and $t_1 \ldots t_n$
being either variables from $x_1 \ldots x_k$ or constants.
A CQ is \emph{Boolean} if it has no free variables.
A Boolean CQ $Q$ \emph{holds} in an instance $I$ exactly when there
is a \emph{homomorphism} of~$Q$ to~$I$: a mapping~$h$ from the variables and
constants of~$Q$ to~$\adom(I)$ which is the identity on constants and which
ensures that,
for every atom $R(x_1 \ldots x_n)$ in~$Q$, the atom $R(h(x_1) \ldots h(x_n))$ is a fact of~$I$.
We let $Q(I)$ be the \emph{output} of $Q$ on $I$, defined in the usual way: if $Q$ is Boolean,
the output is true if the query holds and false
otherwise.
A \emph{union of conjunctive queries} (UCQ) is a disjunction of CQs.

\paragraph{Integrity constraints.}
An \emph{integrity constraint} is a restriction on instances: that is, a
function mapping every instance of a given schema to a Boolean. When we say that
an instance \emph{satisfies} a constraint,
we just mean that the function evaluates to true.
As concrete syntax for constraints  we use fragments of first-order logic
(FO), with the active-domain semantics, and disallowing constants.
The active-domain semantics can be enforced syntactically, e.g., by restricting
first-order logic formulas to always quantify over elements that appear in some
relation. In the few cases in this paper where we talk about  FO integrity constraints,
we will always mean  a constraint in such a restricted fragment. With such a restriction,
the truth value of an FO integrity constraint on an instance is well-defined.
For most of our results we  focus on 
\emph{tuple-generating dependencies} (TGDs) and
\emph{functional dependencies} (FDs), which we now review.

A \emph{tuple-generating dependency} (TGD) is an FO sentence~$\tau$ of the form:
$\forall \vec x ~ (\phi(\vec x) \rightarrow \exists \vec y ~ \psi(\vec x, \vec
y))$
where $\phi$ and $\psi$ are conjunctions of relational atoms:
$\phi$ is the \emph{body} of~$\tau$ while $\psi$ is the \emph{head}.
For brevity, in the sequel, we will omit outermost universal quantifications in TGDs.
The
\emph{exported variables} of~$\tau$ are the variables of~$\vec x$ which occur in the head.
A \emph{full TGD} is one with no existential quantifiers in the head.
A \emph{guarded TGD} (GTGD) is a TGD where $\phi$ is of the form
$A(\vec x) \wedge \phi'(\vec x)$ where $A$ is a relational atom (called the
\emph{guard}) containing 
all free variables of~$\phi'$, while a \emph{frontier-guarded TGD} (FGTGD) is one
where there is a conjunct $A$ of $\phi$ containing all the exported free variables.
An \emph{inclusion dependency} ($\incd$) is a GTGD where both $\phi$ and
$\psi$
consist of a single
atom  with no
repeated variables.
The \emph{width} of an $\incd$ is the number
of exported variables, and an ID is  \emph{unary}
(written $\uincd$) if it has width $1$.
For example,
$R(x, y) \rightarrow \exists z \, w ~ S(z, y, w)$ is a $\uincd$.

A \emph{functional dependency} (FD) is an FO sentence $\phi$ written as
$\forall \vec x \, \vec y ~ (R(x_1 \ldots x_n) \wedge R(y_1 \ldots y_n) \wedge  \left(\bigwedge_{i \in D}
x_i=y_i\right) \rightarrow  x_j=y_j)$,
with $D\subseteq\{1 \ldots n\}$ and
$j \in \{1 \ldots n\}$,
Intuitively, $\phi$ asserts that
position~$j$ is \emph{determined} by the positions of~$D$, i.e., when two
$R$-facts match on the positions of~$D$, they must match on 
position~$j$ as well. We write $\phi$ as $D \determines j$ for brevity.
The positions in $D$ are the \emph{determinant} of~$\phi$
and $j$ the \emph{determined}  position of~$\phi$.

\paragraph{Query and access model.}
We model a collection of Web services as a service schema $\aschema$,
which we simply call a 
\emph{schema}.
It consists of:
\begin{enumerate}
  \item a relational signature $\sign$;
  \item a set of integrity constraints $\Sigma$ given as
    FO sentences; and 
  \item a set of
    \emph{access methods} (or simply \emph{methods}).
\end{enumerate}
Each access method $\mt$ is associated with a relation~$R$ and
a subset of positions of~$R$ called 
the \emph{input positions} of~$\mt$.
The other positions of~$R$ are called
\emph{output positions} of~$\mt$.

In this work, we allow each access method to have an optional \emph{result bound}. 
We study two kinds of result bounds. \emph{Result upper bounds} assert that
$\mt$ returns at most $k$ matching tuples for some $k\in\NN$. \emph{Result
lower bounds} assert that, for some $k\in\NN$, the method $\mt$ returns all matching
tuples if there are no more than $k$ of them, and otherwise returns at least $k$
of the matching tuples. We call $\mt$ a \emph{result-bounded method} associated
to $k\in\NN$ if it has both a result lower bound and a result upper bound for~$k$.
We say that $\mt$ \emph{has no result bound} if it has neither a result lower
bound nor a result upper bound. In the schemas that we will consider, we will
assume that every access method is either result-bounded or has no result bound;
but we will quickly show a technical result asserting that it is sufficient to
consider result lower bounds.

An \emph{access} on an instance $I$ is a method $\mt$ on a relation
$R$ with a \emph{binding} $\accbind$ for~$I$: the binding
is a mapping from the input positions of
$\mt$ to values in~$\adom(I)$.
The \emph{matching tuples}~$M$ of the access $(\mt, \accbind)$ are the tuples
for relation~$R$ in~$I$ 
that match $\accbind$ on the input positions
of~$R$, and an \emph{output} of the access is a subset $J \subseteq M$.
We will sometimes also refer to the \emph{matching facts} of the access, i.e., $R(\vec t)$
where $\vec t$ is a matching tuple.
If the method has  no result bound, 
then
there is only one \emph{valid output} to the access, namely, the 
output $J \colonequals M$
that contains all matching
tuples of~$I$.
If there is a  result bound~$k$ on~$\mt$, then we define the notion of a
\emph{valid output} to the access as
any subset $J \subseteq M$ such that:

\begin{enumerate}[(i)]
\item $J$ has size at most $k$;
\item for any $j \leq k$, if $I$ has
$\geq j$ matching  tuples, then
  $J$ has size~$\geq j$. Formally, if $\card{M} \geq j$ then $\card{J} \geq j$.
\end{enumerate}

If there is a result lower bound of~$k$ on~$\mt$, then a \emph{valid output}
is any subset $J \subseteq M$ satisfying point (ii) above, and
similarly for a result upper bound.

We give specific names to two kinds of methods.
A method is \emph{input-free} if it has no input positions.
A method is \emph{Boolean} if
all positions are input positions. Note that accessing a Boolean method
with a binding $\accbind$ just checks if $\accbind$
is in the relation associated to the method
(and result bounds have no effect).

\paragraph{Plans.}
We use \emph{plans} to describe programs that use the access methods, formalizing
them using the terminology of~\cite{ustods,thebook}. 
In the body of the paper we will deal with \emph{monotone plans}, which are
called this way
because they define transformations that are monotone as the set of facts in an input
instance grows.
In the appendix we will extend our study to so-called \emph{RA plans}, which
define transformations that are not necessarily monotone. RA plans
and their relationship to monotone plans are described in Appendix~\ref{apx:ra}.

The \emph{monotone relational algebra operators} are:
\begin{itemize}
\item  the \emph{product} operator ($\times$), taking as input two relation instances
of arities $m$ and $n$ and returning a relation instance of arity $m+n$;
\item the \emph{union} operator ($\cup$), taking as input two relation instances of the
same arity $m$ and returning a relation instance of arity $m$;
\item the \emph{projection} operators $\pi_{A}$, where
$A$  is a finite set of positions,  taking as input a relation instance of some
fixed arity $m$ and returning the instance containing, for each $m$-tuple
$\vec t$ in the input, the  $|A|$-tuple formed from restricting $\vec t$
to positions in $A$.
\item the \emph{selection} operator ($\sigma_c$), taking as input a relation
of some arity $m$ and returning a relation instance of the same arity~$m$, where $c$
  is an equality or inequality comparing two positions $1 \leq i \leq m$ or
  comparing a position with a constant.
\end{itemize} 
The semantics of these operators are standard \cite{AHV}.
A \emph{monotone relational algebra expression} is a term built up by composing
these operators.
Monotone relational algebra expressions define the same class of queries as
positive first-order logic~---
that is, first-order logic built up from relational atoms and inequalities using the connectives
$\wedge, \vee$ and existential quantification~--- under the active-domain semantics.

A \emph{monotone plan}  $\aplan$
is a 
sequence of \emph{commands} that produce \emph{temporary tables}. There are two
types of commands:

\begin{itemize}
\item \emph{Query middleware commands}, of the form $T~\colonequals~E$, with~$T$ a temporary table
  and $E$ a monotone relational algebra expression over the temporary tables
  produced by previous commands.
\item \emph{Access commands}, written $T \Leftarrow_\outmap \mt \Leftarrow_\inmap E$,
  where $E$ is a monotone relational algebra expression over
  previously-produced temporary tables, $\inmap$ is an \emph{input
  mapping} from the output attributes of~$E$ to the input positions of~$\mt$,
  $\mt$ is a \emph{method} on some relation~$R$, $\outmap$ is an \emph{output
  mapping} from the positions of~$R$ to those of~$T$, and $T$ is a temporary
  table.
We often omit the mappings for brevity.
\end{itemize}

The \emph{output table} $T_0$ of~$\aplan$ is indicated by a special command
$\return~T_0$ at the end, with $T_0$ being a temporary table.

We must now define the semantics of~$\aplan$ on an instance~$I$.
Because of the non-determinism of result-bounded methods,
in this work we will do so relative to an
\emph{access selection} for~$\aschema$ on~$I$, i.e.,
a function $\aselect$ mapping each access $(\mt, \accbind)$ on~$I$ to a
set of facts $J \colonequals \aselect(\mt, \accbind)$ that match the access.
We say that the access selection is \emph{valid} if it
maps every access to a valid output: intuitively, the access selection
describes which
valid output is chosen when
an access to a result-bounded method matches more tuples than
the bound. Note that the definition implies that performing the same access
twice must return the same result;
however, \emph{all our
results still hold without this assumption} (see Appendix~\ref{app:idempotent}).

For every valid access selection~$\aselect$, we can now define
the semantics of each command of $\aplan$ for~$\aselect$ by considering them
in order.
For an access command $T \Leftarrow_\outmap \mt \Leftarrow_\inmap E$ in~$\aplan$,
we evaluate $E$ to get a collection $C$ of tuples. For each tuple $\vec t$
of~$C$, we use
$\inmap$ to turn it into a binding $\accbind$, and we perform the access
on~$\mt$ to obtain $J_{\vec t} \colonequals \aselect(\mt, \accbind)$. We then
take the union $\bigcup_{\vec t \in C} J_{\vec t}$ of all outputs, rename it
according to~$\outmap$, and write it in~$T$.
For a middleware query command $T \colonequals E$, we evaluate $E$ 
and write the result in~$T$.
The \emph{output} of~$\aplan$ on~$\aselect$ is then the set of tuples that are
written to
the output table~$T_0$.

The \emph{possible outputs} of~$\aplan$ on~$I$ are the outputs that can be
obtained with some valid access selection~$\aselect$. Intuitively, when we evaluate
$\aplan$, we can obtain any of these outputs, depending on which valid access
selection $\aselect$ is used.

\begin{exa}
  \label{ex:existencecheckplan}
The plan of Example~\ref{ex:existencecheck}
is as follows:\\[.3em]
\null\hfill$
T \Leftarrow \udaccess \Leftarrow  \emptyset; \qquad
T_0 \colonequals \pi_\emptyset T;
\qquad
\return~ T_0;
$\hfill\null\\[.2em]
The first command runs the relational algebra expression $E = \emptyset$
returning the empty set, giving a trivial binding for~$\udaccess$. The result of
accessing $\udaccess$ is stored in a temporary table~$T$. The second command 
projects~$T$ to the empty set of attributes, and the third command
  returns the result. For every instance~$I$, 
  the plan has only one possible output (no matter the access selection), describing if $\udirectory$ is empty
  in~$I$. We will say that the plan \emph{answers} the query~$Q_2$ of 
  Example~\ref{ex:existencecheck}.
\end{exa}

\paragraph{Answerability.}
Let $\aschema$ be a schema consisting of a relational signature, integrity
constraints, and access
methods, 
and let $Q$  be a CQ over the relational signature  of~$\aschema$.
A monotone plan $\aplan$ 
\emph{answers}~$Q$ 
under~$\aschema$ if the following holds: for all instances $I$ 
satisfying the constraints, $\aplan$ on~$I$ has exactly one possible output,
which is the query output $Q(I)$. 
In other words, this is the standard definition of answerability,
requiring that the output of~$\aplan$ 
on~$I$ is equal to~$Q(I)$, but we have extended it to our setting of result-bounded methods by
requiring that this holds for every valid access selection~$\aselect$.
Of course, $\aplan$ can have a single possible output (and answer~$Q$) even if
some intermediate command of~$\aplan$ has multiple possible outputs. 

We say that~$Q$ is
\emph{monotonically answerable} under schema $\aschema$ 
if there is a monotone plan that answers it.
Monotone answerability generalizes notions of reformulation that have been
previously studied.
In particular, in the absence of constraints and result bounds, it reduces
to the notion of a query having an \emph{executable rewriting with respect to access methods}, 
studied in 
work on access-restricted querying~\cite{access1,access2}.  In the setting where the limited
interfaces simply expose views, monotone answerability
corresponds to the well-known notion of \emph{UCQ rewriting} with respect to views~\cite{lmss}.

\paragraph{Query containment and chase proofs.}
We will reduce answerability to the standard problem of 
\emph{query containment under constraints}.
Query $Q$ is contained in query $Q'$ relative to constraints $\Sigma$
if,
in any instance that satisfies $\Sigma$, the tuples returned by
$Q$ are a subset of the tuples returned by $Q'$.
We write
$Q \subseteq_\Sigma Q'$ to denote this relationship.

In the case where $\Sigma$ consists of dependencies,
query containment under constraints
can be solved
by the well-known method of searching for a 
\emph{chase proof}~\cite{fagindataex}. We now review this notion.

Such a proof starts with
an instance called the \emph{canonical
database of~$Q$} and denoted $\canondb(Q)$: it consists of facts for each atom
of~$Q$, and its elements are the variables and constants of~$Q$. 
The proof then proceeds by \emph{firing dependencies}, as we explain next.

A homomorphism $\trig$ from the body of a dependency $\dep$ into an instance
$I_0$ is called a \emph{trigger} for~$\dep$.
A  \emph{chase step} with dependency $\dep$ and  trigger $\trig$ on $I_0$  transforms
$I_0$ to a new instance
in the following way. If $\dep$ is a TGD, the result of the chase step on~$\trig$ 
for~$\dep$ in~$I_0$ 
is the superinstance
$I_1$ of~$I_0$ obtained by adding new facts
corresponding to an extension of~$\trig$ to the head of~$\dep$, using fresh elements  to instantiate the
existentially quantified variables of the head: we denote these elements as nulls as well.
The remaining elements that occur in these new facts will be said to be \emph{exported in the
chase step}.
If $\dep$ is an FD with~$x_i=x_j$ in the head, then a chase step yields $I_1$
which is the result of identifying $\trig(x_i)$ and $\trig(x_j)$ in~$I_0$. 
A  \emph{chase sequence} for $\Sigma$ 
is a sequence of chase steps with dependencies of $\Sigma$, with the output of each
step being the input of the next. We can thus associate each such sequence with
a sequence of instances $I_0, \ldots$.

We will be particularly interested in sequences that form a  \emph{chase proof}
of~$Q \subseteq_\Sigma Q'$, where $Q$ and $Q'$ have the same free variables $\vec x$.
Recall that for a query with free variables, the free variables become elements within the
canonical database of the  query.
A chase proof of~$Q \subseteq_\Sigma Q'$ is a chase sequence
where we start by the instance $I_0 = \canondb(Q)$,
which has a homomorphism from~$Q$ to~$I_0$ sending each free variable
of~$\vec x$ to itself, and we must finish with an instance which has a homomorphism from $Q'$, 
sending each free variable of $\vec x$ to itself. 

Chase proofs give a sound and complete method for deciding containment under dependencies:
\begin{propC}[{\cite{fagindataex}}] \label{prop:chasecomplete} 
For any CQs $Q$ and $Q'$ and for any collection of dependencies
$\Sigma$, the containment $Q \subseteq_\Sigma Q'$ holds if and only if there is a chase proof
that witnesses the containment.
\end{propC}

In particular, when $\Sigma$ is empty, we obtain the usual characterization for
the containment $Q \subseteq Q'$ without constraints: it holds
if and only if there is a homomorphism from~$Q'$ to~$Q$ which is the
identity on free variables.

The variant of Proposition~\ref{prop:chasecomplete} holds also for chase proofs based
on so-called \emph{restricted chase sequences}, which we now define.
A trigger $\trig$ for a dependency $\dep$ is \emph{active} in an instance~$I$ if it cannot be extended to a homomorphism from the
head of~$\dep$ to~$I$. In other words, an active trigger $\trig$ witnesses the fact
that~$\dep$ does
not hold in~$I$.  A \emph{restricted chase sequence} is one in which all chase steps have active triggers.

If all restricted chase sequences starting with a given initial instance  $I_0$
are finite, we say that the \emph{restricted chase with~$\Sigma$ 
terminates on that instance}.
In this case, we define \emph{the restricted chase  of $I_0$ with
$\Sigma$} as the result of iteratively applying all active triggers according to
some arbitrary order. When $I_0=\canondb(Q)$ we will talk about the restricted chase of~$Q$.

When the phenomenon above occurs for each finite initial instance $I_0$, we
say that $\Sigma$ \emph{has terminating chase}.
If $\Sigma$ has terminating chase we can decide if $Q \subseteq_\Sigma Q'$
by computing the
chase of $Q$ with $\Sigma$ and then
searching for a homomorphism of $Q'$ into the chase.

Even when the chase does not terminate, we define the 
\emph{chase  of $I_0$ with $\Sigma$}
as the infinite fixpoint of applying chase steps following some
arbitrary order which is \emph{fair}, i.e., ensuring that every active trigger will eventually be fired. We similarly define the \emph{restricted chase of $I_0$ with $\Sigma$} in the
same way but with restricted chase steps.  
We can still use the chase and restricted chase to reason about query containment, even though it is an infinite
object that cannot generally be materialized. The result is implicit in  \cite{fagindataex}; see
\cite{onet} for a more detailed exposition.

\begin{prop} 
  For any CQs $Q$ and $Q'$ and for any collection of dependencies $\Sigma$, the
  containment $Q \subseteq_\Sigma Q'$ holds if and only if there is a
  homomorphism from $Q'$ to the chase of~$I_0 = \canondb(Q)$ with~$\Sigma$,  with the homomorphism
being the identity on free variables. The
  same holds for the restricted chase.
\end{prop}

\paragraph{Certain answer problems and TGD implication problems via the chase.}
We say that  a set of first-order sentences $\lambda$ \emph{entails} a first-order sentence
$\rho$, written $\lambda \models \rho$, if every  instance satisfying $\lambda$ also satisfies
$\rho$.
Note that a query containment 
$Q \subseteq_\Sigma Q'$ for Boolean queries $Q$ and~$Q'$ is a special case of an entailment, of the form
$Q \wedge \Sigma \models Q'$.

We will also study another restricted kind of entailment problem, of the form:
\[
\bigwedge_{i \leq n} A_i \wedge \Sigma \models Q
\]
where $\Sigma$ is a set consisting of TGDs and
FDs, each $A_i$ is a fact, and $Q$ is a
CQ\@.  
This is  the problem of \emph{certain answers}~\cite{fagindataex}
 under dependencies for CQs, and we will also use it when discussing the
 implication of some facts from other facts and constraints in
 Section~\ref{sec:linearize}.
We can consider a modification of the definition
of chase proof to solve this problem: this is a chase sequence where we fix the initial instance to  be
$\{ A_1 \ldots A_n \}$, rather than the canonical database of $Q$.
When $\Sigma$ only contains TGDs, there are well-known reductions between the query containment problem 
and the certain answers problem, and in particular the analog of 
Proposition~\ref{prop:chasecomplete} holds for certain answer problems: the
entailment holds iff there is a chase proof witnessing it~\cite{fagindataex}.
Based on these equivalences, we 
freely use known upper and lower complexity bounds stated on the certain answer problem
 (e.g., from~\cite{taming,baget2010walking}) and apply them to query containment under
constraints.

Another special case of entailment is  \emph{entailment of a TGD $\tau$
by a set of TGDs $\Sigma$}. This problem can be reduced to the query
containment problem: we take the body of $\tau$ and see if it is contained
in the head of $\tau$ relative to $\Sigma$. Thus chase proofs
also give a complete method for deciding these entailments.

Note that the reader familiar with  the treatment of chase steps
involving FDs  will find
 our discussion a bit simplified
relative to standard accounts (e.g., \cite{AHV}). In other accounts there is the possibility
that a chase step ``fails'', but in our setting  --- e.g., due to our treatment
on constants and the restriction on their use in constraints ---
 we will not need to consider this.

A set $S$ of elements in an instance is \emph{guarded} if there is a fact of the instance
that contains all these elements. We call such an element a \emph{guard for $S$}.
We note that if $\tau$ is a trigger for a guarded TGD
$\dep$, then the image of $\tau$ must be guarded.

\paragraph{Equivalent formalisms for monotone plans.}
Monotone plans have a number of other presentations.
For instance, if there is only one access per relation,
they are equivalent to  \emph{executable UCQs} which just annotate
each atom of a UCQ with an access method. The semantics is just to execute the method
corresponding to each atom in the order that the atoms are given, accumulating all the bindings.
Executable queries were the
first formalism to
implement queries with access methods \cite{ullmanbook,LiCh00,LiCh01,nash2004edbt,nash2004pods}.
They were considered only in the case of CQs where there is a single access method, without
result bounds, for each relation symbol.
The equivalence with monotone plans is proved
in \cite{thebook}, and extends easily to the presence of multiple access per methods,
and to the presence of result bounds with any of the semantics
we consider in this work.

\begin{exa}
  \label{exa:planandexecutable}
Let us consider the plan mentioned in Example~\ref{exa:plani}.
In our monotone plan syntax it would be written
as 
\begin{align*}
T \Leftarrow \udaccess \Leftarrow \emptyset  ; \qquad
 T_0 \Leftarrow \praccess  \Leftarrow T ; \qquad T_1 \colonequals \pi_{\attrfmt{name}} \sigma_{\attrfmt{salary}=10000}T_0 ; \qquad \return ~ T_1 
\end{align*}
As an executable CQ, this would be expressed simply as
\begin{align*}
\udirectory(\attrfmt{id}), \profinfo(\attrfmt{id}, \attrfmt{name}, 10000)
\end{align*}
\end{exa}

\paragraph{Variations of answerability.}
So far, we have defined monotone answerability. An alternative 
notion
is \emph{RA answerability}, defined using \emph{RA plans} that
allow arbitrary relational algebra expressions in commands.
We think this notion is less natural for CQs and for the class of constraints
that we consider.
Indeed, CQs are monotone: if facts are added to an instance, the output of a
CQ cannot decrease.
Thus the bulk of prior work on implementing CQs over restricted interfaces, both in theory
\cite{lmss,dln,access1,access2,antoinejulien} and in practice 
\cite{candbsigmod14,candbsigrecord,romerocikm}, has focused on monotone 
implementations, often phrased in terms of the executable query syntax mentioned above.
In fact, even in the setting of views,  it was initially assumed that if a CQ can be answered
at all, it must have a monotone plan \cite{SVconf,lmss}. Clever counterexamples
to this fact were only found much later~\cite{NSV}.
In the body of the paper,
we follow the earlier tradition and we focus exclusively on monotone plans. 
Nevertheless, \emph{many of our results extend to answerability with RA plans} (see
Appendix~\ref{apx:ra}). For instance, we can sometimes show that monotone answerability
and RA answerability coincide. We discuss the status of monotone vs
relational algebra plans  further in Section~\ref{sec:conc}.

\section{Reducing to Query Containment} \label{sec:reduce}

We start our study of the monotone answerability problem by
reducing it to \emph{query
containment under constraints},  defined in the previous section.
We explain in this section how this reduction is done. It 
extends the approach
of~\cite{dln,ustods,thebook} to result bounds, and follows the connection
between answerability and determinacy notions of~\cite{NSV,thebook}.
To design this reduction, we will need to show that monotone answerability is
equivalent to a notion of \emph{access monotonic-determinacy}, already studied
in the literature for access methods without result bounds, which we extend to
our setting with result bounds. This characterization (Theorem~\ref{thm:equiv})
will be used many times in the sequel.

\subsection{Access Monotone Determinacy and Equivalence to Monotone Rewritability}
The 
query containment problem corresponding to monotone answerability will capture
the idea that \emph{if 
an instance $I_1$
satisfies a query $Q$ and another instance $I_2$ has more ``accessible data'' than~$I_1$, then $I_2$
should satisfy\/ $Q$ as well}. Here the accessible data means the data
that can be retrieved by iteratively performing accesses. The motivation
is that if we have a monotone plan and the accessible data increases, then
the output of the plan can only increase. 
We will first define accessible data via the notion of
\emph{accessible part}. We
use this to 
formalize the previous idea as access monotonic-determinacy,
and as we claimed we show that it is
equivalent to monotone answerability (Theorem~\ref{thm:equiv}). 
Using access monotonic-determinacy we
show that we can simplify the result bounds of arbitrary schemas, and
restrict to \emph{result lower bounds} throughout this work.
We close the section by showing how to rephrase
access monotonic-determinacy with result lower bounds as
query containment under constraints.

\paragraph{Accessible parts.}
We first formalize the notion of ``accessible
data''. 
Given a schema $\aschema$ with  methods that may have result lower bounds
and also result upper bounds, along with an instance~$I$, an \emph{accessible part} of $I$
is any subinstance obtained by
iteratively making accesses until we
reach a fixpoint.
Formally, we define an accessible part
by choosing an access selection $\aselect$ which is valid for the upper and lower bounds and
 inductively
defining sets of facts $\accpart_i(\aselect, I)$
 and sets of values $\accessible_i(\aselect, I)$ by:
\begin{align*}
  \accpart_0(\aselect, I) & \colonequals \emptyset \mathrm{~~and~~}
  \accessible_0(\aselect,I) \colonequals \emptyset\\
  \accpart_{i+1}(\aselect, I)& \colonequals \!\!\!\!\!\!\!\!\!\!\!\!\!\!\!\!\!\!\!\!\!\!\!\!\!\!\!\!\!\!\!\!\!\!\!\bigcup_{\substack{\mt \mbox{ method},\\[.2em] \accbind
  \mbox{ binding with values in~} \accessible_i(\aselect,I)}} \!\!\!\!\!\!\!\!\!\!\!\!\!\!\!\!\!\!\!\!\!\!\!\!\!\!\!\!\!\!\!\!\!\!\!\!\!\!\!\!\aselect(\mt, \accbind)\qquad \\
   \accessible_{i+1}(\aselect,I) & \colonequals  \adom(\accpart_{i+1}(\aselect, I))
\end{align*}
These equations define by mutual induction the
set of values
($\accessible$)
that we can retrieve by iterating accesses 
and the set of facts
($\accpart$)
that we can retrieve using those values.

The \emph{accessible part} under~$\sigma$, written $\accpart(\aselect,I)$,
is then defined as~$\bigcup_i \accpart_i(\aselect,I)$.
As the equations are monotone, this fixpoint is reached
after finitely many iterations if $I$ is finite,
or as the union of all finite iterations if $I$ is infinite.
When there are no result bounds, there is only one valid access selection~$\aselect$, so
only one accessible part: it intuitively
corresponds to the data that can be accessed using the methods.
In the presence of result bounds, 
there can be many 
accessible parts, depending on~$\aselect$, and thus we refer to ``an accessible part of  instance
$I$'' to mean  an accessible part for some selection function.

\paragraph{Access monotonic-determinacy.}
We now formalize the idea that a query~$Q$ is ``monotone under accessible
parts''.
Let $\Sigma$ be the integrity constraints of~$\aschema$.
We call~$Q$ \emph{access monotonically-determined} in~$\aschema$
(or $\amd$, for short),
 if for any two instances $I_1$, $I_2$ satisfying~$\Sigma$,
 if there is an accessible part of $I_1$
 that is a subset of an accessible part of  $I_2$, then
$Q(I_1) \subseteq Q(I_2)$.
Note that when there are no result bounds, there is a unique accessible part of $I_1$ and of $I_2$, and
$\amd$ says that when the accessible part grows, then $Q$ grows.

In the sequel, it will be more convenient to use an alternative definition
of~$\amd$, based on the notion of \emph{access-valid} subinstances.
A subinstance $I_\acc$ of~$I_1$ is \emph{access-valid in~$I_1$} for~$\aschema$ if,
for any access $(\mt, \accbind)$ performed with a method $\mt$ of~$\aschema$
and with a binding $\accbind$ whose values are in~$I_\acc$,
there is a set $J$ of matching
tuples in $I_\acc$ such that $J$ is a valid output to the access $(\mt,
\accbind)$ in~$I_1$.
In other words, for any access performed on $I_\acc$, we can choose an output
in~$I_\acc$ which is also a valid output to the access in~$I_1$.
We can use this notion to rephrase the definition of~$\amd$ to talk about 
a common subinstance of $I_1$ and $I_2$ that is access-valid:

\newcommand{\propaltdef}{
For any schema $\aschema$ with arbitrary constraints~$\Sigma$  and methods that can have
result lower bounds and result upper bounds, 
a CQ $Q$ is $\amd$
 if and only if the following implication holds: for any two instances $I_1$, $I_2$ satisfying $\Sigma$, if $I_1$
 and $I_2$ have a common subinstance $I_\acc$ that
is access-valid in~$I_1$, then
$Q(I_1) \subseteq Q(I_2)$.
}

\begin{prop}  \label{prop:altdef}
  \propaltdef
\end{prop}

To show this, 
it suffices to show that the two definitions of ``having more accessible data''
agree. Proposition~\ref{prop:altdef} follows immediately from the following
proposition:

\begin{prop}
  \label{prop:equivalence}
The following are equivalent:
 \begin{enumerate}[(i)]
\item $I_1$ and $I_2$ have a common subinstance $I_\acc$ that
is access-valid in~$I_1$.
\item There  are $A_1 \subseteq A_2$ such that $A_1$  is an accessible part
for~$I_1$ and~$A_2$ is an accessible part for~$I_2$.
\end{enumerate}
\end{prop}

\begin{proof}
Suppose  $I_1$ and $I_2$ have a common subinstance $I_\acc$ that is access-valid
  in~$I_1$.   Since $I_\acc$ is access-valid in $I_1$ there is  an  access selection~$\aselect_1$ 
  which maps any access performed with values of~$I_\acc$ to some 
  set of matching tuples in~$I_\acc$, with $\aselect_1$  valid  in~$I_1$. We can extend $\aselect_1$
to be valid in $I_1$ by choosing tuples arbitrarily for accesses with bindings not in $I_\acc$.
  We then extend $\aselect_1$ to an access selection~$\aselect_2$ which returns a superset of the
  tuples returned by $\aselect_1$ for accesses with values of~$I_\acc$, and returns
  an arbitrary set of tuples from~$I_2$ otherwise, such that this output to the
  access is valid in~$I_2$. We only need to modify $\aselect_1$ when
the full set of matching tuples of an access in $I_1$ is below a method's lower bound, but there are more matching
tuples in $I_2$: in this case we just add enough matching tuples from $I_2$ to achieve the upper bound,
or add all the matching tuples in $I_2$ if the number is still within the method's upper bound.
This ensures that
  $\accpart(\aselect_1,I_1) \subseteq\accpart(\aselect_2,I_2)$, so that (i)
  implies (ii).

  Conversely, assuming point (ii), let $\sigma_1$ and $\sigma_2$ be the access
  selections used to define the accessible parts $A_1$ and $A_2$, so that 
$\accpart(\aselect_1, I_1) \subseteq \accpart(\aselect_2, I_2)$.
Let $I_\acc \colonequals \accpart(\aselect_1, I_1)$, and let us show that
$I_\acc$ is a common subinstance of~$I_1$ and~$I_2$ that is access-valid
in~$I_1$.
By definition, we know that $I_\acc$ is a subinstance of~$I_1$, and by
  assumption we have $I_\acc \subseteq A_2 \subseteq I_2$, so indeed $I_\acc$ is
  a common subinstance of~$I_1$ and $I_2$. Now, to show that it is access-valid
  in~$I_1$, consider any access $(\mt, \abind)$
  with values in~$I_\acc$. We know that there is $i$ such
that~$\abind$ is in the domain of $\accpart_i(\aselect_1, I_1)$ --- that is in
$\accessible_i(I_1)$.  So by definition of the
  fixpoint process and of the access selection~$\aselect_1$ there is a valid
  output  that is a subset of the facts within $\accpart_{i+1}(\aselect_1, I_1)$, hence a subset of 
the facts within $I_\acc$.
  Thus, $I_\acc$ is access-valid. This shows the converse
  implication, and concludes the proof.
\end{proof}

The alternative definition of~$\amd$ in Proposition~\ref{prop:altdef} is more convenient, because it only deals
with a subinstance of~$I_1$ and not with accessible parts. Thus, 
\emph{we will use this characterization of monotone answerability in  the rest of
this paper}.
Now, the usefulness of~$\amd$ is justified by the following result:

\newcommand{\thmequiv}{
For any CQ~$Q$ and schema $\aschema$ containing only constraints
in active-domain first-order logic, with access methods
that may have result upper and lower bounds, the following are equivalent:
\begin{enumerate}
\item $Q$ is monotonically answerable \wrt\ $\aschema$.
\item 
$Q$ is $\amd$ over~$\aschema$.
\end{enumerate}
}
\begin{thm}
  \label{thm:equiv}
  \thmequiv
\end{thm}
Without result bounds, this equivalence of monotone answerability and
access monotonic-determinacy is proven in \cite{ustods,thebook}, using a variant of Craig's interpolation theorem.
Theorem~\ref{thm:equiv} shows that the equivalence extends to schemas with
result bounds.

We now begin the proof of Theorem~\ref{thm:equiv}, which will use
Proposition~\ref{prop:altdef}.
We first prove the ``easy direction'':

\begin{prop} \label{prp:plantoproof}
Assume that our schema has arbitrary constraints along with methods that may
have both upper and lower bounds.
If a CQ~$Q$ has a  (monotone) plan $\aplan$ that answers it \wrt\ $\aschema$, then
$Q$ is $\amd$ over~$\aschema$.
\end{prop}
\begin{proof}
  We use the definition of $\amd$ given in Proposition~\ref{prop:altdef}.
Assume that there are two instances $I_1, I_2$ satisfying the constraints
  of~$\aschema$ and that there is a common subinstance $I_\acc$ 
  that is access-valid in~$I_1$. Let us show that $Q(I_1) \subseteq Q(I_2)$.
  As $I_\acc$ is access-valid, let $\aselect_1$ be a valid access selection
  for~$I_\acc$: for any access with values in~$I_\acc$, the access selection~$\aselect_1$ returns an output which is valid in~$I_\acc$. We extend $\aselect_1$
  to a valid access selection for~$I_2$ as in the proof of
  Proposition~\ref{prop:altdef}: for accesses in~$I_\acc$, the access selection~$\aselect_2$ returns a superset of~$\aselect_1$, which is possible because $I_\acc
  \subseteq I_2$, and for other accesses it returns some valid subset of tuples
  of~$I_2$.

We  argue that  
for each temporary table of~$\aplan$, its value
when  evaluated on~$I_1$ with~$\aselect_1$, is contained
in its value when evaluated on~$I_2$ with~$\aselect_2$.
 We prove this by induction on~$\aplan$. As the plan is monotone, the property
  is preserved by query middleware commands, so inductively it suffices to look at an
access command  $T \Leftarrow \mt \Leftarrow E$ with~$\mt$ an access method on
some relation $R$. Let $E_1$ be the value of~$E$ when evaluated on~$I_1$
  with~$\aselect_1$,
and let $E_2$ be the value when evaluated on~$I_2$ with~$\aselect_2$. Then by the monotonicity
of the query $E$ and the induction hypothesis, we have $E_1 \subseteq E_2$.
Now, given a tuple $\vec t$ in~$E_1$,  let  $M^1_{\vec t}$ be the
set of tuples selected by $\aselect_1$ for the access with $\mt$ using $\vec t$
in $I_1$. Similarly let $M^2_{\vec t}$ be the set
selected by $\aselect_2$ in~$I_2$.  By construction of~$\aselect_2$, we have $M^1_{\vec t} \subseteq M^2_{\vec t}$,
and thus $\bigcup_{\vec t \in E_1} M^1_{\vec t} \subseteq \bigcup_{\vec t \in
  E_1} M^2_{\vec t}$,
which  completes the induction.

  Thanks to our
  induction proof, we know that the output of~$\aplan$ on~$I_1$ with~$\aselect_1$ is a
  subset of the output of~$\aplan$ on~$I_2$ with~$\aselect_2$. As we have assumed
  that $\aplan$ answers~$Q$ on~$\aschema$, this means that $Q(I_1) \subseteq
  Q(I_2)$, which is what we wanted to show.
\end{proof}

To prove the other direction of Theorem~\ref{thm:equiv}, we first recall the result that corresponds to
Theorem~\ref{thm:equiv} in the case without result upper and lower bounds:

\begin{thmC}[\cite{thebook,ustods}]
  \label{thm:mdetermandplansclassic}
  For any CQ~$Q$ and schema $\aschema$ (with no result bounds)
  whose constraints $\Sigma$ are expressible
  in active-domain first-order logic, the following are equivalent:
\begin{enumerate}
\item $Q$ has a  monotone plan that answers it over~$\aschema$.
\item $Q$ is $\amd$ over~$\aschema$.
\end{enumerate}
\end{thmC}

The theorem above holds even for more general relational algebra queries, but
we will not require this generality in this work.
Thus, for schemas without result-bounded  methods, the existence
of a monotone plan is the same as~$\amd$, and both can be expressed as a query containment problem.
It is further shown in~\cite{ustods} that a monotone plan can be extracted from
any proof of the query containment for~$\amd$.
This reduction to query containment is what we will now extend to the setting
with result-bounded methods. Specifically,
we will lift the above result to the setting with result-bounded methods via a simple
construction that allows us
to rewrite away the result-bounded methods by expressing them in the
constraints: we call this \emph{axiomatizing} the result-bounded methods.

\paragraph{Replacing result bounds on methods with additional constraints.}
Given a schema $\aschema$ with constraints and access methods, possibly
with result upper and lower bounds, we will define an auxiliary schema $\elimnd(\aschema)$ without
result bounds.
The schema $\elimnd(\aschema)$ includes the relational signature $\aschema$, and for
every method $\mt$ with  result bound~$k$ on relation~$R$  we also have a new relation
$R_\mt$ whose arity agrees with that of~$R$. Informally, $R_\mt$ stores the
tuples returned by the access selection for~$\mt$.
The constraints include all the constraints of~$\aschema$ (on the original
relation names).
In addition, for every method $\mt$ with input positions $i_1 \ldots i_m$
we have the following constraints, which we call \emph{axioms}:

\begin{itemize}
\item An
  axiom stating that~$R_\mt$ is a subset of
  $R$.
\item If $\mt$ has a result upper bound of $k$, we have
an  axiom stating that for any binding of the input positions,
$R_\mt$ has at most $k$ distinct matching tuples.
\item If $\mt$ has a lower bound of $k$ then for each $1 \leq j \leq k$ we have
  a \emph{result lower bound axiom} stating that,
for any values $c_{i_1} \ldots c_{i_m}$, if
$R$
    contains at least $j$ matching tuples (i.e., tuples $\vec c$ that extend
    $c_{i_1} \ldots c_{i_m}$), then $R_\mt$ contains at least $j$  such tuples. 
\end{itemize}

In this schema we adjust the access methods of the original
schema,  \emph{removing any access method
$\mt$ with a  result upper or lower bound over~$R$, and in its place 
adding an access method
with no  result bound over~$R_\mt$}.

Given a query $Q$ over  $\aschema$, we can consider it as a query  over
$\elimnd(\aschema)$ instances by simply ignoring the additional relations.

We claim that, in considering $Q$ over~$\elimnd(\aschema)$ rather than $\aschema$,
we do not change monotone answerability.

\begin{prop} \label{prop:elimresultboundplan}
Let $\aschema$ be a schema with active-domain first order constraints, result upper bounds and result lower bounds.
For any CQ~$Q$ over~$\aschema$, there is a monotone plan that answers~$Q$
  over~$\aschema$ iff there is a monotone plan that answers~$Q$ over~$\elimnd(\aschema)$.
\end{prop}

In other words, we can axiomatize 
result upper and lower bounds, at the cost of including new constraints. 

\begin{proof}
Suppose that there is a monotone plan $\aplan$ over~$\aschema$ that answers $Q$. Let
$\aplan'$ be formed from~$\aplan$ by replacing every access
with method $\mt$ on relation~$R$ with an access to~$R_\mt$ with the
  corresponding method.
We claim that~$\aplan'$ answers $Q$ over~$\elimnd(\aschema)$.
Indeed, given an instance $I'$ for~$\elimnd(\aschema)$, we can drop the relations
$R_\mt$ to get an instance $I$ for~$Q$, and use the relations $R_\mt$ to define
a valid access selection~$\aselect$ for each method of~$\aschema$, and we can
show that~$\aplan$ evaluated with~$\aselect$ over~$I$ gives the same output as
$\aplan'$ over~$I$. Since the former evaluates to~$Q(I)$, so must the latter.

Conversely,
suppose that there is a monotone plan $\aplan'$ that answers $Q$ over~$\elimnd(\aschema)$.
Construct $\aplan$  from~$\aplan'$ by replacing accesses to~$R_\mt$ with accesses to~$R$.
We claim that~$\aplan$ answers $Q$ over~$\aschema$.
To show this, consider
an instance $I$ for~$\aschema$, and a particular valid access selection~$\aselect$,
  and let us show that the evaluation of~$\aplan$ on~$I$ following~$\aselect$
  correctly answers~$Q$.
We 
build an instance $I'$ of~$\elimnd(\aschema)$ by copying~$I$ and interpreting each $R_\mt$ as follows:
  for each tuple $\vec{t}$ such that $R(\vec{t})$ holds in~$I$, project
  $\vec{t}$ on the input positions $i_1 \ldots i_m$
of~$\mt$, and include all of the outputs of this access according to~$\aselect$
  in~$R_\mt$. As the outputs of accesses according to~$\aselect$  must be valid,
  $I'$
  must satisfy the
  constraints of~$\elimnd(\aschema)$.
  We define a valid access selection~$\aselect'$
  from~$\aselect$ so that every access on~$R_\mt$ returns the output of the
  corresponding access on~$R$ according to~$\aselect$.
Since $\aplan'$ answers $Q$,
  we know that evaluating $\aplan'$ on~$I'$ with~$\aselect'$ yields the output~$Q(I')$ of~$Q$
  on~$I'$.
Now, the definition of~$\aselect'$ ensures that the accesses made by $\aplan'$ on~$I'$
  under~$\aselect'$ are exactly the same
as those made by $\aplan$ on~$I$ under~$\aselect$, and that the outputs of these
accesses are the same.
Thus $\aplan$ evaluated 
on~$I$ under~$\aselect$ gives the same result as~$\aplan'$ does on~$I'$
  under~$\aselect'$, namely, $Q(I')$.
Now, $Q$ only uses the original relations of~$\aschema$, so the definition
  of~$I'$ clearly implies that $Q(I') = Q(I)$, so indeed the evaluation
  of~$\aplan$ on~$I$ under~$\aselect$ returns~$Q(I)$. As this holds for any
  valid access selection~$\aselect$, we have shown that~$\aplan$ answers~$Q$
  over~$\aschema$, the desired result.
\end{proof}

The equivalence of a schema $\aschema$ with
result bounds and its variant $\elimnd(\aschema)$ easily
extends  to~$\amd$.

\begin{prop} \label{prop:elimresultbounddet}
For any CQ~$Q$ over~$\aschema$, 
$Q$
is $\amd$ over $\elimnd(\aschema)$ if and only if
$Q$ is $\amd$ over~$\aschema$.
\end{prop}

\begin{proof}
  For the forward direction,
  assume $Q$ that is $\amd$ over~$\elimnd(\aschema)$, and let us show that $Q$ is
  $\amd$ over~$\aschema$.
  We use the characterization of $\amd$ in terms of access-valid
subinstances given in Proposition~\ref{prop:altdef}.
  Let 
  $I_1$ and $I_2$ be instances satisfying the constraints of~$\aschema$, 
  and let $I_\acc$ be a common subinstance of~$I_1$ and~$I_2$ which is
  access-valid in~$I_1$ for~$\aschema$. Let $\aselect_1$ be a valid access selection
  for~$I_\acc$. 
  As in the proof of Proposition~\ref{prop:altdef},
  we can extend it to an access selection $\aselect_2$ for~$I_2$ 
that ensures that every access
  with $\aselect_2$ returns a superset of the tuples obtained with~$\aselect_1$.
  We now extend~$I_1$ into an instance~$I_1'$ for~$\elimnd(\aschema)$ by
  interpreting each~$R_\mt$ as the union of the
outputs given by~$\aselect_1$ over every
  possible access with~$\mt$ on~$I_\acc$, as in the proof of
  Proposition~\ref{prop:elimresultboundplan}. 
We define $I_2'$ from $I_2$ and
  $\aselect_2$ in the same way. As the access outputs given by~$\aselect_1$ and
  $\aselect_2$ must be valid, we know that $I_1'$ and $I_2'$ satisfy the
  new constraints of~$\elimnd(\aschema)$, and clearly they still satisfy the
  constraints of~$\aschema$. Now extend $I_\acc$ to~$I_\acc'$ by adding all
  $R_\mt$-facts of~$I_1'$ for all~$\mt$. Clearly $I_\acc'$ is a subinstance
  of~$I_1'$.  It is access-valid because $I_\acc$ was access-valid. 
It is a subinstance
  of~$I_2'$ because $I_\acc$ is a subinstance of~$I_2'$ and because the
  $R_\mt$-facts in~$I_1'$ also occur in~$I_2'$ by construction of~$\aselect_2$.
  Thus, because $Q$ is $\amd$ over~$\elimnd(\aschema)$, we know that $Q(I_1') \subseteq
  Q(I_2')$. Now, as $Q$ only uses the relations in~$\aschema$, we have 
  $Q(I_1)=Q(I'_1)$ and $Q(I_2)=Q'(I'_2)$, so we have shown that $Q(I_1) \subseteq
  Q(I_2)$, concluding the forward direction.

Conversely, suppose $Q$ is $\amd$ over~$\aschema$
and consider instances  $I'_1$ and $I'_2$ for $\elimnd(\aschema)$
with valid access selections $\aselect'_1$ and $\aselect'_2$ giving accessible parts $A'_1 \subseteq A'_2$.
  We create  an instance $I_1$ for
$\aschema$ from~$I'_1$ by dropping the relations $R_\mt$, and similarly
create $I_2$ from~$I'_2$. Clearly both satisfy the constraints of~$\aschema$.
We modify $\aselect'_1$ to obtain an access selection~$\aselect_1$ for~$I_1$:
for every access on~$I_1$ with a method~$\mt$, the output is that of the
corresponding access with~$\aselect_1'$ on~$R_\mt$.  We do the same to build
$\aselect_2$ from~$\aselect_2'$. 
  By the additional axioms of $\elimnd(\aschema)$,
  it is clear that these access selections are
valid. That is, that they return valid outputs to any access.  
And letting
$A_1$ and~$A_2$ be the corresponding accessible parts of~$I_1$ and~$I_2$, it is
clear that $A_1 \subseteq A_2$. Thus, because $Q$ is $\amd$ over~$\aschema$, we
know that $Q(I_1) \subseteq Q(I_2)$, and again we have $Q(I_1) = Q(I_1')$ and
$Q(I_2) = Q(I_2')$. So we have $Q(I_1') \subseteq Q(I_2')$, which concludes the proof.
\end{proof}

Putting together Proposition~\ref{prop:elimresultboundplan}, Proposition
\ref{prop:elimresultbounddet} and Theorem~\ref{thm:mdetermandplansclassic},
 we have completed the proof of Theorem~\ref{thm:equiv}.

\subsection{Elimination of result upper bounds}
The characterization of monotone answerability in terms of~$\amd$ allows
us to prove a key simplification in the analysis of result bounds.
Recall that a result bound of~$k$
declares both an \emph{upper bound} of~$k$ on the number of returned results, and
a \emph{lower bound} on them: for all $j \leq k$,
if there are $j$ matches, then
$j$ must be returned. 
We can show that the upper bound makes no difference for monotone answerability.
Formally, for a schema $\aschema$  with
integrity constraints and access methods, some of which
may be result-bounded, we define
the schema $\relaxs(\aschema)$. It has the same vocabulary, constraints,
and access methods as in~$\aschema$.  For each access method
$\mt$ in~$\aschema$ with
result bound of~$k$, $\mt$ has instead a \emph{result lower bound} of~$k$ in
$\relaxs(\aschema)$, i.e., $\mt$ does not impose the upper bound.
We can then show:

\newcommand{\elimupper}{
 Let $\aschema$ be a schema with
arbitrary constraints and access methods which may be result-bounded.
A CQ~$Q$ is monotonically answerable in~$\aschema$ if and only if it is monotonically answerable
in~$\relaxs(\aschema)$.
}
\begin{prop} \label{prop:elimupper}
  \elimupper
\end{prop}

\begin{proof}
  We show the result for $\amd$ instead of monotone answerability,
  thanks to Theorem~\ref{thm:equiv}, and use Proposition~\ref{prop:altdef}.
  Consider arbitrary instances $I_1$ and
  $I_2$ that satisfy the constraints, and let us show that any common subinstance
  $I_\acc$ of~$I_1$ and $I_2$ is access-valid in~$I_1$ for~$\aschema$ iff
  it is access-valid in~$I_1$ for~$\relaxs(\aschema)$: this implies the claimed
  result.

  In the forward direction, if $I_\acc$ is access-valid in~$I_1$ for
  $\aschema$, then clearly it is access-valid in~$I_1$
  for~$\relaxs(\aschema)$, as any output of an access on~$I_\acc$ which is
  valid in~$I_1$ for~$\aschema$ is also valid for~$\relaxs(\aschema)$.

  In the backward direction, assume  $I_\acc$ is access-valid in~$I_1$
  for~$\relaxs(\aschema)$, and consider an access $(\mt, \allowbreak\accbind)$
  with values of~$I_\acc$.
  If $\mt$ has no result lower bound, then there is only one possible output for the
  access, and it is also valid for~$\aschema$.
  Likewise, if $\mt$ has a result lower bound of~$k$ and there are $\leq k$ matching
  tuples for the access, then the definition of a result lower bound ensures
  that there is only one possible output, which is again valid for~$\aschema$.
  Finally, if there are $>k$ matching tuples for the access,
  we let $J$ be a set of tuples in~$I_\acc$ which is is a valid output
  to the access in~$\relaxs(\aschema)$, and take any subset $J'$ of~$J$ with
  $k$ tuples; it is clearly a valid output to the access for~$\aschema$. This
  establishes the backward direction, concluding the proof.
\end{proof}

Thanks to this, in our study of monotone answerability in the rest of the paper, 
\emph{we only consider result lower bounds}.

\subsection{Reducing to query containment}
Now that we have reduced our monotone answerability problem to~$\amd$, and
eliminated result upper bounds, we explain how to
restate $\amd$ as a query containment 
problem, which was our original goal in this section.
To do so, we will expand the relational signature.
We let $\accessible$ be a new unary relation, and for
each relation~$R$ of the original signature,
we introduce two copies $R_\acc$ and $R'$ with the same
arity as~$R$.
Letting $\Sigma$  be the integrity constraints
in the original schema, we let $\Sigma'$ 
be formed
by replacing every relation~$R$ with~$R'$. For any CQ $Q$, we define
$Q'$ from~$Q$ in the same way.
Intuitively, $R$ and $R'$ represent the interpretations of the relation~$R$ in~$I_1$ and~$I_2$;
$R_\acc$ represents the interpretation of~$R$ in~$I_\acc$; and $\accessible$ represents the active  domain
of~$I_\acc$.

The \emph{$\amd$ containment for~$Q$ and $\aschema$} is
then the CQ containment $Q \subseteq_\Gamma Q'$ for
constraints $\Gamma$ that we will define shortly.
Intuitively, $\Gamma$ will include the original constraints $\Sigma$, and the
analogue $\Sigma'$ of $\Sigma$ on the relations~$R'$, to enforce that~$I_1$
and $I_2$ both satisfy $\Sigma$.
Further, $\Gamma$ will include additional constraints called \emph{accessibility
axioms}. These axioms will enforce that $I_\acc$ is access-valid
in~$I_1$, i.e., that any access performed with values for~$I_\acc$ returns a
valid output which is in~$I_\acc$; and enforce that $I_\acc$ is a common
subinstance of~$I_1$ and~$I_2$.

Formally, $\Gamma$ includes the original constraints~$\Sigma$, the
constraints $\Sigma'$ on the relations $R'$, and the following 
\emph{accessibility axioms}:

\begin{itemize}
\item For each method $\mt$ that is not result-bounded, letting $R$ be
  the relation accessed by~$\mt$:
  \[\Big(\bigwedge_i \accessible(x_i)\Big) \wedge
  R(\vec x, \vec y) \rightarrow R_\acc(\vec x, \vec y)\]
where $\vec x$ denotes the input positions of~$\mt$ in~$R$.
\item For each method $\mt$ with a result lower bound of~$k$, 
  letting $R$ be the relation accessed by~$\mt$, 
  for all $j \leq k$:
  \[\Big(\bigwedge_i \accessible(x_i)\Big) \wedge \exists^{\geq j} \vec y ~ 
  R(\vec x, \vec y)
  \rightarrow \exists^{\geq j} \vec z ~ R_\acc(\vec x, \vec z)\]
where $\vec x$ denotes the input positions of~$\mt$ in~$R$.
Note that we write
  $\exists^{\geq j} \vec y ~ \phi(\vec x, \vec y)$ for a subformula $\phi$
  to mean that there exist at least $j$ different values of~$\vec y$
  such that $\phi(\vec x, \vec y)$ holds.
\item For every relation $R$ of the original signature:
  \[R_\acc(\vec w) \rightarrow R(\vec w) \wedge R'(\vec w) \wedge \bigwedge_i
  \accessible(w_i).\]
\end{itemize}

The $\amd$ containment above simply formalizes the definition of~$\amd$, via
Proposition~\ref{prop:altdef}.
The first two accessibility axioms enforce that $I_\acc$ is access-valid in~$I_1$: for non-result-bounded
methods, accesses to a method $\mt$ on a relation~$R$ return all the results, while for result-bounded methods it respects the lower bounds.
The last accessibility axiom enforces that~$I_\acc$ is a common subinstance
of~$I_1$ and $I_2$ and
that~$\accessible$ includes the active domain of~$I_\acc$.
Hence, from the definitions and from Theorem~\ref{thm:equiv} and Proposition~\ref{prop:altdef}, we have:

\begin{prop} \label{prop:reduce} 
  Let $Q$ be a CQ, and let $\aschema$ be a schema with constraints
  expressible in active-domain first-order logic and with access methods that may
  have result upper and lower bounds. Then the following are equivalent:
  \begin{itemize}
    \item $Q$ is monotonically answerable with respect to $\aschema$.
    \item $Q$ is $\amd$ over $\aschema$.
    \item The $\amd$ containment for~$Q$ and $\aschema$ holds.
  \end{itemize}
\end{prop}

\begin{proof}
  We know by Theorem~\ref{thm:equiv} that the first two points are equivalent,
  and we can further rephrase them using Proposition~\ref{prop:altdef}: $Q$
  is monotonically answerable iff whenever two instances $I_1, I_2$ satisfying~$\Sigma$ have a common
  subinstance $I_\acc$ which is access-valid in~$I_1$, then we have $Q(I_1) \subseteq
  Q(I_2)$. 
Assuming this, let us show that the query containment holds. Fix an instance
  $J$ satisfying $\Gamma$. We let $I_1$ consist of the facts
  of~$J$ over
the relations in the original schema, and $I_2$ consist of the facts $R'(\vec c)$ for each
fact $R(\vec c)$ of $J$.  Clearly $I_1$ and $I_2$ satisfy $\Sigma$.
We consider the instance $I_\acc$ containing facts $R(\vec c)$ for all facts $R_\acc(\vec c)$
in~$J$. The last class of axioms for $\Gamma$ guarantee that this is a common subinstance of $I_1$ and $I_2$, 
while the first two sets of axioms guarantee that $I_\acc$ is access-valid in $I_1$. We conclude
  that $Q(I_1) \subseteq Q(I_2)$, and this implies that the containment of $Q$ in $Q'$
holds in $J$.

In the other direction we assume the query containment holds, and consider
$I_1, I_2$ with the required $I_\acc$. Build an instance $J$ by defining the relations
  $R$ from $I_1$, relations $R'$ from $I_2$, and relations $R_\acc$ from $I_\acc$. We can
verify that $J$ satisfies~$\Gamma$, and the query 
containment gives $Q(J) \subseteq Q'(J)$.
Tracing back through the definitions  this tells us that $Q(I_1) \subseteq Q(I_2)$.
\end{proof}

Note that, for a schema without result bounds,  the accessibility axioms above
can all be rewritten
as follows (as in~\cite{ustods,thebook}): for each method $\mt$, letting $R$
be the relation accessed by~$\mt$ and $\vec x$ be the input positions
of~$\mt$ in~$R$, we simply have the axiom:
\[
\Big(\bigwedge_i \accessible(x_i)\Big) \wedge
  R(\vec x, \vec y) \rightarrow  R'(\vec x, \vec y) \wedge \bigwedge_i
  \accessible(y_i).\]

\begin{exa} \label{ex:reduce} Let us apply the reduction above to the schema of Example~\ref{ex:simple}
  with the result bound of 100 from Example~\ref{exa:rbound}.
We  see that monotone answerability of a CQ $Q$  is equivalent to $Q
  \subseteq_\Gamma Q'$, for~$\Gamma$ containing:
\begin{itemize}
\item  the referential constraint from $\univdirect$ into $\profinfo$  and from
$\univdirect'$ into $\profinfo'$;

\item  $\accessible(i) \wedge \profinfo(i, n, s) \rightarrow
  \profinfo_\acc(i, n, s)$;
\item the following, for all $1 \leq j \leq 100$:
$\exists \vec y_1 \cdots \vec y_j
    (\bigwedge_{1 \leq p < q \leq j} \vec y_p \neq \vec y_q \wedge
    \univdirect(\vec y_p))
    \rightarrow \exists \vec z_1 \cdots \vec z_j
    (\bigwedge_{1 \leq p < q \leq j} ~  \vec z_p \neq \vec z_q \wedge
    \univdirect_\acc(\vec z_p))$;

\item 
$\profinfo_\acc(\vec w) \rightarrow \profinfo(\vec w) \wedge \profinfo'(\vec w) \wedge 
  \bigwedge_i \accessible(w_i)$
and similarly for $\univdirect$.
\end{itemize}
Note that the  constraint in the third item is  quite complex; it contains
  inequalities and also disjunction, since we write
$\vec y \neq \vec z$ to abbreviate a disjunction $\bigvee_{i \leq |\vec y|} y_i
  \neq z_i$. This makes it challenging to decide if $Q
  \subseteq_\Gamma Q'$ holds.
Hence, our goal in the next section will be to simplify result bounds to avoid
  such complex constraints.
\end{exa}

\paragraph{Bottom line: monotone answerability  and  query containment.}
The results  in this section have allowed us to reduce the analysis of monotone answerability
to a problem concerning containment of queries with integrity constraints.
We will rely on this equivalence throughout the  remainder of the paper, in that all of our
results on expressiveness and complexity will go through transformations and analysis
of the corresponding containment for $\amd$. Note that since we allow constants
that may be equal to one another in our analysis, \emph{it will always be possible to reduce
the setting for non-Boolean queries to that for Boolean queries}: we simply consider
the free variables as constants. With this in mind, \emph{we will state all of our results for
non-Boolean CQs, but in the proofs will assume the Boolean case, relying on this trivial reduction}.

\section{Simplifying result bounds with $\incd$s and FDs}
\label{sec:simplify}
The results in Section~\ref{sec:reduce} allow us to reduce the monotone answerability problem to
a query containment problem. However, for result bounds greater than~$1$, the
containment problem involves complex cardinality constraints, as illustrated in Example~\ref{ex:reduce},
and thus we cannot apply standard results or algorithms on query containment under
 constraints to get decidability ``out of the box''. There is also little hope
 to establish the decidability of query containment for the precise constraints
 that we define.
Hence, to address this difficulty, we study how to \emph{simplify} result-bounded schemas,
i.e., change or remove the result bounds.
We do so in this section, with \emph{simplification} results of the following
form: if we can find a plan for a query on a result-bounded schema,
then we can find a plan in a \emph{simplification} of the schema, i.e.,
a schema with simpler result bounds or
no result bounds at all.

These simplification results have two benefits.
First, they give insight about the use of result bounds, following the examples
in the introduction. For instance, our
results will
show that for most of the common classes of  constraints used in databases, the actual
numbers in the result bounds never matter for answerability.
Secondly, they help us to obtain decidability of the monotone answerability problem.

\paragraph{Existence-check simplification.}
\label{subsec:exist}
The simplest way to use result-bounded methods is to 
check if some tuples exist, as in Example~\ref{ex:existencecheck}.
We will formalize this as the \emph{existence-check simplification}, where we
replace result-bounded methods by Boolean methods that can only do such
existence checks.

Given a schema $\aschema$ with result-bounded methods,
its \emph{existence-check simplification}
$\aschema^\dagger$ is formed as follows:

\begin{itemize}
\item The signature of~$\aschema^\dagger$ is that of~$\aschema$ plus some new relations:
  for each result-bounded method $\mt$, letting $R$ be the relation accessed
  by~$\mt$, we add 
a relation~$\checkview_\mt$ whose arity is the number of input positions
  of~$\mt$.
\item The integrity constraints of~$\aschema^\dagger$ are those of~$\aschema$ plus, for
  each result-bounded method $\mt$ of~$\aschema$,
  two new $\incd$ constraints:
  \begin{align*}
    R(\vec x, \vec y) \rightarrow~ &
    \checkview_\mt(\vec x)\\
    \checkview_\mt(\vec x) \rightarrow~ & \exists
\vec y ~ R(\vec x, \vec y)
  \end{align*}
where $\vec x$  denotes the input positions of~$\mt$ in~$R$.
\item The methods of~$\aschema^\dagger$ are the  methods of~$\aschema$ that have
  no result bounds, plus one new Boolean method $\mt^\dagger$ on each new relation
  $\checkview_\mt$, that has no result bounds either.
\end{itemize}
\begin{exa}
\label{ex:existssimplify}
  Recall the schema $\aschema$ of Examples~\ref{ex:simple} and~\ref{ex:fd}.
  It featured a relation
  $\profinfo(\attrfmt{id}, \attrfmt{name}, \attrfmt{salary})$
  with an access method $\praccess$ having input $\attrfmt{id}$ to obtain
  information about a professor; and featured a relation $\univdirect(\attrfmt{id},
  \attrfmt{address}, \allowbreak \attrfmt{phone})$ with 
  an access method
  $\udaccess_2$ taking an $\attrfmt{id}$ as input and returning the 
  $\attrfmt{address}$ and phone number of tuples with this $\attrfmt{id}$.
  We assumed a result bound of~$1$ on~$\udaccess_2$, and assumed 
  the functional dependency $\phi$: each employee id 
  has exactly one $\attrfmt{address}$ (but possibly
  many phone numbers).

  The existence-check simplification of~$\aschema$ 
  has a signature with relations $\udirectory$, $\profinfo$, and a new relation
  $\udirectory_{\udaccess_2}$ of arity~$1$.
It has two access methods without result bounds: the method $\praccess$
  on~$\profinfo$ like in~$\aschema$, and a Boolean method $\udaccess_2'$ on
  $\udirectory_{\udaccess_2}$. Its constraints are those of~$\aschema$, plus the
  following IDs:
  \begin{align*}
    \udirectory(i, a, p) \rightarrow &~  \udirectory_{\udaccess_2}(i)\\
    \udirectory_{\udaccess_2}(i) \rightarrow &~  \exists a\,p ~ \udirectory(i, a, p)
  \end{align*}
\end{exa}

Clearly, every plan that uses the existence-check simplification
$\aschema^\dagger$ of a schema
$\aschema$ can be converted into a plan using $\aschema$, by replacing the
accesses on the Boolean method of~$\checkview_\mt$ to non-deterministic accesses
with~$\mt$, and only checking whether the result of these accesses is empty.
We want to understand when the converse is true. That is, when a plan on~$\aschema$ can
be converted to a plan on~$\aschema^\dagger$. For instance,
recalling the plan of Example~\ref{ex:existencecheck} that tests
whether~$\udirectory$ is empty simply by accessing $\udaccess_2$,
we could implement it in the existence-check
simplification of this schema.
More generally, we want to identify schemas $\aschema$ for which \emph{any} 
CQ having a monotone plan over~$\aschema$ has a plan 
on the existence-check
simplification~$\aschema^\dagger$.
We say that 
$\aschema$ 
is \emph{existence-check simplifiable} when this holds:
this intuitively means that 
``result-bounded methods of~$\aschema$ are only useful for existence checks''.

\paragraph{Showing existence-check simplifiability.}
We first show that this notion of existence-check simplifiability holds for schemas like Example~\ref{exa:plani} whose constraints consist of
inclusion dependencies:

\newcommand{\thmsimplifyidsexistence}{
  Let $\aschema$ be a schema whose constraints are
$\incd$s, and let $Q$ be a CQ that is
monotonically answerable 
  in~$\aschema$.
Then $Q$ is monotonically answerable in the existence-check simplification
   of~$\aschema$.
}
\begin{thm} \label{thm:simplifyidsexistence}
  \thmsimplifyidsexistence
\end{thm}

This existence-check simplifiability result implies in particular that 
\emph{for schemas with~$\incd$s, monotone answerability  is decidable even
with result bounds}. This is because the existence-check simplification of the
schema features only $\incd$s and no result bounds,
so the query containment problem for~$\amd$ 
only features guarded TGDs, which implies decidability. We will show a finer
complexity bound in the next section.

To prove Theorem~\ref{thm:simplifyidsexistence}, we show that if $Q$ is not $\amd$ 
in the existence-check simplification~$\aschema^\dagger$ of~$\aschema$, then it cannot be $\amd$ 
in~$\aschema$. This
suffices to prove the contrapositive of the result, because 
$\amd$ is equivalent
to monotone answerability  (Theorem~\ref{thm:equiv}).
As in all of our results concerning entailment problems like $\amd$, in the proof
for simplicity we will assume the query $Q$ is Boolean. The general
case is handled  by simply considering free variables as
additional constants.

Let us show, for a Boolean query~$Q$, that $Q$ not being $\amd$ in
$\aschema^\dagger$ implies
that it is not $\amd$ in $\aschema$. To do so,
we introduce a general method of
\emph{blowing up models} that we will reuse in all subsequent simplifiability
results. We assume that~$\amd$ does not hold in the
simplification~$\aschema^\dagger$, and consider
a \emph{counterexample to~$\amd$} for~$\aschema^\dagger$:
two instances $I_1^\dagger,I_2^\dagger$ both satisfying the schema constraints, 
such that~$I_1^\dagger$ satisfies $Q$ while $I_2^\dagger$ satisfies $\neg Q$, and
$I_1^\dagger$ and $I_2^\dagger$ have a
common subinstance $I_\acc^\dagger$ which is access-valid in~$I_1^\dagger$.
We use them to build a counterexample to~$\amd$
for the original schema~$\aschema$: we will always do so by adding more facts
to~$I_1^\dagger$ and~$I_2^\dagger$
and then restricting to the relations of~$\aschema$.
We formalize the sufficiency of such a construction in the following lemma,
whose proof is immediate, and which we state in full generality as we will use
it in multiple places:

\begin{lem}
  \label{lem:enlarge}
  Let $\aschema$ and $\aschema^\dagger$ be 
  schemas and $Q$ a CQ on the common relations of~$\aschema$
  and~$\aschema^\dagger$
  such that~$Q$ is not $\amd$ in~$\aschema^\dagger$.
  Suppose that for some counterexample $I_1^\dagger, I_2^\dagger$ to~$\amd$ for~$Q$
  in~$\aschema^\dagger$
  we can construct instances $I_1$ and $I_2$ over $\aschema$, satisfying the constraints of $\aschema$,
   which have a common subinstance
  $I_\acc$ that is access-valid in~$I_1$ for~$\aschema$, such that $I_2$ has
  a homomorphism to~$I_2^\dagger$, and such that 
  the restriction of~$I_1^\dagger$ to the relations of~$\aschema$ is a subinstance
  of~$I_1$.
Then $Q$ is not $\amd$ in~$\aschema$.
\end{lem}

\begin{proof}
  The instances $I_1$ and $I_2$ satisfy the constraints of~$\aschema$ and they have
  a common subinstance which is access-valid in~$I_1$ for~$\aschema$.
  Recall that, by definition of a counterexample, $I_1^\dagger$ satisfies~$Q$
  and $I_2^\dagger$ does not.
  Now, instance $I_1$ satisfies~$Q$, because $I_1^\dagger$ does and $Q$ only
  mentions the relations of~$\aschema$, and $I_2$ does not
  satisfy~$Q$, because it has a homomorphism to~$I_2^\dagger$ which does not.
  Hence, $I_1, I_2$ is a counterexample showing that~$Q$ is not $\amd$
  in~$\aschema$.
\end{proof}

Using this lemma, we can now prove Theorem~\ref{thm:simplifyidsexistence}:

\begin{proof}
We use the equivalence
  between $\amd$ and monotone plans given by Theorem~\ref{thm:equiv}, and we
  prove the contrapositive of the theorem, using
  Lemma~\ref{lem:enlarge}.
Let $\aschema$ be the original
schema and $\aschema^\dagger$ be the existence-check simplification.
Notice that the query $Q$ is indeed posed on the common relations of~$\aschema$
and~$\aschema^\dagger$, i.e., it does not involve the $R_\mt$ relations added
  in~$\aschema^\dagger$.
  To use Lemma~\ref{lem:enlarge}, suppose that we have  a counterexample
  $(I_1^\dagger, I_2^\dagger)$ to
$\amd$ for~$Q$ and the  simplification~$\aschema^\dagger$, i.e.,
the instances $I_1^\dagger$ and $I_2^\dagger$ satisfy the constraints $\Sigma^\dagger$
  of~$\aschema^\dagger$, the instance $I_1^\dagger$ satisfies $Q$ and
the instance $I_2^\dagger$ violates $Q$, and $I_1^\dagger$ and $I_2^\dagger$
  have a common subinstance $I_\acc^\dagger$ that is
access-valid in~$I_1^\dagger$.
We will show how to ``blow up'' each instance
  to~$I_1$ and $I_2$ which have a common subinstance which is
  access-valid in~$I_1$,
  i.e., we must ensure that each access to a method with
  a result bound in~$I_1$ returns either no tuples or more tuples than the bound.
In the blowup process we will
preserve the constraints $\Sigma^\dagger$ and the properties of the~$I_i$ with respect
to the CQ~$Q$. Intuitively, the blowup process will consider all accesses that
  can be performed with the common subinstance $I_\acc^\dagger$, and instantiate
  infinitely many witnesses to serve as answers for these accesses. We will then
  repair the instances by applying chase steps so that they satisfy the constraints again.

  We now explain formally how $I_1$ and $I_2$ are formed.
The first step is ``obliviously chasing with the existence-check constraints'':
for any existence-check constraint $\dep$
of the form
\[
  \forall x_1 \ldots x_m ~ \checkview_\mt(\vec x) \rightarrow ~ \exists y_1 \ldots y_n ~
  R(\vec x, \vec y)
\]
and any homomorphism $h$ of the variables $x_1 \ldots x_m$ to~$I_\acc^\dagger$,
we extend the  mapping by choosing infinitely many fresh witnesses for~$y_1 \ldots y_n$, naming the~$j^{th}$ value for~$y_i$
 in some canonical way
depending on~$(h(x_1), \ldots h(x_m), \dep, j, i)$, and creating the
  corresponding facts.
We use the term ``obliviously chasing'' to emphasize that the trigger
may not be active.
We let $I_\acc^*$ be $I_\acc^\dagger$ extended with these facts.

  The  second step is ``chasing with the original constraints''.
Recall the definition of ``the chase'' in Section~\ref{sec:prelims}.
  Specifically, we let $I_\acc$ be the %
chase of  $I_\acc^*$
by $\Sigma$.

  We now construct $I_1 \colonequals I_1^\dagger \cup I_\acc$
  and similarly define $I_2 \colonequals I_2^\dagger \cup I_\acc$.
  We also remove all facts from $I_1$, $I_2$, and $I_\acc$ where the underlying relation 
is not in $\aschema$.

  We now show correctness. First observe that
  the restriction of~$I_1^\dagger$ to the relations of~$\aschema$ is a subinstance of
  $I_1$, so that~$I_1$ still satisfies $Q$.
  Further,
  we argue that for all $p \in \{1, 2\}$, the instance~$I_p$ satisfies~$\Sigma$. 
As $\Sigma$ consists only of $\incd$s,
 its triggers consist of single facts, so it suffices to check this
on~$I_p^\dagger$ and on~$I_\acc$ separately. For $I_\acc$, we know that it satisfies $\Sigma$ by
definition of the chase. For $I_p^\dagger$, we know it satisfied $\Sigma^\dagger$ (before the
last step of removing the facts not on relations of~$\aschema$), so it 
satisfies $\Sigma$.
  
  We must now justify that $I_2$
has a homomorphism $h$ to~$I_2^\dagger$, which will imply that it still does not satisfy~$Q$. 
We
first define $h$ to be the identity on~$I_2^\dagger$.
It then suffices to define $h$ as a homomorphism from~$I_\acc$
to~$I_2^\dagger$
which is the identity on~$I_\acc^\dagger$, because $I_\acc \cap
I_2^\dagger = I_\acc^\dagger$. 
We next define $h$ on~$I_\acc^* \setminus I_\acc^\dagger$. Consider a fact
  $F = R(\vec a)$ of~$I_\acc^* \setminus I_\acc^\dagger$ created by 
obliviously chasing
  a trigger on an existence-check constraint $\dep$ on~$I_\acc^\dagger$. 
Let $F' = S(\vec b)$ be the
fact of~$I_\acc^\dagger$ in the image of the trigger: that is, the fact that matches the body of
$\dep$.
   We know that~$\dep$ holds in~$I_2^\dagger$ and thus there is some
 fact  $F'' \colonequals R(\vec c)$ in~$I_2^\dagger$
  that serves as a witness for this. Writing $\arity(R)$ to denote the
  arity of~$R$,
  we define $h(a_i)$
  for each $1 \leq i \leq \arity(R)$ as~$h(a_i) \colonequals c_i$. In this way, the image
of the fact $F$ under $h$ is  $F''$.

   We argue that this is consistent with the stipulation that~$h$ is the identity
  on~$I_\acc^\dagger$. This is
  because whenever $a_i \in \adom(I_\acc^\dagger)$, $a_i$ was
  not a fresh element when firing the trigger that created~$F$. So $c_i$ was not
  fresh either and must have been the same element, i.e., $c_i =
  a_i$.  

Further, we claim that all these assignments are consistent across the facts of
  $I_\acc^* \setminus I_\acc^\dagger$ because all elements of~$I_\acc^* \setminus
  I_\acc^\dagger$ which do not occur in~$\adom(I_\acc^\dagger)$ occur at exactly one position in
  one fact of~$I_\acc^* \setminus I_\acc^\dagger$.

We now define $h$ on facts of~$I_\acc \setminus I_\acc^*$ inductively by extending it
on the new elements introduced throughout the chase.
Whenever we create a fact $F = R(\vec a)$ in~$I_\acc$ for
a trigger $\tau$ mapping to~$F' = S(\vec b)$ for an $\incd$ $\dep$ in~$I_\acc$, we explain how to
extend $h$ to the nulls introduced in~$F$. Consider the fact $h(F')
= S(h(\vec
b))$ in~$I_2^\dagger$. The body of~$\dep$ also matches this fact, and
as~$I_2^\dagger$ satisfies $\Sigma^\dagger$
there must be a fact $F'' = R(\vec c)$ in~$I_2^\dagger$ which extends this match to the
head of~$\dep$,
since  $\dep$ holds in~$I_2^\dagger$.
For the elements $a_i$ that are not nulls created when firing~$\tau$, the image
$h(a_i)$ of~$a_i$ by~$h$ is already defined, and more precisely we must have
$h(a_i) = c_i$, by the same reasoning as when we defined~$h$ on~$I_\acc^*
\setminus I_\acc^\dagger$. Now, for the $a_i$'s that are nulls, noting that all
of them are distinct, we simply set $h(a_i) \colonequals c_i$. This ensures
that $h(F) = F''$, so $F$ has a homomorphic image. Hence, performing this
process inductively indeed creates a homomorphism.

This
concludes the proof of the fact that there is a homomorphism from~$I_2$ to
$I_2^\dagger$.

It remains to justify that the common subinstance $I_\acc$ in~$I_1$ and
$I_2$ is
access-valid in~$I_1$. Consider one access in~$I_1$ performed with some
method $\mt$ of a relation~$R$, with a binding $\accbind$ of values
in~$I_\acc$, and let us show that we can define a valid output to this
access in~$I_\acc$.
It is clear by definition of~$I_\acc$ that, if some value of~$\accbind$ is not
in the domain of~$I_\acc^\dagger$, it must be a null introduced in the chase to create
$I_\acc$, in the first or in the second step.  In this case the only
possible matching facts in~$I_1$ are the facts containing such a null, i.e.,
the facts in~$I_\acc \setminus I_\acc^\dagger$, so these facts are all
in~$I_\acc$
and there is nothing to show as they can all be returned. 

We thus focus on  the case when all values of~$\accbind$ are in~$I_\acc^\dagger$. If $\mt$ is not a
result-bounded access, then we can simply use the fact that~$I_\acc^\dagger$ is access-valid
in~$I_1^\dagger$ to know that all matching tuples in~$I_1^\dagger$ were
in~$I_\acc^\dagger$, so the
matching tuples in~$I_1$ must be in~$I_\acc^\dagger \cup (I_1
\setminus I_1^\dagger)$, hence
in~$I_\acc$. If $\mt$ is a result-bounded access, then consider the access
on~$\mt^\dagger$ with the same binding. Either this access returns nothing or it tells
us that there is a fact $\checkview_\mt$ containing the values of~$\accbind$.
In the first case, as $I_\acc^\dagger$ is access-valid in~$I_1^\dagger$, we know
that~$I_1^\dagger$
contains no matching tuple, hence the constraints of $\aschema^\dagger$ imply
that~$I_1^\dagger$ does not contain any $R$-fact which matches~$\accbind$ in the input
positions of~$\mt$. This means that any matching tuple in~$I_1$ for the access
on~$\mt$ must be in~$I_1 \setminus I_1^\dagger$, so they are
in~$I_\acc$ and we can
define a valid output to the access in~$I_\acc$. This covers the first case.

In the second case, the $\checkview_\mt$-fact of~$I_1^\dagger$ implies by construction
that~$I_\acc^*$, hence $I_\acc$, contains infinitely many suitable
facts matching the access. Letting $k$ be the result bound
of~$\mt$, we choose $k$ facts among those, and obtain a valid output to the
access with~$\accbind$ on~$\mt$ in~$I_1$. Hence, we have shown
that~$I_\acc$ is
access-valid in~$I_1$.

Hence, we have shown the conditions of Lemma~\ref{lem:enlarge}. Using
this lemma, we have completed the proof of Theorem~\ref{thm:simplifyidsexistence}.
\end{proof}

\newcommand{\simplifyfddef}{
\item The signature of~$\aschema^\dagger$ is that of~$\aschema$ plus some new
relations: for each result-bounded
method $\mt$, letting $R$ be the relation accessed by~$\mt$,
we add a relation~$R_\mt$ whose arity is
$\left|\detby(\mt)\right|$.
\item The integrity constraints of~$\aschema^\dagger$ are those of~$\aschema$ plus, 
for each result-bounded method~$\mt$ of~$\aschema$,
two new $\incd$ constraints:
\begin{align*}
  R(\vec x, \vec y, \vec z) \rightarrow ~ & R_\mt(\vec x, \vec y)\\
  R_\mt(\vec x, \vec y) \rightarrow ~ & \exists \vec z ~ R(\vec x, \vec y, \vec z)
\end{align*}
where $\vec x$ denotes the input positions of~$\mt$ and $\vec y$ denotes the
other positions of~$\detby(\mt)$.
\item The methods of~$\aschema^\dagger$ are the methods of~$\aschema$ that have
no result bounds, 
 plus the following: for each result-bounded method
$\mt$ on relation~$R$ in~$\aschema$,
a method $\mt^\dagger$ on~$R_\mt$ that has no result bounds and whose input positions
are the positions of~$R_\mt$ corresponding to input positions of~$\mt$.
}

\paragraph{FD simplification.}
When our constraints include functional dependencies, we can hope for another kind of simplification,
generalizing
the idea of Example~\ref{ex:fd}:
an FD can force the output of a result-bounded method to be deterministic on a
projection of the output positions.
We will define the \emph{FD simplification} to formalize this intuition.

Given a set of constraints $\Sigma$, a relation~$R$ that occurs in~$\Sigma$, and a
subset $P$
of the positions of
$R$, we write $\detby(R,P)$ for the set of positions \emph{determined} by $P$, i.e.,
the set of positions~$i$ of~$R$ such that~$\Sigma$ implies the FD
$P \determines i$. In particular, we have $P \subseteq \detby(R,P)$.
For any access method $\mt$, letting $R$ be the relation that it accesses,
we let $\detby(\mt)$ denote $\detby(R,P)$ where $P$ is the set of input
positions of~$\mt$. 
Given a schema $\aschema$ with result-bounded methods, we can now define its \emph{FD simplification} 
$\aschema^\dagger$ as follows:

\begin{itemize}
  \simplifyfddef
\end{itemize}

Note that the FD simplification is the same as the existence-check
simplification when the integrity constraints~$\Sigma$ do not imply any FD\@.
Further observe that, even though the methods of~$\aschema^\dagger$ have no result
bounds, any access to a new method $\mt^\dagger$ of~$\aschema^\dagger$
is guaranteed to return at most one result. This is thanks to the FD on the
corresponding relation~$R$, and thanks to the constraints that relate
$R_\mt$ and~$R$.

\begin{exa} \label{ex:fdsimplify}
Recall the schema $\aschema$ of Example~\ref{ex:fd} and the FD $\phi$
  on~$\univdirect$. In the FD simplification of~$\aschema$, we
  add a relation $\univdirect_{\udaccess_{2}}(\attrfmt{id}, \attrfmt{address})$,
  we replace $\udaccess_2$ by a method $\udaccess_2$ on~$\univdirect_{\udaccess_{2}}$ whose
  input attribute is~$\attrfmt{id}$, and we add the $\incd$s 
  $\univdirect(i, a, p) \rightarrow \univdirect_{\udaccess_{2}}(i, a)$ and
  $\univdirect_{\udaccess_{2}}(i, a) \rightarrow \exists p ~ \univdirect(i, a, p)$.
  The method $\udaccess_2'$ has no result bound, but the $\incd$s above and the
  FD $\phi$ 
  ensure that it always returns at most one result.

  The point of the FD simplification is that it has no result-bounded methods, 
  so that, like for the existence-check simplification, 
  the query containment problem for the schema of the simplification
will not  
  use any  complex cardinality constraints. This is in contrast to the query
  containment problem obtained in Example~\ref{ex:reduce}.
\end{exa}

A schema $\aschema$ is \emph{FD simplifiable} if every CQ having a monotone plan
over~$\aschema$
has one over the FD simplification of~$\aschema$. As for existence-check,
if a schema is FD simplifiable, 
 we can decide  monotone answerability by reducing to the same  problem in 
a schema
without result bounds.

We use a variant of our ``blowup process'' to show that schemas 
with only FD constraints are FD simplifiable:

\newcommand{\fdsimplify}{
  Let $\aschema$ be a schema whose constraints are FDs, and let $Q$ be a CQ
  that is
   monotonically answerable in~$\aschema$. Then $Q$ is monotonically answerable
   in the FD simplification $\aschema^\dagger$ of~$\aschema$.
}
\begin{thm} \label{thm:fdsimplify}
  \fdsimplify
\end{thm}

\begin{proof}
We will again show the contrapositive of the statement.
Assume that we have a counterexample $I_1^\dagger,I_2^\dagger$ to
$\amd$ for
$\aschema^\dagger$, with $Q$ holding in $I_1^\dagger$, with $Q$
  not holding in $I_2^\dagger$,
and with $I_1^\dagger$ and $I_2^\dagger$ having a common subinstance
  $I_\acc^\dagger$ that is access-valid
  in~$I_1^\dagger$ under~$\aschema^\dagger$.
We will  upgrade
  these to $I_1, I_2, I_\acc$ having the same property for $\aschema$, by
  blowing up accesses one after the other. To do so, we initially set
  $I_1$ to
  be the restriction of~$I_1^\dagger$ to the relations of~$\aschema$, i.e., all
  relations but the $R_\mt$ relations. We define~$I_2$
  from~$I_2^\dagger$ and
  $I_\acc$ from~$I_\acc^\dagger$ in the same way.
  We fix some valid access selection $\sigma_1$ for~$I_1^\dagger$ that always returns tuples
  from~$I_\acc^\dagger$ when performing accesses with values of~$I_\acc^\dagger$.
  We consider all possible accesses in parallel,
  performing for each access a process described just below.

  We consider all the (non-result-bounded) access methods $\mt^\dagger$
  introduced in~$\aschema^\dagger$ --- not including the access methods $\mt$ of
  $\aschema^\dagger$ which are simply those without result bounds
  in~$\aschema$.  Given such an $\mt^\dagger$, we write $\mt$ for the corresponding
  (result-bounded) access method in~$\aschema$.
  We call $\mt^\dagger$ (and also $\mt$) \emph{non-dangerous} if the input positions of~$\mt$
  determine all positions of the accessed relation. Equivalently, $\detby(\mt)$
  contains all positions of~$R$. Or again equivalently, $R_\mt$ and
  $R_\mt^\dagger$ have the same arity. We call $\mt$ and $\mt^\dagger$ \emph{dangerous} otherwise.
  The blowup process we provide for each access will differ depending on whether
 the method is dangerous or non-dangerous.

  First, we handle the non-dangerous methods, and simply copy
  in~$I_\acc$
  the results of accesses on these methods.
  Consider every non-dangerous method $\mt^\dagger$ of~$\aschema^\dagger$.
  We again write $\mt$ the corresponding method in $\aschema$ and use $R$ to denote the
  relation of the access. We consider every possible access $(\mt^\dagger,
  \abind)$ on~$I_1^\dagger$ with
  values in~$I_\acc^\dagger$ such that one tuple (and, by the FDs, exactly one tuple) is
  returned. The $\incd$s from $R_\mt$ to~$R$ imply there is an $R$-fact with
  exactly the same elements in~$I_1^\dagger$ and~$I_2^\dagger$. We then add this one fact
  to~$I_\acc$.
  We do this for all the non-dangerous methods and accesses using these methods.

  Second, we blow up the dangerous methods, which  is the complicated step of the
  construction.
  Consider every dangerous method $\mt^\dagger$ of~$\aschema^\dagger$.
  Write $\mt$ for the method in $\aschema$ corresponding to $\mt^\dagger$ and~$R$ for the relation
of $\mt$.
  Consider every possible access $(\mt^\dagger, \abind)$ on~$I_1^\dagger$ with
  values in~$I_\acc^\dagger$.
  There are two possibilities: either this access returns nothing, or, by the
  FDs of~$\aschema$ and the constraints introduced in~$\aschema^\dagger$, 
  it returns exactly one tuple, which must be in~$I_\acc^\dagger$ because
  $I_\acc^\dagger$ is
  access-valid in~$I_1^\dagger$.

  In the first case, we do nothing. Intuitively, we know that there are no
  matching tuples in~$I_1^\dagger$.  Now, considering the $\incd$ constraint in the FD
  simplification that goes from~$R$ to~$R_\mt$, we infer that there is no
  $R$-fact in~$I_1^\dagger$ whose projection to the input positions of~$\mt$
  matches~$\abind$.  Thus we will have no problem building a valid answer to this
  access.

  In the second case, consider the fact $M_1'$ that was returned, following the
  access selection $\sigma_1$
  in response to the access $(\mt^\dagger,
  \abind)$. Recall that there is an  $\incd$ constraint in~$\aschema^\dagger$ that
  goes from~$R_\mt$ to~$R$. This constraint allows us to infer
from the existence of~$M_1'$ 
  that there must be some \emph{witnessing facts} in~$I_1^\dagger$
  and in~$I_2^\dagger$, namely,
  $R$-facts whose projection to $\detby(\mt)$ matches~$M_1'$. In this case, we
  perform a modification that we refer to as \emph{blowing up the access}. Specifically,
  let $X$ be the positions of~$R$ that
  are \emph{not} in $\detby(\mt)$. By our assumption that $\mt^\dagger$ is
  dangerous, $X$ is nonempty.
  Construct infinitely many $R$-facts
  with all positions in $\detby(\mt)$ agreeing with $M_1'$,
  and with all positions in~$X$ filled using fresh values
  that are different from each other and from other values in $I_1^\dagger \cup
  I_2^\dagger$.
  We add these \emph{duplicate facts} to~$I_1$, to~$I_2$, and
  to~$I_\acc$.

  Performing this process for all accesses in~$I_\acc^\dagger$ on all dangerous access
  methods that return a tuple, we have finished the definition of~$I_1$,
  $I_2$, and $I_\acc$.

  Having completed our construction, we now check that the conditions are satisfied.
  It is clear that the restriction of $I_1^\dagger$ to the relations of~$\aschema$ is a
  subset of~$I_1$.
  We see that
  $I_\acc \subseteq I_1$ and $I_\acc \subseteq I_2$, because these two
  last inclusions are true initially and all tuples added to~$I_\acc$ are also
  added to~$I_1$ and~$I_2$, or in the case of non-dangerous accesses are already
  present in~$I_1^\dagger$ and~$I_2^\dagger$. Further,
  $I_2$ has a homomorphism back to~$I_2^\dagger$: we can define it as the 
  identity on~$I_2^\dagger$, and as mapping the fresh elements of every duplicate tuple
  of~$I_2 \setminus I_2^\dagger$ to
  a witnessing fact in~$I_2^\dagger$. This only collapses fresh values
  of~$\adom(I_2)$ to values of~$\adom(I_2^\dagger)$, and is the identity on constants
  of~$I_2^\dagger$.

  We must justify that $I_1$ and~$I_2$ still satisfy the FD constraints
  of~$\aschema$.
  This is true because, whenever we add a set of duplicate facts
  to~$I_1$ and~$I_2$,  each fact in the set
  contains fresh values at the positions of the set~$X$, and must match the witnessing
  facts already present in~$I_1^\dagger$ and~$I_2^\dagger$ at the other positions. Hence, if
  adding such a duplicate fact violated an FD, the left-hand-side of the FD could not
  contain a position in~$X$, as elements at these positions are fresh. So the
  left-hand-side would be contained in~$\detby(\mt)$.  Thus the right-hand-side
  would also be contained in~$\detby(\mt)$, because $\detby(\mt)$ is closed
  under the FDs. We conclude that if adding the new fact violated an FD, then the
  witnessing facts also did, breaking the assumption that $I_1^\dagger$
  and~$I_2^\dagger$
  satisfied the FDs before.

  We must now show that~$I_\acc$ is access-valid in~$I_1$. 

  To do so, consider a method
  $\mt$ of~$\aschema$ and binding $\accbind$. 
If $\accbind$ contains values
  from~$\adom(I_\acc) \setminus \adom(I_\acc^\dagger)$, then  we know that these values
  occur only in tuples from~$I_\acc\setminus I_\acc^\dagger$.  Thus
  matching tuples in~$I_1$ are all in~$I_\acc$ and there is nothing to show.
  Hence, we focus on the case where $\accbind$ consists of values
  of~$\adom(I_\acc^\dagger)$.

  We first focus on the subcase where~$\mt$ is not result-bounded.
In this subcase,
  when performing the same access~$(\mt^\dagger, \accbind)$ in~$I_1^\dagger$, the valid access
  selection $\sigma_1$ that we fixed
  returns all matching tuples in~$I_1^\dagger$, and these tuples
  must be part of~$I_\acc^\dagger$ because $I_\acc^\dagger$ is access-valid.
  Considering all matching tuples for the access $(\mt, \accbind)$ in
  the larger structure $I_1$,
 we see they are of
  two kinds.
  There are those that were already present in~$I_1^\dagger$, which are
  in~$I_\acc^\dagger$
  because as we explained it is access-valid, so
  so they are in~$I_\acc$.
  The second kind  are those that were added in $I_1\setminus I_1^\dagger$, and they were
  added to~$I_\acc$ as well. So in both cases all matching tuples
  in~$I_1$
  are in~$I_\acc$.

  Now, if~$\mt$ is result-bounded but not dangerous, then performing the access $(\mt^\dagger, \accbind)$
  on~$R_\mt$ in~$I_1^\dagger$ either returned a single
  matching tuple or no tuples. If it returned
  no matching tuple, then the $\incd$ from~$R$ to~$R_\mt$ implies that there was no matching tuple
  to the access $(\mt^\dagger, \accbind)$ in~$I_1^\dagger$, hence there is still none
  in~$I_1$ except potentially those of~$I_\acc \setminus
  I_\acc^\dagger$.
  So~$I_\acc$ can be used to construct a valid response. If there is a
  matching tuple, then the $\incd$ from~$R_\mt$ to~$R$ implies that there is a
  matching tuple for the access $(\mt^\dagger, \accbind)$ in~$I_1^\dagger$, which we added
  to~$I_\acc$ at the end of the construction. So the matching tuple is
  in~$I_\acc$ and it is still a valid response to the access $(\mt, \abind)$
  in~$I_1$ (recall that the FDs imply that this single tuple is the only
  possible matching tuple).

  We can thus focus on the case where~$\mt$ is result-bounded and dangerous. 
  In this case, when we considered the access $(\mt, \accbind)$ in the blow-up
  process above for~$\mt$, either we blew up the access or we did not. If we
  did not, then we know that there were no matching tuples in~$I_1^\dagger$ for the
  access, and the $\incd$ from~$R$ to~$R_\mt$ implies that $I_1^\dagger$ contains no
  matching tuple for the access $(\mt^\dagger, \accbind)$, so all matching tuples to
  the access $(\mt, \abind)$ in $I_1$ are in $I_\acc$.
  If we did blow the access up, then we know that
  $I_\acc$ contains infinitely many matching tuples to the access that we can use as a
  valid response to the access in~$I_1$. Thus, $I_\acc$ is indeed
  access-valid. 

Thus,
  Lemma~\ref{lem:enlarge} implies that $Q$ is not $\amd$ in~$\aschema$,
  concluding the proof of Theorem~\ref{thm:fdsimplify}.
\end{proof}

\section{Decidability of Monotone Answerability\texorpdfstring{\\}{ }using Existence check and FD simplification}
\label{sec:complexity}

Thus far we have seen a general way to  reduce monotone answerability problems with result bounds to query containment
problems (Section~\ref{sec:reduce}). We have also seen schema  simplification
results for both FDs and $\incd$s, which give us insight
into how result-bounded methods can be used (Section~\ref{sec:simplify}).
We now show that for these two classes of constraints,
the reduction to containment and simplification results combine to
give decidability results, along with tight complexity bounds.

\subsection{Decidability for FDs}
We first consider schemas whose constraints consist of FDs. 
We start with an analysis of monotone answerability in the case \emph{without result
bounds}:

\newcommand{\fdclassic}{
We can decide whether a CQ is monotonically answerable with respect to a
  schema without result bounds whose constraints
  are FDs. The problem is $\np$-complete.
}
\begin{prop} \label{prop:decidmdetfdclassic}
\fdclassic
\end{prop}

\begin{proof}
The lower bound already holds
  without result bounds or constraints \cite{access2}, so it suffices to show the upper
  bound.
   We know that, by  Theorem~\ref{thm:equiv}
  and Proposition~\ref{prop:reduce},
the problem reduces to the $\amd$ query
  containment problem $Q \subseteq_\Gamma Q'$ for~$\aschema$. As $\aschema$ has
  no result bounds, we can define $\Gamma$ using the rewriting of the accessibility axioms given
  after Proposition~\ref{prop:reduce}. 
  The constraints $\Gamma$
  thus consist of FDs and of full TGDs of the form:
\[\Big(\bigwedge_i \accessible(x_i)\Big) \wedge
  R(\vec x, \vec y) \rightarrow  R'(\vec x, \vec y) \wedge \bigwedge_i
  \accessible(y_i).\]

Since this is a query containment problem with FDs and TGDs, by Proposition~\ref{prop:chasecomplete}
it can be solved by computing the chase.
  As the TGDs are full, we know that
  we do not create fresh values when computing the chase.
Further, because there are no TGD constraints with primed relations in their
body, once $\accessible$ does not change, the entire chase process has
terminated.
Besides, when adding values to $\accessible$ we must reach a fixpoint in
  linearly many chase steps 
since $\accessible$ is unary. 
  Thus the chase with~$\Gamma$ terminates in linearly many steps. Thus, we can decide
  containment by checking in~$\np$ whether $Q'$ holds on the chase result,
  concluding the proof.
\end{proof}

We now return to the situation \emph{with result bounds}. We know that
schemas with FDs are FD simplifiable.
 From this we get a reduction to query containment with no result bounds, but introducing
new axioms.
We can show that the additional axioms involving $R_{\mt}$ and $R$ do not harm chase
termination,
so that~$\amd$ 
is decidable; in fact, it is NP-complete, i.e., no harder than CQ evaluation:

\newcommand{\thmdecidfd}{
  We can decide whether a CQ is monotonically answerable with respect to a schema with
  result bounds whose constraints are FDs. The problem is $\np$-complete.
}
\begin{thm} \label{thm:decidfd} 
  \thmdecidfd
\end{thm}

\begin{proof}
By Theorem~\ref{thm:fdsimplify} it suffices to deal with the  FD simplification, meaning that we can 
reduce to a schema of the following form:

\begin{itemize}
  \simplifyfddef
\end{itemize}
By Proposition~\ref{prop:reduce}, we then reduce $\amd$ to query containment.
The resulting query containment problem
involves two copies of the constraints above, on primed and unprimed
copies of the schema, along with accessibility axioms for each
access method (including the new methods~$R_\mt$).
We can observe a few obvious simplifications of these constraints, when working
  with the restricted chase:
\begin{itemize}
\item The ``unprimed method-to-regular constraint'',  $R_\mt(\vec x, \vec y) \rightarrow \exists \vec z ~
  R(\vec x, \vec y, \vec z)$ will never fire, since a fact $R_\mt(\vec a, \vec b)$
is always generated by a corresponding fact  $R(\vec a, \vec b, \vec c)$. 
\item 
  When firing a constraint of the form $\tau: R'(\vec x, \vec y,
  \vec z) \rightarrow R'_\mt(\vec x, \vec y)$ to create a fact $F_2 = R'_\mt(\vec a,
    \vec b)$ from a fact $F_1 = R'(\vec a, \vec b, \vec c)$, 
    the fact $F_2$ will not be a trigger for any rule firing. Indeed, the only rule
    applicable to a $R'_\mt$-fact is the reverse constraint $R'_\mt(\vec x, \vec y)
    \rightarrow \exists \vec z ~ R'(\vec x, \vec y, \vec z)$, for which $F_1$
    witnesses that $F_2$ is not an active trigger. What is more, the
    $R'_\mt$-facts created by constraints of the form $\tau$ cannot help make
    the query true, as the query does not mention relations of the form $R'_\mt$. For
    this reason, we can disregard unprimed method-to-regular constraints 
 without changing the query containment problem.
  \end{itemize}
  So the constraints that remain in addition to the FDs are:
  \begin{itemize}
    \item The $\incd$ constraints $R(\vec x, \vec y, \vec z) \rightarrow
      R_\mt(\vec x, \vec y)$ where $\vec x$ denotes the input positions of~$\mt$
      and $\vec y$ denotes the other positions of~$\detby(\mt)$;
    \item For every access method $\mt$ on a relation $S$,
      the accessibility axioms which are of the form $\left(\bigwedge_i \accessible(x_i) \right)
      \land S(\vec x) \rightarrow S_\acc(\vec x)$ and $S_\acc(\vec w)
      \rightarrow S(\vec w) \land S'(\vec w) \land \bigwedge_i
      \accessible(w_i)$. Note that $S$ may be one of the original relations, or
      one of the relations $R_\mt$, depending on whether $\mt$ originally had
      result bounds or not.
    \item The $\incd$ constraints $R'_\mt(\vec x, \vec y) \rightarrow \exists z
      ~
      R'(\vec x, \vec y, \vec z)$, where $\vec x$ denotes the input positions of~$\mt$
      and $\vec y$ denotes the other positions of~$\detby(\mt)$.
  \end{itemize}

  The only non-full TGDs in these constraints are those of the last bullet
  point: 
  these are the only rules that create new values, and these values
will never propagate back to the unprimed relations. Further,
whenever a primed fact $F$ is created containing  a null using such a rule,
the only further chase  steps that can apply to $F$ are FDs, and these will only
merge elements in $F$. 
Thus the chase will terminate in polynomially many steps
as in the proof of Proposition~\ref{prop:decidmdetfdclassic},
  which establishes the NP upper bound and concludes the proof of
  Theorem~\ref{thm:decidfd}.
\end{proof}

\subsection{Decidability for $\incd$s}
Next we consider schemas whose constraints consist of~$\incd$s.
As we already mentioned, Theorem~\ref{thm:simplifyidsexistence} implies
decidability for such schemas. We now give the precise complexity bound:

\begin{thm} \label{thm:decidids}
  We can decide whether
a CQ is monotonically answerable
with respect to  a schema with result bounds whose  constraints are
$\incd$s. Further, the problem is $\exptime$-complete.
\end{thm}
\begin{proof}
  Hardness already holds without result bounds~\cite{bbbicdt}, so we focus on
  the upper bound.
  By Theorem~\ref{thm:simplifyidsexistence}, we can equivalently replace
  the schema $\aschema$ with
  its existence-check simplification $\aschema^\dagger$, and $\aschema^\dagger$ does not
  have result bounds. Further, it is easy to see that $\aschema^\dagger$ consists only of
  $\incd$s, namely, those of~$\aschema$ plus the $\incd$s added in the simplification. 
Note that the resulting query containment problem only involves guarded TGDs, and thus we can conclude
that the problem is in $\twoexp$ from~\cite{datalogpm}. However, we can do better:
\cite{bbbicdt} showed that the monotone answerability problem for schemas where
the constraints are $\incd$s is in $\exptime$, and thus we conclude the proof.
\end{proof}

\subsection{Complexity for Bounded-Width $\incd$s and Special Properties of the Query Containment for Access Methods}
Up until now we have seen a reduction of answerability to query answering.
We can see that the query answering problem involves adding auxiliary
constraints  ---
the ``transfer'' axioms that capture properties of an access --- and
these are of a very special form. Our goal now is to illustrate how the restricted shape
of these axioms can be used to get lower complexity bounds, compared
to what we can get by appealing to coarser classes like guarded or frontier-guarded TGDs.

We illustrate this in an important case for $\incd$s, those whose \emph{width} --- the number of
exported variables, i.e., of variables shared between the body and the head --- is bounded by a constant. Recall that this
includes in particular $\uincd$s, which have width~$1$. For bounded-width $\incd$s, it was shown by Johnson and Klug~\cite{johnsonklug}
that query containment under constraints is $\np$-complete. This result showed
that the width  parameter plays an important role in lowering the complexity of the containment problem.
A natural question is whether the same holds for monotone
answerability.
We accordingly conclude the section by showing
the following, which is new even in the setting without result bounds:

\newcommand{\npidsbounds}{
  It is $\np$-complete to decide whether a CQ is monotonically answerable with respect to a schema
  with result bounds whose constraints are bounded-width $\incd$s.
}
\begin{thm} \label{thm:npidsbounds}
  \npidsbounds
\end{thm}

To show this result, we will again use the fact that $\incd$s are
existence-check simplifiable (Theorem~\ref{thm:simplifyidsexistence}).
Using  Proposition~\ref{prop:reduce} we 
reduce to a query containment problem with guarded TGDs. But this is not enough
to get an $\np$ bound.
The reason is that the query containment problem includes accessibility axioms, which
are not~$\incd$s.
So we cannot hope to conclude directly 
using~\cite{johnsonklug}. 

The rest of this section will be devoted to the proof
of Theorem~\ref{thm:npidsbounds}. As mentioned
above, this will require a finer-grained analysis
of the query containment problem produced from our reduction.
In fact, we will note a particular property of these containment problems that can
be exploited: \emph{they
involve constraints that are $\incd$s and $\gtgd$s  that are ``close to $\incd$s'':
involving only guards and a fixed set of relations}, specifically, the $\accessible$
relation. Our results give evidence
that looking at other parameters in query answering problems for tame classes
of dependencies can yield new insights, despite the wealth of results already
present in this area \cite{datalogpmj,gmp}.

We begin with  the case \emph{without result bounds}, and then extend
to support result bounds.

\paragraph{Proving Theorem~\ref{thm:npidsbounds} without result bounds.}
 In the absence of result bounds, recall that the~$\amd$ query containment problem
$Q \subseteq_\Gamma Q'$ can be expressed as follows: 
$\Gamma$ contains the bounded-width $\incd$s $\Sigma$ of the schema, their
primed copy $\Sigma'$, and for
each access method $\mt$ accessing relation~$R$ with input positions~$\vec x$
there is an accessibility axiom:
\[\Big(\bigwedge_i \accessible(x_i)\Big) \wedge
  R(\vec x, \vec y) \rightarrow R'(\vec x, \vec y) \wedge \bigwedge_i
  \accessible(y_i).\]
For each method~$\mt$, we can rewrite the accessibility axiom above by splitting its head, and
obtain the following pair
of axioms, where the truncated accessibility axioms only create the
$\accessible$
facts (hence the name), and the transfer axioms create the primed facts:

\begin{itemize}
\item (Truncated Accessibility): $\left(\bigwedge_i \accessible(x_i)\right) \wedge
  R(\vec x, \vec y) \rightarrow \bigwedge_i
  \accessible(y_i)$.
\item (Transfer): $\left(\bigwedge_i \accessible(x_i)\right) \wedge R(\vec x, \vec y)
  \rightarrow  R'(\vec x, \vec y)$.
\end{itemize}

We let $\Delta$ be the set of the truncated accessibility axioms and
transfer axioms that we obtain for all the methods~$\mt$.

The constraints of~$\Delta$ are TGDs but not IDs. However, we will take
advantage of their structure to \emph{linearize} $\Delta$ together with
$\Sigma$, i.e., construct a set~$\Sigma^\lift$
of $\incd$s
that ``simulate''
the chase by $\Sigma$ and~$\Delta$.
To define $\Sigma^\lift$ formally, we will change the signature.
Let $\sign$ be the signature of the relations used in~$\Sigma$, not including
the special unary relation $\accessible$ used in~$\Delta$; and let $w\in\NN$ be
the constant bound on the width of the $\incd$s in~$\Sigma$.
We expand~$\sign$ to the signature $\sign^{\lift}$ as follows.
For each relation~$R$ of arity~$n$ in~$\sign$,
we consider each subset $P$ of the positions
of~$R$ of size at most~$w$.
For each such subset $P$,
we add a relation
$R_{P}$ of arity $n$ to~$\sign^{\lift}$. 
Intuitively, an $R_P$-fact denotes an $R$-fact where the elements
in the positions of~$P$ are 
accessible.

Remember that our goal is to \emph{linearize}~$\Sigma$ and~$\Delta$ to a set of
$\incd$s $\Sigma^\lift$ which emulates the chase by~$\Sigma$ and~$\Delta$.
If we could ensure that $\Sigma^\lift$ has bounded width, we could then conclude
using the result of~\cite{johnsonklug}. We will not be able to enforce this, but $\Sigma^\lift$
will instead satisfy a notion of \emph{bounded semi-width} that we now define.

The \emph{basic position graph} of a set of TGDs 
$\Sigma$
is the directed graph whose nodes are the positions of relations
in~$\Sigma$
with an edge 
from position~$i$ of a relation~$T$ to position~$j$ of a relation~$U$
if and only
if the following is true:
there is a dependency $\dep \in \Sigma$ whose body contains an atom~$A$ using
 relation $T$,
whose head atom~$A'$ uses relation $U$, and with an exported variable $x$
that occurs at position~$i$ of~$A$ and at position $j$ of~$A'$.

We say that
$\Sigma^\lift$ has \emph{semi-width} bounded
by $w$ if it can be decomposed as~$\Sigma^\lift = \Sigma^\lift_1 \cup \Sigma^\lift_2$
where $\Sigma_1^\lift$ has width bounded by~$w$ and
the basic position graph of~$\Sigma_2^\lift$ is acyclic.
The bound on the semi-width of $\Sigma^{\lift}$ then implies an
$\np$ bound on query containment, thanks 
to the following easy generalization of the result of Johnson and Klug~\cite{johnsonklug}:
\newcommand{\semiwidthclassic}{%
For any fixed  $w \in \NN$,
there is an $\np$ algorithm for
containment under $\incd$s of
semi-width at most~$w$.
}%
\begin{prop} \label{prop:semiwidthclassic}
  \semiwidthclassic
\end{prop}

This is proven by a slight modification of Johnson and Klug's argument,
so we defer it to 
Appendix~\ref{apx:semiwidthproof}.

Having defined semi-width, we can now state our linearization result:

\newcommand{\linearizeaccessids}{
  For any fixed~$w \in \NN$,
  given a set $\Sigma$ of $\incd$s of width~$w$ and a set
  $\Delta$ of truncated accessibility and transfer axioms, and given
  a set of facts~$I_0$,
  we can compute in $\ptime$ a set of $\incd$s $\Sigma^\lift$ of semi-width~$w$ 
  and a set of facts $I_0^\lift$ satisfying the following:
  for any  Boolean CQ
  $Q^*$ over the primed signature using constants from $I_0$
  and existentially quantified variables,
  $Q^*$ is entailed from~$I_0$, $\Sigma$,
and~$\Delta$ iff
$Q^*$ is entailed from $I_0^\lift$ 
 and $\Sigma^\lift$.
}
\begin{prop} \label{prop:linearizeaccessids}
  \linearizeaccessids
\end{prop}

The proof of this proposition is our main technical challenge, and it
is deferred to 
Section~\ref{sec:linearize}. The special form of the
constraints is crucial in getting an efficient linearization that leads
to linear TGDs of small semi-width.

These two results allow us to decide in NP whether the query containment for $\amd$ 
holds. Indeed, first rewrite $I_0 \colonequals \canondb(Q)$, along with $\Sigma$ and $\Delta$
to obtain $I_0^\lift$ and $\Sigma^\lift$, using
Proposition~\ref{prop:linearizeaccessids}.
Then, recalling that $\Sigma^\lift$ has semi-width~$w$,
let $\Gamma_{\bounded}$ consist of the primed copy~$\Sigma'$ of the constraints,
along with the $\incd$s
of~$\Sigma^\lift$ that have width $\leq w$;
and let
$\Gamma_{\acyclic}$ consist of the rules of~$\Sigma^\lift$ that do not have
width bounded by $w$. By assumption, these rules have an acyclic position
graph.
It is clear that~$\Gamma^\lift \colonequals \Gamma_{\bounded} \cup \Gamma_{\acyclic}$ also has
semi-width~$w$. Now, the following is clear:

\begin{clm}
  \label{clm:amdequiv}
Let the instance $I_0^\lift$ and constraints $\Gamma^\lift$ 
be defined from $I_0$ and $\Gamma$ as above.

Then $Q$ is $\amd$ with respect to $\aschema$
if and only if  the chase of  $I_0^\lift$ by  
$\Gamma^\lift$  satisfies~$Q'$.
\end{clm}

\begin{proof}
We know that~$\amd$ is equivalent to the containment $Q \subseteq_\Gamma Q'$ with 
$\Gamma =
  \Sigma \cup \Sigma' \cup \Delta$. This in turn is equivalent
to the existence of a chase proof of $Q'$ starting with $I_0$ using chase steps
from $\Gamma$. Thus what we need to show is how to convert  such a chase
proof to a chase proof of $Q'$ starting from $I_0^\lift$ using steps
of $\Gamma^\lift$, and vice versa.

We start with the converse direction.
It is easy to see that any chase proof of~$Q'$ formed from~$I_0^\lift$
  using~$\Gamma^\lift$ can be converted to a chase proof from~$I_0$ using~$\Gamma$. Indeed,
  Proposition~\ref{prop:linearizeaccessids} ensures that
any Boolean CQ over the primed signature that is derivable in~$I_0^\lift$
  using~$\Sigma^\lift$ can be derived in~$I_0$ using~$\Sigma$ and~$\Delta$. %
  And the Boolean CQs that are derivable using~$\Gamma^\lift$ are  those
  that can be derived by first applying chase steps
using $\Sigma^\lift$ to produce some set $S$ of primed facts,
and then applying chase steps involving constraints of $\Sigma'$  to the facts of~$S$. By converting $S$ to a Boolean CQ
we see that we can derive a homomorphic image of $S$ using
$\Sigma$ and $\Delta$. Thus  all Boolean CQs that
can be obtained from~$I_0^\lift$ using~$\Gamma^\lift$ can also be obtained using
  $\Sigma$, $\Delta$, and $\Sigma'$.

For the forward direction
  we consider a chase proof of~$Q'$ formed from~$I_0$ using $\Gamma$: that is,
using $\Sigma$, $\Sigma'$, and
  $\Delta$. We will show how to obtain  a chase proof of $Q'$ from  $I_0^\lift$ using steps of~$\Gamma^\lift$.
We observe that in our input chase proof
  we can assume that we first fire rules 
  of~$\Sigma \cup \Delta$ to get
a set~$S$ of primed facts,
and then fire rules of~$\Sigma'$ to get $I'$ containing all the facts of the
chase proof.
Now, from Proposition~\ref{prop:linearizeaccessids} we know that 
a homomorphic image~$S^\lift$ of the facts
  of~$S$ can all be derived from~$I_0^\lift$ using $\Gamma^\lift$. As
  $\Gamma^\lift$ contains $\Sigma'$, we can also derive 
a homomorphic image of the facts of~$I'$
  from $S^\lift$, and thus derive homomorphic
images of the facts in the chase proof of $Q'$. This justifies that $Q'$ is also
entailed by~$I_0^\lift$ and $\Gamma^\lift$, concluding the proof.
\end{proof}

Claim~\ref{clm:amdequiv} implies that
to solve the $\amd$ problem, it suffices
determine  whether the set of primed facts corresponding to~$Q'$ can be
derived from $I_0^\lift$ by applying chase steps with $\Gamma^\lift$. This in turn can be
determined using Proposition~\ref{prop:semiwidthclassic}. This concludes the
proof of Theorem~\ref{thm:npidsbounds} in the case without result bounds.

\paragraph{Proving Theorem~\ref{thm:npidsbounds} with result bounds.}
We now conclude the proof of Theorem~\ref{thm:npidsbounds} by handling the case
with result bounds. This will require only slight changes to the prior argument.
By Theorem~\ref{thm:simplifyidsexistence}, for any schema $\aschema$ whose constraints $\Sigma$ are $\incd$s,
we can reduce the monotone answerability problem to the same problem for the
existence-check simplification $\aschema^\dagger$ with no result bounds, by
replacing each result-bounded method $\mt$ on a  relation~$R$  with
a non-result-bounded access method $\mt^\dagger$ on a new relation  $\checkview_\mt$, and expanding
$\Sigma$ to a larger set of constraints $\Sigma^\dagger$,
adding new constraints that capture the semantics of the ``existence-check views'' $\checkview_\mt$:
\begin{itemize}
  \item (Relation-to-view): $R(\vec x, \vec y) \rightarrow 
    \checkview_\mt(\vec x)$;
  \item (View-to-relation): $\checkview_\mt(\vec x) \rightarrow \exists \vec
    y ~ R(\vec x, \vec y)$.
\end{itemize}
Note that these IDs do not have bounded width, hence we cannot simply reduce to
the case without result bounds that we have just proved. We will explain how to
adapt the proof to handle these IDs, namely, 
linearizing using Proposition~\ref{prop:linearizeaccessids} to $\incd$s of
bounded semi-width.

Let us consider the query containment problem for the monotone answerability problem of~$\Sigma^\dagger$.
This problem is of the form
$Q \subseteq_\Gamma Q'$, where
$\Gamma$ contains $\Sigma^\dagger$, its copy $(\Sigma^\dagger)'$, and the accessibility
axioms. These axioms can again be rephrased.  For each access method $\mt$ on a
relation~$R$, letting $\vec x$ denote the input positions of~$\mt$, we have the
following two axioms:
\begin{itemize}
  \item  (Truncated Accessibility): $\left(\bigwedge_i \accessible(x_i)\right) \wedge R(\vec x, \vec y) \rightarrow \bigwedge_i
  \accessible(y_i)$;
\item (Transfer): $\left(\bigwedge_i \accessible(x_i)\right) \wedge R(\vec x, \vec y)
  \rightarrow  R'(\vec x, \vec y)$.
\end{itemize}
In the above two items the relation $R$ can be any of the relations
of~$\Sigma^\dagger$, including relations of the
original signature and relations of the form $\checkview_\mt$. For relations in the original schema,
$\mt$ is an access method of~$\aschema$ that did not have a result
bound. For  the new relations, $\mt$ is a method of the form $\mt^\dagger$ introduced in the
existence-check simplification $\aschema^\dagger$ for a result-bounded method
of~$\aschema$, so $\mt^\dagger$ has no output positions: this means that, in this case,
the (Truncated Accessibility) axiom is vacuous and the (Transfer) axiom further
simplifies to:
\[
  \text{(Simpler Transfer):~}\Big(\bigwedge_i \accessible(x_i) \Big) \wedge \checkview_\mt(\vec x) \rightarrow \checkview'_\mt(\vec x).
\]

We first observe that in~$\Gamma$ we do not need to include 
the view-to-relation constraints of~$\Sigma^\dagger$:
facts over~$\checkview_\mt$ can only be formed from the corresponding
$R$-fact with the relation-to-view constraint, so triggers of the
view-to-relation constraints will never be active in the chase, and
we know that we can decide $Q \subseteq_\Gamma Q'$ by looking at restricted
chase sequences  --- i.e., where non-active triggers are never fired ---  hence removing
view-to-relation constraints makes no difference.
Similarly, we do not need  to include the relation-to-view constraints
of~$(\Sigma^\dagger)'$. These rules could fire to produce a new 
$\checkview'_\mt$-fact, but such a fact could only trigger the corresponding
view-to-relation constraint of~$(\Sigma^\dagger)'$, resulting in a state of the chase
that has a homomorphism to the one before the firing of the relation-to-view
constraint. Thus such firings can not lead to new matches.
Thus, $\Gamma$ consists now of~$\Sigma$, of~$\Sigma'$,
of (Truncated Accessibility) and
(Transfer) axioms for each method $\mt$ having no result bound in~$\aschema$,
and for each method $\mt$ with a result bound in~$\aschema$ we have
a relation-to-view
constraint from~$R$ to~$\checkview_\mt$ that comes from~$\Sigma^\dagger$,
a view-to-relation constraint from~$\checkview_\mt'$ to~$R'$
that comes from $(\Sigma^\dagger)'$, and a 
(Simpler Transfer) axiom.

We next note that we can normalize chase proofs with~$\Gamma$ so that the relation-to-view constraints
are applied only prior to (Simpler Transfer). 
Thus, for each result-bounded method $\mt$ of~$\aschema$,
we can merge the relation-to-view rule
from~$R$ to~$\checkview_\mt$, the
(Simpler Transfer) axiom from~$\checkview_\mt$ to~$\checkview_\mt'$, and
the view-to-relation rules from~$\checkview_\mt'$ to~$R'$, into an axiom of the
following form, where $\vec x$ denotes the input positions of~$\mt$:
\[
  \mbox{(Result-bounded Fact Transfer) } \Big(\bigwedge_i \accessible(x_i) \wedge R(\vec x, \vec y)\Big) \rightarrow  \exists \vec z ~ R'(\vec x, \vec z).
\]
To summarize, the resulting axioms 
consist of:
\begin{itemize}
  \item The original constraints $\Sigma$ of the schema;
  \item Their primed copy $\Sigma'$;
  \item The (Truncated Accessibility) and (Transfer) axioms for each access method
    without result bounds;
  \item The (Result-bounded Fact Transfer) axioms for access methods with result
    bounds.
\end{itemize}
In other words, the only difference with the setting without result bounds is
the last bullet point corresponding to (Result-bounded Fact Transfer). 

We will now need 
an extension of  the linearization result,
Proposition~\ref{prop:linearizeaccessids}, to handle these additional
constraints:

\begin{prop} \label{prop:linearizeaccessidsbounds}
  For any fixed~$w \in \NN$,
  given a set $\Sigma$ of $\incd$s of width~$w$ and a set
  $\Delta$ of truncated accessibility, transfer, and Result-bounded fact transfer axioms,
  and given
  a set of facts~$I_0$,
  we can compute in $\ptime$ a set of $\incd$s $\Sigma^\lift$ of semi-width~$w$
  and a set of facts $I_0^\lift$ satisfying the following:
  for any  Boolean CQ
  $Q^*$ over the primed signature using constants from $I_0$
  and existentially quantified variables,
  $Q^*$ is entailed from~$I_0$, $\Sigma$,
and~$\Delta$ iff
$Q^*$ is entailed from $I_0^\lift$
 and $\Sigma^\lift$.
\end{prop}

The proof of this will be a variation of the argument for Proposition~\ref{prop:linearizeaccessids}. It will be explained at the end of Section~\ref{sec:linearize}.

Thus we follow the same route as before: linearization, 
noting that the special form of our constraints results in linear constraints of bounded
semi-width.
This completes the proof of Theorem~\ref{thm:npidsbounds} in the case with
result bounds.

\subsection{Proof of the Linearization Results
(Proposition~\ref{prop:linearizeaccessids} and~\ref{prop:linearizeaccessidsbounds})}
\label{sec:linearize}

We now turn to the missing element in the proof of Theorem~\ref{thm:npidsbounds}, which are our
linearization results, Proposition~\ref{prop:linearizeaccessids} and its
analog for result-bounded methods,
Proposition~\ref{prop:linearizeaccessidsbounds}.

We first give the intuition on how our linearization works. In the tree-like
chase for guarded TGDs, we have steps that create new nodes, and also \emph{propagation
steps}, that replicate facts across tree nodes. An intermediate goal will be to show
that for guarded TGDs we can get a similar chase where we do not need to propagate
across tree nodes. This will be the \emph{shortcut chase}, defined later, where
we only grow the tree and fire full rules at a given node, with no propagation.
Note that the point of the shortcut chase is not to actually perform or
construct it, but to reason about it.
Once we have defined the shortcut chase and shown
it is complete, it will be easy to perform linearization.
The shortcut chase will make use of  full GTGDs that we derive from our
original set of GTGDs. The saturation process that creates these GTGDs will
be a first step.

First we will
review the  notion of tree-structured chase proof that is well-known for guarded TGDs 
\cite{datalogpmj}, and  show that we can further enforce
the  \emph{downward-free} property, where facts only propagate back from a child to its
ancestors in the tree.
This is a step towards simplifying propagation
in the chase.
Second we will need to define a more general notion of \emph{truncated accessibility axioms}, 
and give a $\ptime$ algorithm for generating the ones that are small enough: this
will give us the full GTGDs that we will need.
Finally we present shortcut chase proofs, where these dependencies
are fired in an even more specific order, and show that this definition of the
chase is still complete.
Lastly we use these tools to prove
Proposition~\ref{prop:linearizeaccessids}.

\paragraph{Tree-like chase proofs and the downward-free property.}
As a step towards our linearization result,
we now present a general result about
chase proofs with \emph{single-headed GTGDs}, i.e., GTGDs having a single atom
in the head. This will be applicable in particular to our analysis
of the chase with $\incd$s and candidate truncated accessibility axioms.

For any  chase sequence $I_0 \ldots I_n$ using single-headed GTGDs, we can associate
a \emph{tree-like chase sequence}, i.e., a sequence $T_0 \ldots T_{n'}$  of \emph{chase trees}.
 A chase  tree
  $T_i$ in such a sequence consists of a tree structure with a function $\factsof_i$ that maps
each node  of $T_i$ to a collection of facts. 
Each $T_i$ is associated to the instance formed by
unioning all the facts in its nodes, i.e., the union of $\factsof_i(v)$ across
all nodes~$v$.
Further, if $v$ is not the root, there is a fact $F$ in $\factsof_{i}(v)$, the
\emph{birth fact of~$v$},  which serves as a guard for the elements  in the
facts of~$\factsof_i(v)$.  This fact $F$ for the node~$v$ will never change throughout the sequence, so
we denote it by $\birthfact(v)$.

In a tree-like chase sequence $T_0  \ldots T_{k'}$, consecutive chase trees will be linked
by two kinds of steps.
First, there will be \emph{chase steps}, which add a fact to the tree, possibly in a
new node.
If $T_{i+1}$ is produced from
$T_i$ by a chase step, this will correspond
to a valid chase step for the two corresponding instances.
Second, a step from
$T_i$ to $T_{i+1}$  can be a \emph{propagation step}, which  does not change the underlying instance, but
just 
copies facts from one node to another, i.e., it modifies $\factsof$ while maintaining the other components.
Both steps are described in detail below.

For the case of \emph{chase steps}, when we perform a chase step to transform $T_i$ to~$T_{i+1}$, we will always
require it to be \emph{tree-friendly}, i.e., we require that
the image of the trigger lies in
$\factsof_i(v)$ for some node $v$.
When a chase step fires a trigger for a GTGD $\tau$ to create a
fact~$F$, we choose one such node $v$ in which the image lies. 
If $\tau$ is not full, then we extend the  sequence to
$T_{i+1}$ by adding to $T_i$ a new node $v'$ as a child of~$v$,
setting $\birthfact(v') \colonequals F$, and setting $\factsof_{i+1}(v')$ to be~$F$ along with any facts in $v$ that
are guarded by $F$ --- these facts are \emph{forward propagated} from~$v'$ to~$v$.
If $\tau$ is full, then we extend the sequence by defining
$T_{i+1} \colonequals T_i$ but changing the function $\factsof_{i+1}$.
We set $\factsof_{i+1}(v)$ to be $\factsof_{i}(v) \cup \{F\}$ to create
the new fact. 

For the case of \emph{propagation steps}, such a step can only take place if the
preceding step was a chase step
with a full GTGD~$\tau$.
Letting $F = R(\vec c)$ be the newly created fact, we
consider the set $B_{\vec c}$ of all nodes that contain a guard for~$\vec c$. Our process ensures that
$B_{\vec c}$ will form a connected subtree of the chase tree $T_i$,  and it contains 
the node~$v$ in which the chase step was performed.
We choose a subset $B'$ of $B_{\vec c}$, and propagate the new fact to all
these nodes: for every node~$v' \in B'$, we add~$F$ to $\factsof_{i+1}(v')$.
Note that, unlike forward propagation, propagation
steps allow us to propagate a fact upwards (from descendants to ancestors) as
well as downwards (from descendant to ancestor). Further, it is optional, i.e.,
we can choose not to propagate. %

It is clear that every tree-like chase sequence $T_0 \ldots T_{n'}$ induces
a chase sequence in the usual sense of instances $I_0 \ldots I_n$: propagation steps in the $T_i$ do not result in any
change to the instance, thus $n$ may be less than $n'$.
We say that such a sequence is a \emph{tree-like chase proof} of some entailment
if the resulting sequence $I_0 \ldots I_n$ is, and in addition
$T_0$ consists of only  a single node
with $\factsof_0(r)=I_0$.
Note that any chase proof with single-headed GTGDs
can be made into a tree-like chase proof: we can always choose to propagate facts
everywhere they are guarded, and then the  restriction to tree-friendly chase steps
is without loss of generality.

An example is given in Figure~\ref{fig:exmpl_tree_chase_seq},
and explained in more detail in Example~\ref{exmpl:tree_like_chase_sequence} just below.
Note that as the chase proceeds, we only add nodes to the chase tree and add facts
to existing nodes. In other words, given a tree node $v$  associated to some instance $I_i$ in a chase
proof, $v$ will exist at each later stage $I_j$, but may have additional facts.

We will be particularly interested in
the case of proofs with $\incd$s and candidate derived truncated accessibility
axioms. In this case  the full TGDs include the full $\incd$s, as well as the
candidate derived truncated accessibility axioms,
which generate new accessibility facts.

\begin{exa}\label{exmpl:tree_like_chase_sequence}
We use an example from \cite{kevinarxiv}.
We consider the initial instance $I_0=\{R(c,d)\}$ and a set of single-headed GTGDs $\Sigma$
\begin{align*}
R(x_1,x_2)&\rightarrow\exists y\,S(x_1,y),&
R(x_1,x_2)&\rightarrow\exists y\,T(x_1,x_2,y),\\
T(x_1,x_2,x_3)&\rightarrow\exists y\,U(x_1,x_2,y),&
U(x_1,x_2,x_3)&\rightarrow P(x_2),\\
T(x_1,x_2,x_3)\land P(x_2)&\rightarrow M(x_1),&
S(x_1,x_2)\land M(x_1)&\rightarrow\exists y\, N(x_1,y),
\end{align*}
the sequence
\begin{align*}
I_0&=\{R(c,d)\},I_1=I_0\cup\{S(c,d_1)\},I_2=I_1\cup\{T(c,d,d_2)\},I_3=I_2\cup\{U(c,d,d_3)\},\\
I_4&=I_3\cup\{P(d)\},I_5=I_4\cup\{M(c)\},I_6=I_5\cup\{N(c,d_4)\}
\end{align*}
is a chase sequence for $I_0$ and $\Sigma$.
A corresponding tree-like chase sequence $T_0, \ldots, T_8$ is depicted in
  Figure~\ref{fig:exmpl_tree_chase_seq}.

\begin{figure}[ht]
        \centering
        \begin{tikzpicture}
\tikzset{node distance=0.5cm,
	r-0/.pic={\node (r) {$R(c,d)$};},
	I0/.pic={\pic{r-0};},
	cl-1/.pic={\node (cl) {$S(c,d_1)$};},
	I1/.pic={%
          \pic[trigger]{r-0};
		\pic[newnode,below=of r]{cl-1};
		\path (r) edge (cl);
	},
	cr-2/.pic={\node (cr) {$T(c,d,d_2)$};},
	I2/.pic={%
          \pic[trigger]{r-0};
		\pic[below=of r,xshift=-1.1cm]{cl-1};
		\pic[newnode][right=of cl]{cr-2};
		\path (r) edge (cl)
		(r) edge (cr);
	},
	crl-3/.pic={\node (crl) {$U(c,d,d_3)$};},
	I3/.pic={%
		\pic{r-0};
		\pic[below=of r,xshift=-1.1cm]{cl-1};
		\pic[trigger,right=of cl]{cr-2};
		\pic[newnode][below=of cr]{crl-3};
		\path (r) edge (cl)
		(r) edge (cr) 
		(cr) edge (crl);
	},
        crl-4/.pic={\node (crl) {$U(c,d,d_3),P(d)$};},
        crl-4b/.pic={\node (crl) {$U(c,d,d_3),\textcolor{red}{P(d)}$};},
        crl-4c/.pic={\node (crl) {$U(c,d,d_3),\textcolor{blue}{P(d)}$};},
	I4-/.pic={%
		\pic{r-0};
		\pic[below=of r,xshift=-1.1cm]{cl-1};
		\pic[right=of cl]{cr-2};
		\pic[trigger,below=of cr]{crl-4b};
		\path (r) edge (cl)
		(r) edge (cr) 
		(cr) edge (crl);
	},
	r-4/.pic={\node (r) {$R(c,d),P(d)$};},
        r-4b/.pic={\node (r) {$R(c,d),\textcolor{red}{P(d)}$};},
	cr-4/.pic={\node (cr) {$T(c,d,d_2),P(d)$};},
        cr-4b/.pic={\node (cr) {$T(c,d,d_2),\textcolor{red}{P(d)}$};},
	I4/.pic={%
		\pic{r-4b};
		\pic[below=of r,xshift=-1.5cm]{cl-1};
		\pic[right=of cl]{cr-4b};
		\pic[below=of cr]{crl-4c};
		\path (r) edge (cl)
		(r) edge (cr) 
		(cr) edge (crl);
	},
	cr-5/.pic={\node (cr) {$T(c,d,d_2),P(d),M(c)$};},
        cr-5b/.pic={\node (cr) {$T(c,d,d_2),P(d),\textcolor{red}{M(c)}$};},
        cr-5c/.pic={\node (cr) {$T(c,d,d_2),P(d),\textcolor{blue}{M(c)}$};},
	I5-/.pic={%
		\pic{r-4};
		\pic[below=of r,xshift=-1.5cm]{cl-1};
		\pic[trigger,right=of cl]{cr-5b};
		\pic[below=of cr]{crl-4};
		\path (r) edge (cl)
		(r) edge (cr) 
		(cr) edge (crl);
	},
	r-5/.pic={\node (r) {$R(c,d),P(d),M(c)$};},
	crl-5/.pic={\node (crl) {$U(c,d,d_3),P(d),M(c)$};},
	cl-5/.pic={\node (cl) {$S(c,d_1),M(c)$};},
        r-5b/.pic={\node (r) {$R(c,d),P(d),\textcolor{red}{M(c)}$};},
        crl-5b/.pic={\node (crl) {$U(c,d,d_3),P(d),\textcolor{red}{M(c)}$};},
        cl-5b/.pic={\node (cl) {$S(c,d_1),\textcolor{red}{M(c)}$};},
        cl-5d/.pic={\node (cl) {$S(c,d_1),\textcolor{blue}{M(c)}$};},
	I5/.pic={%
		\pic{r-5b};
		\pic[below=of r,xshift=-2cm]{cl-5b};
		\pic[right=of cl]{cr-5c};
		\pic[below=of cr]{crl-5b};
		\path (r) edge (cl)
		(r) edge (cr) 
		(cr) edge (crl);
	},
        cll-6-/.pic={\node (cll) {$N(c,d_4), M(c)$};},
	I6-/.pic={%
		\pic{r-5};
		\pic[trigger,below=of r,xshift=-2cm]{cl-5d};
		\pic[newnode,below=of cl]{cll-6-};
		\pic[right=of cl]{cr-5};
		\pic[below=of cr]{crl-5};
		\path (r) edge (cl)
		(r) edge (cr) 
		(cl) edge (cll)
		(cr) edge (crl);
	},
	%cll-6/.pic={\node (cll) {$N(c,d_4),M(c)$};},
	%I6/.pic={%
	%	\pic{r-5};
	%	\pic[below=of r,xshift=-2cm]{cl-5};
	%	\pic[below=of cl]{cll-6};
	%	\pic[right=of cl]{cr-5};
	%	\pic[below=of cr]{crl-5};
	%	\path (r) edge (cl)
	%	(r) edge (cr) 
	%	(cl) edge (cll)
	%	(cr) edge (crl);
	%},
}
\node[scope] (I0) {
	\begin{tikzpicture}
		\pic{I0};
		\node[lnode,above=0.1cm of r]{$T_0$};
	\end{tikzpicture}
};
\node[scope,right=of I0] (I1) {
	\begin{tikzpicture}
		\pic{I1};
		\node[lnode,above=0.1cm of r]{$T_1$};
	\end{tikzpicture}
};
\node[scope,right=of I1] (I2) {
	\begin{tikzpicture}
		\pic{I2};
		\node[lnode,above=0.1cm of r]{$T_2$};
	\end{tikzpicture}
};
\node[scope,right=of I2] (I3) {
	\begin{tikzpicture}
		\pic{I3};
		\node[lnode,left=of r]{$T_3$};
	\end{tikzpicture}
};
\pic{drawv=I0/I1/I3/I3};
\pic{drawv=I1/I2/I3/I3};
\pic{drawvr=I2/I3};
\node[scope,right=of I3] (I4-) {
	\begin{tikzpicture}
		\pic{I4-};
		\node[lnode,left=of r]{$T_4$};
	\end{tikzpicture}
};
\pic{drawv=I3/I4-/I3/I3};
\node[scope,below=1.7cm of I0,xshift=1.5cm] (I4) {
	\begin{tikzpicture}
		\pic{I4};
		\node[lnode,left=of r]{$T_5$};
	\end{tikzpicture}
};
\node[scope,right=of I4] (I5-) {
	\begin{tikzpicture}
		\pic{I5-};
		\node[lnode,left=of r] {$T_6$};
	\end{tikzpicture}
};
\pic{drawv=I4/I5-/I4/I4};
\node[scope,right=of I5-] (I5) {
	\begin{tikzpicture}
		\pic{I5};
		\node[lnode,left=of r] {$T_7$};
	\end{tikzpicture}
};
\pic{drawv=I5-/I5/I4/I4};
\node[scope,below=0.8cm of I4,xshift=7cm] (I6-) {
	\begin{tikzpicture}
		\pic{I6-};
		\node[lnode,left=of r] {$T_8$};
	\end{tikzpicture}
};
% \node[scope,right=of I6-] (I6) {
% 	\begin{tikzpicture}
% 		\pic{I6};
% 		\node[lnode,left=of r] {$T_9$};
% 	\end{tikzpicture}
% };
%\pic{drawvr=I6-/I6};
%\pic{I6-};
\pic{drawh=I3/I4/I4/I5};
\pic{drawh=I4/I6-/I4/I5};
\end{tikzpicture}
        \caption{Tree-like chase for
        Example~\ref{exmpl:tree_like_chase_sequence}.}
        \label{fig:exmpl_tree_chase_seq}
\end{figure}

  The representation is the following.
  When we have performed a chase step with a non-full GTGD (e.g., from $T_0$
  to~$T_1$), the new created node is represented in red, and the node containing
  the image of the trigger (its parent node) has a blue border. When such a
  step performs forward propagation of facts (e.g., from $T_7$ to $T_8$), the
  forward propagated fact are written in blue in the parent node, and in red in
  the newly created node. When we have performed a chase step with a full GTGD
  (e.g., from $T_3$ to $T_4$), the node containing the image of the trigger
  again has a blue border, and the new fact (created in the same node) is in
  red. When we have performed a propagation step (e.g., from $T_4$ to~$T_5$), the fact being propagated is
  written in blue in the node from where we propagate it --- namely, the node
  where it was created in the previous step by a full chase step, and is
  written in red in the nodes where it is propagated. Note that, in these
  examples, propagation steps always propagate new facts everywhere they are
  guarded.
\end{exa}

Our linearization result will rely on the fact that chase proofs can be
normalized to ensure that the propagation of facts in propagation steps only
happens in the ``upwards'' direction, that is, towards the root of the tree.
Note that this does not affect the propagation of facts to child nodes when
firing a chase step with a non-full GTGD\@.

\begin{defi}[Downward-free chase sequence]\label{defn:one_pass_chase}
We say that a tree-like chase sequence $T_0 \dotsc T_n$ with single-headed GTGDs is  \emph{downward
-free}
if propagation steps always propagate a fact $F$ to ancestors of the node where
  it is created.
  A downward-free
chase proof of an entailment is just a downward-free chase sequence that is a chase proof.
\end{defi}

\begin{exa}\label{exmpl:one_pass_chase_sequence}
  We continue with Example~\ref{exmpl:tree_like_chase_sequence}, taken from \cite{kevinarxiv}.
The chase depicted in Figure~\ref{fig:exmpl_tree_chase_seq} is not
  downward-free. Indeed, the propagation step from~$T_6$ to~$T_7$ propagates the
  fact $M(c)$ to the left child of the rood, which is not an ancestor of the
  node where it was created.
However, if we do not propagate it to this node, we can no longer perform the
  chase step from~$T_7$ to~$T_8$ to create $N(c, d_4)$.

Instead, we can redo steps in the chase to make it downward-free. We design a
  tree-like chase sequence $T_0', \ldots, T_{10'}$, with $T_i = T_i'$ for all $0
  \leq i \leq 6$. We depict $T_5' = T_5$, $T_6' = T_6$, and $T_7', \ldots, T_{10}'$ in 
Figure~\ref{fig:exmpl_one_pass_chase_seq}.
\begin{figure}[ht]
	\centering
	\begin{tikzpicture}
\tikzset{node distance=0.5cm,
	cl-1/.pic={\node (cl) {$S(c,d_1)$};},
	crl-4/.pic={\node (crl) {$U(c,d,d_3),P(d)$};},
        crl-4c/.pic={\node (crl) {$U(c,d,d_3),\textcolor{blue}{P(d)}$};},
	r-4/.pic={\node (r) {$R(c,d),P(d)$};},
        r-4b/.pic={\node (r) {$R(c,d),\textcolor{red}{P(d)}$};},
	cr-4/.pic={\node (cr) {$T(c,d,d_2),P(d)$};},
        cr-4b/.pic={\node (cr) {$T(c,d,d_2),\textcolor{red}{P(d)}$};},
	I4/.pic={%
		\pic{r-4b};
		\pic[below=of r,xshift=-1.5cm]{cl-1};
		\pic[right=of cl]{cr-4b};
		\pic[below=of cr]{crl-4c};
		\path (r) edge (cl)
		(r) edge (cr) 
		(cr) edge (crl);
	},
	cr-5/.pic={\node (cr) {$T(c,d,d_2),P(d),M(c)$};},
        cr-5b/.pic={\node (cr) {$T(c,d,d_2),P(d),\textcolor{red}{M(c)}$};},
        cr-5c/.pic={\node (cr) {$T(c,d,d_2),P(d),\textcolor{blue}{M(c)}$};},
        cr-5cc/.pic={\node (cr) {$T(c,d,d_2),\textcolor{blue}{P(d)},\textcolor{blue}{M(c)}$};},
	I4'-1/.pic={%
		\pic{r-4};
		\pic[below=of r,xshift=-1.5cm]{cl-1};
		\pic[trigger,right=of cl]{cr-5b};
		\pic[below=of cr]{crl-4};
		\path (r) edge (cl)
		(r) edge (cr) 
		(cr) edge (crl);
	},
	r-5/.pic={\node (r) {$R(c,d),P(d),M(c)$};},
        r-5b/.pic={\node (r) {$R(c,d),P(d),\textcolor{red}{M(c)}$};},
        r-5c/.pic={\node (r) {$R(c,d),P(d),\textcolor{blue}{M(c)}$};},
	I4'/.pic={%
		\pic{r-5b};
		\pic[below=of r,xshift=-1.5cm]{cl-1};
		\pic[right=of cl]{cr-5c};
		\pic[below=of cr]{crl-4};
		\path (r) edge (cl)
		(r) edge (cr) 
		(cr) edge (crl);
	},
	crr-5/.pic={\node (crr) {$U(c,d,d_4),P(d),M(c)$};},
	I4''/.pic={%
		\pic{r-5};
		\pic[below=of r,xshift=-2cm]{cl-1};
		\pic[trigger,right=of cl]{cr-5cc};
		\pic[below=of cr,xshift=-2cm]{crl-4};
		\pic[newnode,right=of crl]{crr-5};
		\path (r) edge (cl)
		(r) edge (cr) 
		(cr) edge (crl)
		(cr) edge (crr);
	},
	cnr-5/.pic={\node (cnr) {$S(c,d_5),M(c)$};},
	I5/.pic={%
                \pic[trigger]{r-5c};
		\pic[below=of r]{cr-5};
		\pic[left=of cr]{cl-1};
		\pic[below=of cr,xshift=-2cm]{crl-4};
		\pic[right=of crl]{crr-5};
		\pic[newnode][right=of cr]{cnr-5};
		\path (r) edge (cl)
		(r) edge (cr) 
		(cr) edge (crl)
		(cr) edge (crr)
		(r) edge (cnr);
	},
	cnrl-6-/.pic={\node (cnrl) {$N(c,d_6)$};},
	I6-/.pic={%
		\pic{r-5};
		\pic[below=of r,xshift=-0cm]{cr-5};
		\pic[left=of cr]{cl-1};
		\pic[below=of cr,xshift=-2cm]{crl-4};
		\pic[right=of crl]{crr-5};
		\pic[trigger,right=of cr]{cnr-5};
		\pic[newnode,below=of cnr,xshift=1.5cm]{cnrl-6-};
		\path (r) edge (cl)
		(r) edge (cr) 
		(cr) edge (crl)
		(cr) edge (crr)
		(r) edge (cnr)
		(cnr) edge (cnrl);
	},
	cnrl-6/.pic={\node (cnrl) {$N(c,d_6),M(c)$};},
	I6/.pic={%
		\pic{r-5};
		\pic[below=of r,xshift=-0cm]{cr-5};
		\pic[left=of cr]{cl-1};
		\pic[below=of cr,xshift=-2cm]{crl-4};
		\pic[right=of crl]{crr-5};
		\pic[right=of cr]{cnr-5};
		\pic[below=of cnr,xshift=2cm]{cnrl-6};
		\path (r) edge (cl)
		(r) edge (cr) 
		(cr) edge (crl)
		(cr) edge (crr)
		(r) edge (cnr)
		(cnr) edge (cnrl);
	},
}
\node[scope] (I4) {
	\begin{tikzpicture}
		\pic{I4};
		\node[lnode,right=of r] {$T_5'$};
	\end{tikzpicture}
};
\node[scope,right=of I4] (I4'-1) {
	\begin{tikzpicture}
		\pic{I4'-1};
		\node[lnode,right=of r] {$T_6'$};
	\end{tikzpicture}
};
\pic{drawvr=I4/I4'-1};
\node[scope,right=of I4'-1] (I4') {
	\begin{tikzpicture}
		\pic{I4'};
		\node[lnode,right=of r] {$T_7'$};
	\end{tikzpicture}
};
\pic{drawvr=I4'-1/I4'};
\node[scope,below=of I4,xshift=2cm] (I4'') {
	\begin{tikzpicture}
		\pic{I4''};
		\node[lnode,right=of r] {$T_8'$};
	\end{tikzpicture}
};
\node[scope,right=of I4''] (I5) {
	\begin{tikzpicture}
		\pic{I5};
		\node[lnode,right=of r] {$T_9'$};
	\end{tikzpicture}
};
\pic{drawvr=I4''/I5};
\node[scope,below=of I4'',xshift=5cm] (I6-) {
	\begin{tikzpicture}
		\pic{I6-};
                \node[lnode,right=of r] {$T_{10}'$};
	\end{tikzpicture}
};
%\node[scope,right=of I6-] (I6) {
%	\begin{tikzpicture}
%		\pic{I6};
%		\node[lnode,right=of r] {$I_6$};
%	\end{tikzpicture}
%};
%\pic{drawvr=I6-/I6};
\pic{drawh=I4''/I6-/I4''/I5};
\pic{drawh=I4/I4''/I4''/I5};
\end{tikzpicture}
	\caption{Modification of the chase from Figure~\ref{fig:exmpl_tree_chase_seq} to obtain a
downward-free chase.
        }  \label{fig:exmpl_one_pass_chase_seq}
\end{figure}
\end{exa}

One can think of downward-free chase sequences as a chase version which is more
similar to the chase with linear TGDs, because it is never useful to propagate
facts when chasing with such TGDs --- further chase steps cannot use the
propagated fact to fire a rule.
Another advantage of downward-free proofs is that if we follow
the evolution of a subtree of some node $v$ within a proof, what we see
happening in that subtree
is a self-contained proof of all facts derived in the subtree, using
only the initial facts of $v$. Indeed, the chase steps performed afterwards
outside of the subtree will never modify the contents of the subtree.

\begin{prop}\label{prop:one_pass_value}
Let $T_0 \ldots T_j$ be a downward-free chase proof. Assume that, in moving
to $T_i$, we perform a chase step that creates a node $v$.
Let $T'_i \ldots T'_m$ be the tree-like chase sequence obtained from $T_i \ldots T_j$ by
restricting to the subtree rooted at~$v$ in $T_i \ldots T_j$, eliminating
  duplicate consecutive chase trees.
Then $T'_i \ldots T'_m$ is a downward-free chase proof of the facts of~$T'_m$ from
  the facts $\factsof_i(v)$ of~$v$ in~$T_i$.
\end{prop}

\begin{proof}
  Note that $T'_i$ consists of a single node containing the facts of~$v$
  in~$T_i$.
  By the definition of the downward-free chase, the triggers in all chase steps within $T_i \ldots T_j$
  used only facts that were generated within the subtree of $v$ earlier in the same
sequence in a node in the subtree of~$v$. Thus, by an immediate
  induction on~$m-i$, all these steps can also
  be triggered in~$T'_i \ldots T'_m$. So $T'_i \ldots T'_m$ is indeed a chase
  sequence proving the required facts, and it is downward-free because the original sequence was.
\end{proof}

This is  a variation of
Corollary~3.1.5 of \cite{kevinarxiv}. Similar statements appear in our earlier
work~\cite{oldarxiv}. 

We now show the key claim that we can always restrict to downward-free chase sequences:
\newcommand{\thmgtgdonepass}{
For every tree-like chase sequence using single-headed GTGDs  $T_0 \ldots T_n$, %
there is a downward-free tree-like chase sequence
$T_0 = \overline{T_0}, \ldots \overline{T_m}$
such that there is a homomorphism $h$ from the instance of $T_n$ to the
instance of $\overline{T_m}$
with $h(c)=c$ for any values $c$ in the domain of the instance of~$T_0$.
}
\begin{thm}\label{thm:gtgd_one_pass} 
  \thmgtgdonepass
\end{thm} 

In particular, if we have a
 chase proof of a UCQ~$Q$ from an instance $I_0$ using GTGDs $\Sigma$, then
we have downward-free  proof of~$Q$, starting from $T_0$ consisting of a single node containing $I_0$, applying chase
steps via $\Sigma$.
The proof is based on the idea in Figure~\ref{fig:exmpl_one_pass_chase_seq}. 
It is inspired by the conference version of this paper,
and by Proposition~3.1.6 in
\cite{kevinarxiv}. It is presented in Appendix~\ref{sec:onepass}. Note that, in
this downward-free chase, we may need to fire triggers that are not active.

We emphasize that the downward-free chase is never used as an algorithm
to get better complexity bounds directly.
We will only use it to justify steps in our linearization process.

\paragraph{Generalized  truncated accessibility axioms and saturation.} 
Having presented the downward-free chase, we return to the proof of our first
linearization result (Proposition~\ref{prop:linearizeaccessids}). Recall that
this result 
applies to constraints
formed of $\incd$s $\Sigma$ of width~$w$ and a set $\Delta$ of truncated accessibility and transfer
axioms. Recall that a transfer axiom is of the form:
\[\Big(\bigwedge_i \accessible(x_i)\Big) \wedge R(\vec x, \vec y)
  \rightarrow  R'(\vec x, \vec y).\]

In the first step towards linearization, we 
 will  perform a construction that enlarges the truncated accessibility axioms to certain
TGDs that have a similar shape,
which we call  \emph{candidate
derived truncated accessibility axioms}. By this  we mean
any TGD of the following form:
\[\Big(\bigwedge_{i \in P} \accessible(x_i)\Big) \wedge R(\vec x) \rightarrow
\accessible(x_j)\]
where $R$ is a relation and~$P$ is a subset of the positions of~$R$. 
Notice that the axioms of the form (Truncated Accessibility) defined earlier
can indeed be rewritten to be of this form:  the only difference from their original
form is that we have rewritten them
further to ensure that the head always contains a single accessibility fact.

Intuitively, such an axiom tells us that, when a subset of the elements of
an~$R$-fact are accessible, then another element of the fact becomes accessible
(by performing an access). We will start by considering what we call the
\emph{original truncated accessibility axioms}: these are
simply the (Truncated Accessibility) axioms in the set~$\Delta$, which are in
the form above, i.e., with a single 
$\accessible$ fact in the head.
For these axioms, the set $P$ is the set of input
positions of some method~$\mt$ on~$R$.
We will also study  candidate derived truncated
accessibility axioms that  are not necessarily given in~$\Delta$, but which 
are semantically \emph{entailed} by the original
truncated accessibility axioms in~$\Delta$ and by the constraints in~$\Sigma$.
By entailment, we always mean the semantic notion discussed in Section~\ref{sec:prelims},
i.e.,
entailments witnessed
by a chase proof starting with facts in the body of the  dependency, concluding
with an instance having a suitable homomorphism from the head of the dependency.
The candidate derived truncated accessibility axioms that are entailed are
simply called the \emph{derived truncated accessibility axioms}.

There can be exponentially many derived truncated accessibility axioms, but we will not
need to compute all of them: it will suffice to compute those of small
\emph{breadth}. Formally,
the \emph{breadth} of a candidate derived truncated accessibility axiom is  the size of~$P$.
Note that the  number of possible candidate derived truncated accessibility axioms of  breadth
$b$ is at most $r \cdot a^{b+1}$, where $r$ is the number of relations
in the signature and  $a$ is the maximal arity of a relation. We show that we
can efficiently
compute the derived truncated accessibility axioms of a given breadth, by
introducing a \emph{truncated accessibility axiom saturation algorithm}.

The algorithm iteratively builds up a set $O$ of triples $(R,\bar{p},j)$ with~$\bar{p}$ a
set of positions of~$R$ of size at most $w$ and $j$ a position of~$R$. Each such
triple represents the following candidate derived truncated accessibility axiom of breadth $\leq
w$:
\[
  \Big(\bigwedge_{i \in \bar{p}} \accessible(x_i)\Big) \wedge R(\vec x) \rightarrow
\accessible(x_j).
\]

The first step of the algorithm is to set $O \colonequals \{(R, \bar{p}, j) \mid j \in \bar{p}\}$, representing
trivial axioms.
The algorithm then repeats the steps below:
\begin{itemize}
  \item ($\incd$):
    If we have an $\incd$ 
$\forall \vec x ~ R(\vec x) \rightarrow \exists \vec y ~ S(\vec z)$, where 
     $x_{j_1}, \dots, x_{j_{m'}}, x_j$ (with $m'+1 \leq w$) are
     exported variables that appear respectively
in positions $k_1 \ldots k_{m'}, k$ within the head atom $S(\vec z)$,
     and if we have $(S,\{{k_1} \ldots {k_{m'}}\},k) \in O$,
     then we add the tuple $(R,\{j_1 \ldots j_{m'}\},j)$ to~$O$.

     The intuition for the ($\incd$) step is that derived truncated accessibility axioms
     that hold on the target relation $S$ can be ``propagated upwards'' to~$R$, i.e., if
     an $\accessible$ fact is created using the $S$-fact, then the same creation
     can happen using the $R$-fact.
   \item (Transitivity): If there exists a relation $R$, a set of positions
     $\bar{p}$ of~$R$,  and a set
     of positions $\{t_1 \ldots t_m\}$ of~$R$ with
  $m\leq w$ such that we have $(R, \bar{p}, t_i) \in O$ for all $1 \leq i \leq m$,
    and we have $(R, \bar{r}, t') \in O$ with
    $\bar{r} \subseteq \bar{p} \cup \{t_1 \ldots t_m\}$, then we add $(R, \bar{p}, t')$
to~$O$.

The intuition for (Transitivity) is that we add triples that
result from the natural entailment relation on triples,  provided that the number of intermediate values
that are projected away is not more than $w$.
\item (Access): If we have a method $\mt$ on~$R$ with input positions $j_1
  \ldots j_m$
and a set $\bar{p}$ of at most $w$ positions such that  $(R,\bar{p}, j_i) \in O$ for all
$1 \leq i \leq m$, then we add $(R,\bar{p}, j)$ to~$O$ for all $j$ between~$1$
  and the arity of~$R$.

To understand (Access), notice that we cannot add triples corresponding to all access methods,
since the number of inputs to an access method might be above the breadth bound. Thus (Access)
is actually a special kind of transitivity calculation that adds triples of low breadth
that can result from composing derived truncated accessibility axioms with an access method.
\end{itemize}
We continue the algorithm until we reach a fixpoint.

Note that  the maximal number of triples produced is
$r \cdot a^{w+1}$, with~$r$ the number of relations in the schema and $a$
  the maximal arity of a relation. Thus
a fixpoint must be reached in this number of steps.
Thus, for fixed~$w$, it is clear that the algorithm runs
  in polynomial time in~$\Sigma$ and in the set of access methods. 

We will
  show that this algorithm correctly computes all derived truncated accessibility
  axioms satisfying the breadth bound:

\newcommand{\derivedaccessids}{
For any fixed $w \in \NN$,
  given as input a set of~$\incd$s of width~$w$ and a set of access methods,
the truncated accessibility saturation algorithm
 computes all derived truncated accessibility axioms of breadth
  at most~$w$.
}
\begin{prop} \label{prop:derivedaccessids}
  \derivedaccessids
\end{prop}

Our proof of  Proposition~\ref{prop:derivedaccessids} is where we
use the downward-free chase defined earlier.  

\begin{proof}
For one direction, it is straightforward to see that all rules  obtained by this process
are in fact derived truncated accessibility axioms.
Conversely, we claim that, for all derived truncated accessibility axioms of
  breadth $\leq w$ 
\[ \accessible(x_{s_1}) \wedge \ldots \wedge \accessible(x_{s_l}) \wedge
R(\vec x) \rightarrow \accessible(x_i),
\]
   the corresponding triple $(R, \{s_1 \ldots s_l\}, i)$ is added to~$O$.
   We write $\bar{p} = \{s_1 \ldots s_l\}$; note that $\card{\bar{p}} \leq w$.

  Remember
  that, by the completeness of the chase (see Section~\ref{sec:prelims}) this
semantic entailment is always witnessed by a chase proof, and remember
   by Theorem~\ref{thm:gtgd_one_pass} that we can assume without loss of
   generality that it is a downward-free tree-like chase proof.
  We prove the claim by induction on the length of a downward-free tree-like chase proof
of the fact~$\accessible(c_i)$ from~$I_0 = \{R(\vec c)\} \cup
  \{\accessible(c_{j}) \mid j \in \bar{p})$. Here $I_0$ is the canonical
  database of the left-hand side of the implication, where we have used $\vec c$ for the variables
to emphasize that they are being treated as elements of the canonical database.

  If the proof is trivial, i.e., the fact $\accessible(c_i)$ is one of the
  $\accessible(c_{s_j})$, then clearly $(R, \bar{p}, i) \in O$ by the initialization of~$O$.
If it is non-trivial then some accessibility axiom provided the final firing to produce $\accessible(c_i)$,
and we can fix a guard atom $F$ and accessibility facts $F_1 \ldots F_l$ that were hypotheses
  of the chase step. If $F$ is the fact $R(\vec c)$, then each $F_j$ 
is of the form $\accessible(c_{t_j})$ for some index $t_j$ and by induction we
  have $(R, \bar{p}, t_j) \in O$ for each $i$. 
  Now by (Access)
  we deduce that $(R, \bar{p}, i) \in O$.

  Otherwise, the guard $F$ is the birth fact of some non-root tree node $v$.
  Consider the child $v'$ of the root node which is an ancestor of~$v$;
  potentially $v' = v$. Let $S(\vec d)$ be the birth fact of~$v'$; we know that
  $v'$ was created by firing an $\incd$ $\dep$ on the root node. Let $\bar{q}$
  be the subset of positions $j \in \{1, \ldots, |\vec d\,|\}$ for which the
  fact $\accessible(d_j)$ was propagated from the root node when $v'$ was
  created.
  This propagation witnesses that, in the $\incd$ $\dep$, each position
  in~$\bar{q}$ in the head atom contains an exported variable, i.e., a variable
  that also occurs in the body: we call this
  an \emph{exported head position}. We denote by $\bar{r}$ the corresponding set
  of \emph{exported body positions}, i.e., the positions in the body atom
  of~$\dep$ contains an exported variable. By definition
  $\card{\bar{r}} = \card{\bar{q}}$, and further
  $\card{\bar{q}} \leq w$.
  For each $j \in \bar{q}$, letting $j'$ be the index of~$\bar{r}$ such that
  $c_{j'} = d_j$, we know that the chase up to the creation of~$v'$
  provided a strictly shorter downward-free tree-like chase proof of
  $\accessible(c_{j'})$ from~$I_0$.  
Thus, by the induction hypothesis, we have $(R,
  \bar{p}, j') \in O$ for each $j' \in \bar{r}$.

  As $v$ and the root node both contain the value $c_i$, the fact
  $S(\vec d)$ must also contain this value. Let $i'$ be an index such that $d_{i'}=c_i$.
  By the downward-free property and Proposition~\ref{prop:one_pass_value},
  we know that the chase within the subtree of~$v'$ 
  provides a proof of
  $\accessible(d_{i'})$ from $S(\vec d)$ conjoined with $\accessible(d_j)$ for~$j \in
  \bar{q}$.
  Again, this is a strictly shorter downward-free tree-like chase proof, so by induction hypothesis we have that $(S,
  \bar{q}, i') \in O$. We know that the positions of $\bar{q}$ are exported head
  positions of~$\dep$, and we know that $i'$ is also an exported head position head
  position because $c_i$ appears in~$v$. Thus, we
  know by the (ID) axiom that $(R, \bar{r}, i) \in O$. Now, putting this
  together with the
  conclusion of the previous paragraph, we conclude using the (Transitivity) axiom
  that $(R, \bar{p}, i) \in O$.
\end{proof}

Thus we have shown that we can compute in $\ptime$ the implication closure of truncated
accessibility axioms of bounded breadth under bounded-width $\incd$s.

\paragraph{Shortcut chase and completeness.}
We know from Theorem~\ref{thm:gtgd_one_pass}
that for any set of single-headed GTGDs
we can do a downward-free chase, where propagation of facts
is restricted to be descendant-to-ancestor. %
We used this in the context of the GTGDs generated from  answerability
problems with bounded-depth $\incd$s, 
to justify a saturation algorithm in which certain derived
TGDs are added.
With this in place, we are now ready to simplify the chase process 
further, arriving at a tree-like
chase process that does \emph{no propagation at all}: the chase only grows the tree structure
and adds facts within a node of the tree. 
Instead of applying
chase steps with truncated
accessibility axioms, which would have required propagation to ancestors, we will show that 
we can create the same facts by firing derived axioms of small breadth in a ``greedy
fashion''.
 To connect this to our final goal of linearization, we note
that in doing a tree-like chase with linear TGDs, we do not need to propagate facts at all,
since no rule bodies care about multiple facts. Thus a tree-like
chase without propagation is bringing us closer to our goal of a chase with linear TGDs.

Recall that $\Sigma$ consists of $\incd$s of width $w$ and that we have a set
$\Delta$ of
truncated accessibility axioms (and transfer axioms, which we do not consider at
this stage).
Remember that we can use Proposition~\ref{prop:derivedaccessids} on~$\Sigma$ and~$\Delta$
to compute in PTIME the set of all derived
truncated accessibility axioms of breadth at most~$w$, which we denote by
$\Delta^+$.

A \emph{shortcut chase proof} on an initial instance $I_0$ with~$\Sigma$
and~$\Delta$ will be  a variation of the notion of tree-like chase proof 
defined earlier, but specific to $\incd$s and derived truncated
accessibility axioms.
A shortcut chase proof will alternate
between two kinds of steps:

\newcommand{\shortcutsteps}{
  \item  \emph{$\incd$-steps}, where we fire
an $\incd$ on a trigger $\trig$ to generate a fact $F$: we put $F$ in a new node~$n$
    which is a child of the node~$n'$ containing the fact of~$\trig$; and we
    copy in~$n$ all facts of the form $\accessible(c)$ that held in~$n'$
about any element $c$ that was exported when firing~$\trig$.
  \item \emph{Breadth-bounded saturation steps}, where we consider a newly-created node~$n$ and 
apply all derived truncated accessibility axioms of breadth at most~$w$ on that
    node, i.e., those of~$\Delta^+$, until we reach a fixpoint and there are no more violations of these
    axioms on~$n$.
}

\begin{itemize}
  \shortcutsteps
\end{itemize}

We continue
this process until a fixpoint is reached.
  Any stage in the  proof is thus associated with a tree structure, as in tree-like chase
proofs.  Each node in the tree corresponds to the application of an $\incd$,
which created the \emph{birth fact} for the node; and each node
may additionally contain accessibility facts in addition to the birth fact.
The name ``shortcut'' intuitively indicates that
we shortcut certain derivations that could have been performed by moving up and
down in the chase tree: instead, we apply a derived
truncated accessibility axiom.  Figure~\ref{fig:shortcut} illustrates the notion.

\begin{figure}
  \begin{tikzpicture}[xscale=1.5]
        \tikzstyle{every node}=[font=\footnotesize]
    \node[draw] (n1) at (.5, 0) {$R(c)$};
    \node[draw] (n2) at (2, 0) {$R(c)$};
    \node[draw] (n3) at (4, 0) {$R(c)$};
    \node[draw] (n4) at (6, 0) {$R(c)$};
    \node[draw] (n5) at (9, 0) {$R(c)$};
    \node[draw=none] at (7.5, -.5) {...};
    \node[draw,align=left,anchor=north] (a2) at (2, -.7) {$A(c, d_1 \ldots d_n)$};
    \node[draw,align=left,anchor=north] (a3) at (4, -.7) {$A(c, d_1 \ldots d_n),$\\$\accessible(c)$};
    \node[draw,align=left,anchor=north] (a4) at (6, -.7) {$A(c, d_1 \ldots d_n),$\\$\accessible(c),$\\$\accessible(d_1)$};
    \node[draw,align=left,anchor=north] (a5) at (9, -.7) {$A(c, d_1 \ldots d_n),$\\$\accessible(c),$\\$\accessible(d_1),$\\$\ldots$\\$\accessible(d_n)$};
    \draw[->] (n2) -- (a2);
    \draw[->] (n3) -- (a3);
    \draw[->] (n4) -- (a4);
    \draw[->] (n5) -- (a5);
  \end{tikzpicture}
  \par\noindent\rule{\textwidth}{0.4pt}
  \null\hfill
  \begin{tikzpicture}[xscale=1.5]
        \tikzstyle{every node}=[font=\footnotesize]
    \node[draw=none] (n2b) at (2,.5) {Saturation step};
    \node[draw=none] (n4b) at (6,.5) {Saturation step};
    \node[draw,anchor=north] (n1) at (.5, 0) {$R(c)$};
    \node[draw,align=left,anchor=north] (n2) at (2, 0) {$R(c),$\\$\accessible(c)$};
    \node[draw,align=left,anchor=north] (n3) at (4, 0) {$R(c),$\\$\accessible(c)$};
    \node[draw,align=left,anchor=north] (n4) at (6, 0) {$R(c),$\\$\accessible(c)$};
    \node[draw,align=left,anchor=north] (a3) at (4, -1.4) {$A(c, d_1 \ldots d_n),$\\$\accessible(c)$};
    \node[draw,align=left,anchor=north] (a4) at (6, -1.4) {$A(c, d_1 \ldots
    d_n),$\\$\accessible(c),$\\$\accessible(d_1),$\\$\ldots$\\$\accessible(d_n)$};
    \draw[->] (n3) -- (a3);
    \draw[->] (n4) -- (a4);
  \end{tikzpicture}\hfill\null
\caption{A chase proof (top), and a corresponding shortcut chase proof (bottom)
with saturation steps indicated}
  \label{fig:shortcut}
\end{figure} 

We show that this process correctly
generates everything that the usual chase would generate:

\newcommand{\normalizationaccessids}{
  Let $\Sigma$  be a set of $\incd$s of width~$w$ and
$\Delta$ a set of truncated accessibility axioms.
  Let $I_0$ be  a set of facts, and $I$
be  produced from $I_0$ as the final instance in a  chase proof using $\Sigma$ and $\Delta$.
Let $\Delta^+$ be, as above,  the set of derived truncated accessibility axioms
of breadth at most $w$.
  Lastly, let $I^+_0$ be the set of facts entailed by $I_0$  and 
  $\Delta^+$.

Then
there is  $I^\shortcut$ produced by a shortcut
chase proof based on $\Sigma$ and $\Delta$, from initial
instance $I^+_0$, and a homomorphism from~$I$ to~$I^\shortcut$
which is the identity on~$I_0$.
}
\begin{lem} \label{lem:normalizationaccessids}
\normalizationaccessids
\end{lem}

To prove this lemma,
we start with an observation about the closure properties of shortcut chase proofs.
\begin{lem} \label{lem:normalizedclosureaccessids}
  Let $I_0^+$ be an initial instance closed under~$\Delta^+$,
 and suppose that a shortcut chase proof on~$I_0^+$ with~$\Sigma$ and~$\Delta$
has a breadth-bounded saturation step on node $n$
  producing a fact $\accessible(a)$.
Then $a$ is not in $\adom(I_0^+)$, and $n$
was created by the $\incd$-step where $a$ is generated.
\end{lem}

\begin{proof}
Let $I^-$ be the instance just before the breadth-bounded saturation step
  generating $\accessible(a)$.
  First, notice that the breadth-bounded saturation step in question must 
  apply to a node which is not the root,
  as~$I_0^+$ is closed under~$\Delta^+$ so no new $\accessible$ facts can be
  created at the root node. 
  Thus, it applies to a node~$n_1$ created at an $\incd$-step. Let $\tau$ be
  the $\incd$ that we fired in this step, $n_0$ be the node on which we fired $\tau$.
  Let $E$ be the birth fact of $n_0$, recalling that this is the sole fact over a relation other than $\accessible$ holding
in the node. 
  Let $F$ be the birth fact of the node $n_1$. 
  It suffices to show that $E$ does not contain the element~$a$, as $a$ will
  then have been introduced in the $\incd$-step that creates node~$n_1$, in
  particular it is not in $\adom(I_0^+)$ because $n_1$ is not the root node.

  To show this that $E$ does not contain~$a$,
  let us assume by contradiction that it does.
  Let $R$ be the relation of~$E$, $\bar p$ be the positions of~$E$ containing
  the elements which occur in $F$ and for which the relation
  $\accessible$ holds in $I^-$.
Let $j$ be the position of~$a$
  in~$E$. By considering the subtree rooted at the node~$n_0$, which contains $E$, we see that
  a shortcut chase proof starting with the fact~$E$ and with the elements at
  positions $\bar p$ being accessible would also derive that  the element at
  position $j$ is accessible: it would do so with an $\incd$-step firing $\tau$
  to create a child node, and doing
  the same breadth-bounded saturation step on that node as the one that creates
  $\accessible(a)$ in~$I^-$. 
  Reusing the triple notation from the saturation
  algorithm,
  this implies that $(R, \bar p, j)$ is a derived truncated accessibility axiom. But then this axiom should
  have been fired in the breadth-bounded saturation step just 
  after the node $n_0$ was
  created, contradicting the assumption that $\accessible(a)$ is created at node~$n_1$.
  Hence, the fact~$E$ cannot contain~$a$, which concludes the proof.
\end{proof}

We now are ready to finish the proof of 
Lemma~\ref{lem:normalizationaccessids}, which shows completeness
of the  shortcut chase: 
\begin{proof}
  The instance $I$ is produced from $I_0$ as the final instance $I_f$ of a chase proof $I_1 \ldots I_f$ using $\Sigma$ and
  $\Delta$. We can extend this sequence to an infinite one $I_1 \ldots I_f \ldots$ in which every
trigger of $\Sigma \cup \Delta$ in some instance is eventually fired. Thus letting  $I_\infty$
denote the union of these instances, we have $I_\infty$ satisfies $\Sigma \cup \Delta$ and $I_0$ embeds as 
a subinstance of
$I_\infty$. 
  Likewise, we can continue the shortcut chase process indefinitely, letting $I^\shortcut_\infty$ be the resulting facts.
It suffices to show that $I^\shortcut_\infty$ satisfies the constraints of~$\Sigma$
  and~$\Delta$. Indeed, by universality of the chase \cite{fagindataex}, we
  would then know that there is a homomorphism
from $I_\infty$ into $I^\shortcut_\infty$ that is the identity on $I_0$.
The same function clearly serves as a homomorphism from $I$ into
 $I^\shortcut_\infty$, as required by the lemma.

  It is clear, thanks to the $\incd$-steps, that $I^\shortcut_\infty$ satisfies the
  constraints of~$\Sigma$.
  We claim that~$I^\shortcut_\infty$ also satisfies the constraints of~$\Delta$. 
Assume by contradiction that there is an
  active trigger in~$I^\shortcut_\infty$ for an axiom of~$\Delta$, 
  with facts $\left(\bigwedge \accessible(c_{m_j})\right) \wedge R(\vec c)$, whose firing would have produced fact $\accessible(c_i)$.
Consider the node~$n$ where $R(\vec c)$ occurs in the shortcut chase proof. 

We first observe that $n$ cannot be the root node corresponding to~$I_0^+$.
  Indeed, let us assume that it is. Then, the 
  facts $\accessible(c_{m_j})$ used in the firing
are facts about elements
  of~$I_0^+$, and Lemma~\ref{lem:normalizedclosureaccessids} implies that they
  cannot have been generated in a breadth-bounded saturation step. So they must
  already be in $I_0^+$. But $I_0^+$ then contains all the facts required to
  fire the active trigger, contradicting the fact that $I_0^+$ is closed under the axioms
  of~$\Delta$. Thus, $n$ is not the root node.

  Now, if the node~$n$ is not the root, then consider each fact
  $\accessible(c_{m_j})$ in the trigger of the firing. Let us show that~$n$
  contains all these facts. If $\accessible(c_{m_j})$ is a fact of~$I_0^+$, then
  it has been propagated from the root node to~$n$, so it is in~$n$. Otherwise,
  $\accessible(c_{m_j})$ was created by firing some breath-bounded saturation
  step. By Lemma~\ref{lem:normalizedclosureaccessids}, the node~$n_j$ where this
  step was fired is the node where $c_{m_j}$ is generated. Now, as the node $n$
  contains $c_{m_j}$, it must be a descendant of the node $n_j$ where $c_{m_j}$
  is generated. Now, as the fact $\accessible(c_{m_j})$ was created in~$n_j$, it
  must have been propagated downwards until we created node~$n$, so it must also
  be in~$n$. Hence, in all cases, the node $n$ contained all the facts required
  to fire the active trigger, so the trigger should have been fired at the
  breadth-bounded saturation step at~$n$. This again yields a contradiction.

  We conclude that in fact there cannot be an active trigger in~$I^\shortcut_\infty$, so
  $I^\shortcut_\infty$ satisfies the constraints of~$\Delta$, concluding the proof.
\end{proof}

\paragraph{Concluding the proof of Proposition~\ref{prop:linearizeaccessids}.}
Recall that our goal in Proposition~\ref{prop:linearizeaccessids} is to simulate
the chase with bounded-width $\incd$s $\Sigma$ and truncated accessibility and transfer
axioms $\Delta$.
We will now present our definition of the set of $\incd$s $\Sigma^{\lift}$ that
will do so. Thanks to what precedes
(Lemma~\ref{lem:normalizationaccessids}), we know that it suffices to simulate
the shortcut chase, %
and conversely it is obvious that the shortcut chase is sound in the sense that
any shortcut chase step 
could be replaced in a derivation of a CQ by a sequence of ordinary chase steps.

We let $\Delta^+$ be the set of derived
truncated accessibility axioms 
of breadth $\leq w$
calculated using
the truncated accessibility saturation algorithm
on~$\Sigma$ and~$\Delta$, including the axioms already present in~$\Delta$.
To define the linearized axioms, we first need some notation.
For a relation $R$, a subset $P$ of the
positions of~$R$, and a position $j$ of~$R$, we will say that $P$
\emph{transfers} $j$ if $\Delta^+$ contains the following derived truncated
accessibility axiom: \[\Big(\bigwedge_{i\in P} \accessible(x_i)\Big) \wedge
R(\vec x) \rightarrow \accessible(x_j).\]
Reusing the triple notation from the saturation
algorithm, this axiom corresponds to the triple $(R, P, j)$.

We now define $\Sigma^{\lift}$. Recall that it is defined over the signature
$\sign^\lift$ where we added a relation $R_P$ for every relation $R$ and subset
$P$ of positions of size at most~$w$, intuitively standing for an $R$-fact where
the elements at position~$P$ are accessible. The rules of~$\Sigma^\lift$ are:

\begin{itemize}
  \item (Lifted Transfer): Consider each relation $R$, and subset $P$ of positions of~$R$ of
    size at most~$w$. Let $P'$ be the set of positions transferred by~$P$.
    If $P'$ contains the set of input positions of some access method on~$R$,
    then 
    we add the full $\incd$:
    \[
      R_P(\vec x) \rightarrow R'(\vec x).
    \]

  \item (Lift):
  Consider each $\incd$  $\dep$ of~$\Sigma$, 
  \[R(\vec u) \rightarrow \exists \vec z ~ S(\vec z,\vec u).\]
    For every subset $P$ of positions of~$R$ of size at most~$w$, we let $P'$
    be the set of positions transferred by $P$.
We let $P''$ be the intersection
    of~$P'$ with the positions of $R$ that carry an exported variable in the atom $R(\vec u)$
within the body of~$\dep$.  Finally, we let $P'''$
    be the subset of the exported positions in the head of~$\dep$ that
    corresponds to~$P''$.  Then we add the dependency:
    \[
    R_P(\vec u) \rightarrow \exists \vec z ~ S_{P'''}(\vec z,\vec u).
    \]
\end{itemize}

We also need to define the instance $I_0^\lift$ from $I_0$, to account for the 
effect of~$\Sigma$ and $\Delta$ when we start the
chase. 
We recall that $\sign$ denotes the signature of the schema,  the
constraints of $\Sigma$ are expressed on~$\sign$, and
 the constraints $\Sigma'$ are expressed on a primed copy $\sign'$
of~$\sign$. Lastly, recall that $\Delta$ consists of truncated accessibility axioms and
transfer axioms that are expressed on~$\sign$, $\sign'$, and the unary
relation~$\accessible$.
Given a CQ~$Q$, let $I_0 \colonequals \canondb(Q)$ be its canonical database, and
let $I_0^\lift$ be formed by adding atoms to $I_0$ as follows. 

\begin{itemize}
\item Apply all of the truncated accessibility axioms of~$\Delta^+$
  to~$I_0$ to obtain~$I_0^+$. 
\item 
  Initialize $I_0^\lift \colonequals I_0^+$.
    Now,
  consider every relation~$R$ of the signature~$\sign$, and
  every fact $R(a_1 \ldots a_n)$ of~$I_0'$. Let $P$ be the
  set of the~$i \in \{1 \ldots n\}$ such that~$\accessible(a_i)$ holds
  in~$I_0^+$.  For every $P' \subseteq P$ of size at most~$w$, 
  add to~$I_0^\lift$ the fact $R_{P'}(a_1 \ldots a_n)$. Further, if
    $\accessible(a_i)$ holds for each $1 \leq i \leq n$, then add the fact
    $R'(a_1 \ldots a_n)$ to~$I_0^\lift$.
\end{itemize}

It is now easy to see that $\Sigma^\lift$ and $I_0^\lift$
satisfy the required conditions: for every Boolean CQ over the primed facts~$I$ 
derived using a chase proof
from $I_0$ with $\Sigma$ and~$\Delta$, we can derive the same CQ
from~$I_0^\lift$ via a chase proof  with $\Sigma^\lift$.
Indeed, applying chase steps with the (Lift) rules creates a tree of facts that
corresponds  to a shortcut chase proof, up to an $I_0$-preserving
isomorphism: when we create an $R_P$-fact,
the $P$ subscript denotes exactly the set of positions of the new facts that
contain exported elements that are accessible. Further,  the full
(Lifted Transfer) rules create a primed copy of these facts exactly when they can be transferred
by applying some method. 
As for the converse direction, it is clear that applying chase steps with $\Sigma^\lift$ on
$I_0^\lift$ only generates primed facts that correspond to what would be
generated by the shortcut chase, so that whenever we derive a Boolean CQ over
the primed facts then we generate a match of the same CQ, up to an
$I_0$-preserving isomorphism, in the shortcut chase. This justifies that the CQ
is also entailed from~$I_0$ by applying chase steps with~$\Sigma$ and~$\Delta$, first to
create the facts of~$I_0^\lift$, and then to derive the CQ\@.

The only thing left to do is to notice that $\Sigma^\lift$ has bounded semi-width,
but this is because the rules (Lift) have bounded width
and the rules (Lifted Transfer) clearly have an acyclic position graph. This
concludes the proof of Proposition~\ref{prop:linearizeaccessids}.

\paragraph{Handling Result-bounded Fact Transfer axioms and the proof of
Proposition~\ref{prop:linearizeaccessidsbounds}.}
Recall that to prove Theorem~\ref{thm:npidsbounds} in the case \emph{with} result bounds, 
we need only to prove the corresponding linearization result,
Proposition~\ref{prop:linearizeaccessidsbounds}, which extends
Proposition~\ref{prop:linearizeaccessids}. 

In the presence of result bounds, the reduction to query containment additionally created axioms called
(Result-bounded Fact Transfer), of the following form:
\[
  \Big(\bigwedge_i \accessible(x_i)\Big) \wedge R(\vec x, \vec y) \rightarrow  \exists \vec z ~ R'(\vec x, \vec z).
\]

We can extend the proof of Proposition~\ref{prop:linearizeaccessids} to prove
Proposition~\ref{prop:linearizeaccessidsbounds}.
Our result on restricting to downward-free chase proofs,
Theorem~\ref{thm:gtgd_one_pass}, can be used as-is, because it is stated for
single-headed GTGDs. The truncated accessibility saturation algorithm proved in
Proposition~\ref{prop:derivedaccessids} can also be used as-is, as it only
considers the truncated accessibility axioms.
We can then define the shortcut chase as before, and its completeness result
Lemma~\ref{lem:normalizationaccessids}, again because this only considers
$\incd$s and truncated accessibility axioms. Now, we only change the last step
of the proof of Proposition~\ref{prop:linearizeaccessids} and change our
definition of the $\incd$s $\Sigma^\lift$ by 
adding the following rules to our rewriting:

\begin{itemize}
  \item (Lifted Result-bounded Fact Transfer): For each relation $R$ and subset $P$ of
    positions of~$R$ of size at most~$w$ containing all input positions of some access method $\mt$ on~$R$
    with a result bound, we add the $\incd$:
    \[R_P(\vec x, \vec y) \rightarrow \exists \vec z ~ R'(\vec x, \vec z)\]
    where $\vec x$ denotes the input positions of~$\mt$.
\end{itemize}

It is clear that adding these axioms, and using them when defining 
$I_0^+$ from~$I_0$ in the definition of~$I_0^\lift$ above, ensures that the same primed facts are
generated as in the shortcut chase, and the resulting axioms still have
bounded semi-width: the (Lifted Result-bounded Fact Transfer) axioms are grouped in the
acyclic part together with the (Lifted Transfer) axioms, and they still have an acyclic
position graph. Hence, this establishes 
Proposition~\ref{prop:linearizeaccessidsbounds}, which was all we needed to conclude the proof
of Theorem~\ref{thm:npidsbounds} in the case with result bounds.

\section{Schema simplification for expressive constraints}
\label{sec:simplifychoice}

We have presented in Section~\ref{sec:simplify} the 
two kinds of simplifications anticipated in the introduction:
\emph{existence-check simplification} (using
result-bounded methods to check for the existence of tuples, as
in Example~\ref{ex:existencecheck}); and \emph{FD
simplification} (using them to retrieve
functionally determined information, as 
in Example~\ref{ex:fd}). A natural question is then to understand
whether these simplifications capture \emph{all} the ways in which result-bounded
methods can be useful, for integrity constraints expressed in more general
constraint languages. It turns out that this is not the case when we move even
slightly beyond $\incd$s:
\begin{exa} \label{ex:noexistsimplify}
Consider a schema $\aschema$ with TGD
  constraints
$T(y) \wedge S(x) \rightarrow T(x)$ and
$T(y) \rightarrow  \exists x ~ S(x) $.
We have an input-free access method $\mt_S$ on~$S$ with result bound~$1$
and a Boolean access method $\mt_T$ on~$T$.
Consider the query $Q= \exists y ~ T(y)$.
Note that the constraints imply that $Q$ is equivalent to $\exists x ~ T(x) \wedge S(x)$.

The following monotone plan answers $Q$:\\[.3em]
$T_1 \Leftarrow {\mt_S} \Leftarrow \emptyset$; \hfill
$T_2 \Leftarrow {\mt_T} \Leftarrow T_1$;  \hfill
  $T_3 \colonequals \pi_{\emptyset} T_2$; \hfill
  $\return~ T_3$;
  \\[.3em]
That is, we access $S$
and return true if the result is in~$T$.

On the other hand, consider the existence-check simplification $\aschema^\dagger$
of~$\aschema$. It has an existence-check method on~$S$, 
but 
  we can only test if $S$ is non-empty, giving
 no indication whether $Q$ holds. So
  $Q$ is not answerable in $\aschema^\dagger$. 
  The same holds for the FD simplification of~$\aschema$, because
  $\aschema$ implies no FDs, so the FD simplification and existence-check
  simplification are the same.
\end{exa}

Thus, existence-check simplification and FD simplification no longer suffice for
more expressive constraints. In this section, we introduce a new notion of
simplification, called \emph{choice simplification}. We will show that it allows
us to simplify
schemas with very general constraint classes, in particular TGDs as in
Example~\ref{ex:noexistsimplify}. In the next section, we will 
combine this simplification with our query containment reduction (Proposition~\ref{prop:reduce})
to show decidability of monotone answerability for much more expressive constraints.
Intuitively, choice simplification changes the
\emph{value} of all result bounds, replacing them by one; this
means that the number of tuples returned by result-bounded methods is not
important, provided that we obtain at least one if some exist.
We formalize the definition in this section, and show choice simplifiability for
two constraint classes: equality-free first-order logic
(which includes in particular TGDs), and $\uincd$s and FDs.
We study the decidability and complexity consequences of these results in the
next section.

\paragraph{Choice simplification.}
Given a schema $\aschema$ with result-bounded methods,
its \emph{choice simplification} $\aschema^\dagger$ is defined by keeping the relations and constraints
of~$\aschema$, but changing every result-bounded  method  to have
bound~$1$. That is, every result-bounded method of~$\aschema^\dagger$ returns~$\emptyset$ if there are no matching
tuples for the access, and otherwise selects and returns one matching tuple.
We call~$\aschema$ \emph{choice simplifiable}
if any CQ having a monotone plan over~$\aschema$ has one over~$\aschema^\dagger$.
This implies that the value of the result bounds never matters.

Choice simplifiability is weaker than existence-check or FD simplifiability,
in the sense that existence-check simplifiability or FD simplifiability imply
choice simplifiability.
Still, choice simplifiability
has a dramatic impact on the resulting query containment problem:

\begin{exa} \label{ex:choice}
  Recall the schema $\aschema$ in Example~\ref{ex:simple}
and its na\"ive axiomatization in Example~\ref{ex:reduce}.
  As $\aschema$ is choice simplifiable, we can axiomatize its
  choice simplification instead, and 
the problematic axiom in the third bullet item becomes a simple $\incd$:
    $\univdirect(\vec y) \rightarrow \exists \vec z ~ \univdirect_\acc(\vec z)$.
\end{exa}

\paragraph{Showing choice simplifiability.}
We now give a  result showing that choice simplification holds for a huge class of constraints: all first-order
constraints that do not involve equality. This result implies, for instance,
that choice simplification holds for integrity constraints expressed as
TGDs:

\newcommand{\thmsimplifychoice}{
 Let $\aschema$ be a schema with constraints in
  equality-free first-order logic (e.g., TGDs),  and 
  let $Q$ be a CQ that is monotonically answerable in~$\aschema$.
Then $Q$ is monotonically answerable in the choice simplification of~$\aschema$.
}
\begin{thm} \label{thm:simplifychoice}
  \thmsimplifychoice
\end{thm}

\begin{proof}
  We will again use the equivalence between monotone answerability and~$\amd$,
  and use the
  ``blowing-up'' construction of Lemma~\ref{lem:enlarge}. Note that, this time,
  the schema of~$\aschema$ and~$\aschema^\dagger$ is the same, so we simply need to
  show that
  $I_1^\dagger$ is a subinstance of~$I_1$ for each $p \in \{1, 2\}$.

  Consider a counterexample $I_1^\dagger, I_2^\dagger$ to~$\amd$ for~$Q$ in the choice
  simplification:
  we know that~$I_1^\dagger$ satisfies $Q$, that $I_2^\dagger$ violates $Q$, that
  $I_1^\dagger$ and $I_2^\dagger$ satisfy the equality-free first order constraints 
  of~$\aschema$,
  and that $I_1^\dagger$ and $I_2^\dagger$ have a common subinstance
  $I_\acc^\dagger$ which is access-valid
  in~$I_1^\dagger$ in the choice simplification of~$\aschema$.
  We will expand them to~$I_1$ and $I_2$ that have a common subinstance which
  is access-valid in~$I_1$ for~$\aschema$.

  For each element $a$ in the domain of~$I_1^\dagger$, introduce infinitely many
  fresh elements $a^j$ for~$j \in \NN_{>0}$, and identify $a^0 \colonequals a$.
  Now, define $I_1 \colonequals \mathrm{Blowup}(I_1^\dagger)$, where 
  $\mathrm{Blowup}(I_1^\dagger)$ is the instance with facts $\{R(a_1^{i_1} \ldots a_n^{i_n}) \mid R(\vec
  a) \in I_1^\dagger, \vec i \in \NN^n\}$. Define $I_2$
  from~$I_2^\dagger$ in the same way; it
  clearly has a homomorphism to~$I_2^\dagger$. %

  We will now show correctness of this construction.
  We claim that~$I_1^\dagger$ and $I_1$ agree on all equality-free first-order
  constraints, which we show using a variant of the
  standard Ehrenfeucht-Fra\"iss\'e game without
  equality~\cite{casanovas1996elementary}.
  In this game there are pebbles on both structures.
 The play proceeds by Spoiler placing a new pebble on some element in one structure, and Duplicator
must respond by placing a pebble with the same name in the other structure. Duplicator
loses if the mapping given by the pebbles does not preserve all relations
of the signature.  If Duplicator has a strategy  that never loses, then one can show by induction
that the two structures agree on all equality-free first-order sentences.

  Duplicator's strategy  will maintain the following invariants:
\begin{enumerate}
\item  if a pebble is on some element
  $a^j \in I_1$, then the corresponding pebble in~$I_1^\dagger$ 
 is on~$a$;
\item if a pebble is on some element $a$ in~$I_1^\dagger$, then the corresponding pebble
  in~$I_1$ is on some element $a^j$ for~$j \in \NN$.
\end{enumerate}
These invariants will guarantee that the strategy is winning.
  Duplicator's  response to a move by Spoiler in~$I_1$ is determined by the strategy 
above. In response to a move by Spoiler placing a pebble on an element $b$
  in~$I_1^\dagger$,
  Duplicator
places the corresponding pebble on
  $b^0 $ in~$I_1$.  

Clearly the same claim can be shown for~$I_2^\dagger$ and $I_2$.
In particular this shows that~$I_1^\dagger$ still satisfies $Q$ and $I_2^\dagger$ still
violates $Q$.

All that remains is to construct the common subinstance.
Let $I_\acc \colonequals \mathrm{Blowup}(I_\acc^\dagger)$. As
$I_\acc^\dagger$ is a common
subinstance of~$I_1^\dagger$ and~$I_2^\dagger$, clearly $I_\acc$ is a common subinstance
of~$I_1$ and~$I_2$. To see why $I_\acc$ is
access-valid in~$I_1^\dagger$, 
given an input tuple in
$I_\acc$, let $\vec t$ be the corresponding tuple in~$I_\acc^\dagger$.
If $\vec t$ has no matching tuples in~$I_1^\dagger$, then clearly the same is true  in
$I_1$.
If $\vec t$ has at least one matching tuple $\vec u$ in~$I_1^\dagger$, then such a
tuple exists in~$I_\acc^\dagger$ because it is access-valid in~$I_1^\dagger$, and hence
sufficiently many copies exist in~$I_\acc$ to satisfy the original result
bounds, so that we can find a valid output for the access in~$I_\acc$.
Hence $I_\acc$ is access-valid in~$I_1$, which completes the proof.
\end{proof}

\paragraph{Choice simplifiability with $\uincd$s and FDs.}
The previous result does not cover FDs.
However, we can also show a choice simplifiability result for 
FDs, and also add $\uincd$s, i.e., $\incd$s that only export a single element:

\newcommand{\thmsimplifychoiceuidfd}{
Let $\aschema$ be a schema whose constraints are
$\uincd$s and arbitrary FDs, and $Q$ be a CQ that is monotonically answerable in~$\aschema$.
Then $Q$ is monotonically answerable in the choice simplification of~$\aschema$.
}
\begin{thm} \label{thm:simplifychoiceuidfd}
\thmsimplifychoiceuidfd
\end{thm}

Our high-level strategy to prove Theorem~\ref{thm:simplifychoiceuidfd} is to
use a ``progressive'' variant of the
process of Lemma~\ref{lem:enlarge},
a variant where we ``correct'' one access at a time.
Remember that Lemma~\ref{lem:enlarge} said
that, if a counterexample to~$\amd$ in~$\aschema^\dagger$ can be
expanded to a counterexample in~$\aschema$, then $Q$ being $\amd$  in
$\aschema$ implies the same in~$\aschema^\dagger$. The next lemma makes a weaker
hypothesis: it assumes that for any counterexample in~$\aschema^\dagger$, for any
choice of access $(\mt, \abind)$, we can expand to a counterexample
in~$\aschema^\dagger$ in which we have corrected this access, i.e., there
is an output to $(\mt, \abind)$  which is valid for~$\aschema$. We must ensure
that correcting an access does not break the accesses that we previously corrected:
specifically, we must ensure that every access that previously had a 
valid output for~$\aschema$ still has such an output after we expand. Let us
formally define the process:

\begin{defi}
  \label{def:sablow}
  Let $\aschema$ be a schema and $\aschema^\dagger$ be its choice simplification,
  and let $\Sigma$ be a set of constraints.

  Consider two instances $I_1^\dagger, I_2^\dagger$ that satisfy~$\Sigma$, and a common subinstance
  $I_\acc^\dagger$ which is access-valid in~$I_1^\dagger$ for~$\aschema^\dagger$.
  Let $(\mt, \accbind)$ be an access in~$I_\acc^\dagger$

  A \emph{single-access blowup} of $I_1^\dagger, I_2^\dagger$
  and~$I_\acc^\dagger$ for $(\mt, \accbind)$ is a pair of
  instances $I_1, I_2$ that 
  satisfy $\Sigma$,
  such that $I_1$ is a superinstance of~$I_1^\dagger$,
  $I_2$ has a homomorphism to~$I_2^\dagger$,
  $I_1$ and $I_2$ have a common subinstance $I_\acc$ which is 
  access-valid in~$I_1$ for~$\aschema^\dagger$, 
  and we have:
  \begin{enumerate}
    \item $I_\acc$ is a superinstance of~$I_\acc^\dagger$;
    \item there is an output to the access $\mt, \accbind$ in~$I_\acc$ which is valid
in~$I_1$
  for~$\aschema$;
    \item for any access in~$I_\acc^\dagger$ having an output
      in~$I_\acc^\dagger$ which is valid
      for~$\aschema$ in~$I_1^\dagger$, there is an output to this access
      in~$I_\acc$ 
      which is valid for~$\aschema$ in~$I_1$;
  \item for any
    access in~$I_\acc$ which is not an access in~$I_\acc^\dagger$, there is an
      output in~$I_\acc$ which is valid for~$\aschema$ in~$I_1$.
  \end{enumerate}
\end{defi}

We will now state a lemma making an assumption 
that we can repair the counterexample from~$\aschema^\dagger$ to~$\aschema$ by working one
access at a time, using single-access blow-ups as above. We show that this is sufficient to reach the same conclusion
as with Lemma~\ref{lem:enlarge}:

\begin{lem}
  \label{lem:enlargeprog}
  Let $\aschema$ be a schema, $\aschema^\dagger$ be its choice simplification,
  and $\Sigma$ be a set of constraints.

  Assume that, for any CQ~$Q$ which is not $\amd$ in~$\aschema^\dagger$, for any
  counterexample $I_1^\dagger, I_2^\dagger$ of~$\amd$ for~$Q$ and
  $\aschema^\dagger$ with a common subinstance $I_\acc^\dagger$ which is access-valid
  in~$I_1^\dagger$ for~$\aschema^\dagger$, for any access $\mt, \accbind$
  in~$I_\acc^\dagger$,
  we can construct a single-access blowup of~$I_1^\dagger, I_2^\dagger$
  and~$I_\acc^\dagger$ for~$(\mt,
  \accbind)$.

  Then any CQ which is $\amd$ in~$\aschema$ is also
  $\amd$ in~$\aschema^\dagger$.
\end{lem}

\begin{proof}
  As in some of our prior arguments, we will prove the contrapositive. Let $Q$ be a query which is not
  $\amd$ in~$\aschema^\dagger$, and let $I_1^\dagger, I_2^\dagger$ be a
  counterexample, with~$I_\acc^\dagger$ the common subinstance of~$I_1^\dagger$
  and~$I_2^\dagger$ which
  is access-valid in~$I_1^\dagger$ for~$\aschema^\dagger$.

  Enumerate the
  accesses in~$I_\acc^\dagger$ as a sequence $(\mt^1, \accbind^1), \ldots,
  (\mt^n, \accbind^n), \ldots$: by the definition of~$I_\acc^\dagger$, all of them have an
  output in~$I_\acc^\dagger$ which is valid in~$I_1^\dagger$ for $\aschema^\dagger$, 
  but initially we do not assume that any of these outputs are valid
  for~$\aschema$ as well.
  We then build
  an infinite sequence $(I_1^\dagger, I_2^\dagger) = (I_1^1, I_2^1), \ldots, (I_1^n, I_2^n), \ldots$ with
  the corresponding common subinstances $I_\acc^\dagger = I_\acc^1, \ldots, I_\acc^n, \ldots$, with each
  $I_\acc^i$ being a common subinstance of~$I^i_1$ and $I^i_2$ which is access-valid
  in~$I^i_1$, by 
  performing the single-access blowup in succession to
  $(\mt^1, \accbind^1), \ldots, \allowbreak (\mt^n, \accbind^n), \ldots$. In particular, note
  that whenever $(\mt^i, \accbind^i)$ already has an output in~$I_\acc^i$ which
  is valid in~$I^i_1$ for~$\aschema$, then we
  can simply take $I^{i+1}_1, I^{i+1}_2, I^{i+1}_\acc$ to be respectively equal
  to~$I^i_1, I^i_2, I^i_\acc$, without even having to rely on the hypothesis of
  the lemma.

  It is now obvious by induction that, for all $i \in \NN$, $I^i_1$ and $I^i_2$
  satisfy the constraints $\Sigma$, we have $I_1 \subseteq I^i_1$ so $I^i_1$
  satisfies~$Q$, we have that $I^i_2$ has a
  homomorphism to~$I_2$ so $I_2^i$ does not satisfy~$Q$, and $I^i_\acc$ is a common subinstance of~$I^i_1$ and
  $I^i_2$ which is access-valid in~$I^i_1$ for~$\aschema^\dagger$, where the
  accesses $(\mt^1, \accbind^1), \ldots, \allowbreak (\mt^i, \accbind^i)$ additionally 
  have an output in~$I_\acc^i$ which is valid in~$I_1^i$ 
  for~$\aschema$, and where all the accesses in~$I^i_\acc$ which are not accesses
  of~$I_\acc$ also have an output in~$I^i_\acc$ which is valid in~$I_1^i$ for~$\aschema$. Hence,
  considering the infinite result $(I_1^\infty, I_2^\infty), I_\acc^\infty$ of this
  process, we know that all accesses in~$I^\infty_\acc$ have an output
  in~$I^\infty_\acc$ which is valid in~$I_1^\infty$ for~$\aschema$. Thus
  $I_\acc^\infty$ is actually a common subinstance of~$I_1^\infty$ and
  $I_2^\infty$,
  and $I_\acc^\infty$
  is access-valid in~$I_1^\infty$ for~$\aschema$. So
  $I_1^\infty, I_2^\infty$ is a counterexample to 
  $\amd$ of~$Q$ in~$\aschema$, which concludes the proof.
\end{proof}

  Thanks to Lemma~\ref{lem:enlargeprog}, we can now prove
  Theorem~\ref{thm:simplifychoiceuidfd} by arguing that we can correct each
  individual access. The rest of the section is devoted to this argument.

\begin{proof}[Proof of Theorem~\ref{thm:simplifychoiceuidfd}]
  Let $\aschema$ be the schema, let $\aschema^\dagger$ be its choice simplification, and
  let $\Sigma$ be the set of constraints.

  We explain how we perform the single-access blowup, to fulfil the requirements of  Lemma~\ref{lem:enlargeprog}. 
Let $Q$ be a CQ and assume that it is not
  $\amd$ in~$\aschema^\dagger$, and let $I_1^\dagger, I_2^\dagger$, be a 
  counterexample to~$\amd$, with~$I_\acc^\dagger$ being a common
  subinstance of~$I_1^\dagger$ and $I_2^\dagger$ which is access-valid in~$I_1^\dagger$
  for~$\aschema^\dagger$.
  Let $(\mt, \accbind)$ be an access on some relation~$R$ in~$I_\acc^\dagger$:
  we know that there is an output to the access in~$I_\acc^\dagger$ which is valid
  for~$\aschema^\dagger$ in~$I_1^\dagger$, but this output is not necessarily valid
  for~$\aschema$, i.e., the output may be returning only one tuple whereas the
  result bound in the original schema is higher.
  Our goal is to build a superinstance $I_1$ of~$I_1^\dagger$ and
  $I_2$ of~$I_2^\dagger$ 
  such that $I_2$ homomorphically maps to~$I_2^\dagger$.
  We want both $I_1$ and $I_2$ to satisfy  $\Sigma$, and want
  $I_1$ and
  $I_2$ to have a
  common subinstance $I_\acc$ which is access-valid in~$I_1$,
  where $\accbind$ now has an output which is valid for~$\aschema$ (i.e., not only for the choice
  simplification), all new accesses (the ones using new elements) also have an output which is valid for~$\aschema$, and
  no other accesses are affected.
  At a high level, we will do the same blow-up as in the proof of
  Theorem~\ref{thm:fdsimplify}, except that we will need to chase afterwards to
  make the $\uincd$s true. Even before this chasing phase, the presentation of the blow-up is also a bit
  different relative to the proof of Theorem~\ref{thm:fdsimplify}, for two main
  reasons. First, we are
  working with the choice simplification in the present proof, so there are no
  new relations $R_\mt$. Second, we are blowing up one access after another in
  the present proof, so we will not discuss ``dangerous'' and ``non-dangerous''
  methods, instead we will focus on the single access $(\mt, \accbind)$ that we
  are blowing up.

  First observe that, if there are no matching tuples in~$I_1^\dagger$ for the access
  $(\mt, \accbind)$, then the empty set is already an output in~$I_\acc^\dagger$ to the
  access which is valid in~$I_1^\dagger$ for~$\aschema$
  so there is nothing to do, i.e., we can just take $I_1 \colonequals
  I_1^\dagger$,
  $I_2 \colonequals I_2^\dagger$, and $I_\acc \colonequals
  I_\acc^\dagger$.
  Further, note that if there
  is only one matching tuple in~$I_1^\dagger$ for the access, as~$I_\acc^\dagger$ is access-valid for
  the choice simplification, then this tuple is necessarily in
  $I_\acc^\dagger$ also, so again there is nothing to do. Hence, it suffices to study
  the case where there is
  strictly more than one matching tuple in~$I_1^\dagger$ for the access $(\mt,
  \accbind)$; as~$I_\acc^\dagger$ is access-valid for~$\aschema^\dagger$, then it contains at least
  one of these tuples, say $\vec t_1$, and as~$I_\acc^\dagger \subseteq I_2^\dagger$,
  then $I_2^\dagger$ also contains $\vec t_1$. Let $\vec t_2$ be a second matching tuple
  in~$I_1^\dagger$ which is different from~$\vec t_1$. Let $C$ be 
  the non-empty set of  positions of~$R$ where $\vec t_1$ and $\vec t_2$
  disagree. Note that, since $I_1^\dagger$ satisfies the constraints,
  the constraints cannot imply an FD from the complement of~$C$
  to a position~$j \in C$, as otherwise $\vec t_1$ and $\vec t_2$
  would witness that~$I_1^\dagger$ violates this FD\@. Note also that $C$ cannot
  include input positions of~$\mt$. In fact, in the terminology of the proof of
  Theorem~\ref{thm:fdsimplify}, $C$ witnesses that $\mt$ is dangerous.

  We form an infinite collection of facts $R(\vec o_i)$
  where $\vec o_i$ is constructed from~$\vec t_1$
  by replacing the values at positions in~$C$ by fresh values. In particular we choose values distinct from
  those in   other positions in~$R$ and in other $\vec o_j$'s.
  Let $N \colonequals  \{R(\vec o_1), \ldots R(\vec o_n), \ldots\}$.
  We claim that
  $I_1^\dagger \cup N$ does not violate any FD implied by
  the schema.
  The argument is similar to that of the proof of Theorem~\ref{thm:fdsimplify}.
  Let us proceed by contradiction and assume that there is a violation of a
  FD~$\phi$. The violation
  $F_1, F_2$ must involve some new fact $R(\vec o_i)$, as~$I_1^\dagger$ on its own
  satisfies the constraints. We know that the left-hand-side of~$\phi$ cannot include a
  position of~$C$, as all elements in the new facts $R(\vec o_i)$ at these
  positions are fresh. Hence, the left-hand-side of~$\phi$ is included in the
  complement of~$C$, but recall that we argued above that then the
  right-hand-side of~$\phi$ cannot 
  be in~$C$. 
  Hence, both the left-hand-side and right-hand-side of~$\phi$ are in the complement of~$C$. But
  on this set of positions the facts of the violation~$F_1$ and $F_2$ agree
  with the existing fact $\vec t_1$ and~$\vec t_2$ of~$I_1^\dagger$, a contradiction.
  So we know that~$I_1^\dagger \cup N$ does not violate the FDs.
  The same argument shows that~$I_2^\dagger \cup N$ does not violate the FDs.

  So far, the argument was essentially the same as in the proof of
  Theorem~\ref{thm:fdsimplify}, but now we explain the additional chasing phase.
  This is analogous to the chasing done in the proof of 
  Theorem~\ref{thm:simplifyidsexistence}, but we define it differently
to
  avoid introducing FD violations.
  Formally, let $W$ be the infinite fixpoint of applying restricted chase steps to 
 $N$ with the $\uincd$s, 
but
  ignoring triggers whose exported element occurs in~$\vec t_1$. The process is
  illustrated in Figure~\ref{fig:blowupp}.
  We have argued that~$I_1^\dagger \cup N$ and $I_2^\dagger \cup N$ satisfy the FDs. 
We want to show that both the $\uincd$s and FDs hold in~$I_1^\dagger \cup W$ and
$I_2^\dagger \cup W$.
The key argument to use for this is that every 
element which is both in the domain of $I_1^\dagger$ and $N$ and which occurs at a certain position~$(R,i)$ in~$N$ must also 
occur at position~$(R,i)$
  in~$I_1^\dagger$, and likewise for~$I_2^\dagger$, namely:

\begin{figure}
  \begin{tikzpicture}[scale=.7]
    \tikzstyle{every node}=[font=\footnotesize]
    \draw[rounded corners=5pt] (-2.5,-.5) rectangle ++(13,2.5);
    \node[draw=none] (a) at (0, .25) {$R(a, b)$};
    \node[draw=none] (aa) at (0, 1.25) {$R(a, b')$};
    \node[draw=none] (b) at (3, .25) {$S(b, c)$};
    \node[draw=none] (c) at (9.5,1.5) {$I_1^\dagger$};
    \node (b1) at (-1, -2) {$R(a, b_1)$};
    \node (c1) at (-1, -3.5) {$S(b_1, c_1)$};
    \node[draw=none] (d1) at (-1, -4.5) {$\cdots$};
    \node[draw=none] (xx) at (-1.5, .25) {$\vec{t_1}$};
    \node[draw=none] (yy) at (-1.5, 1.25) {$\vec{t_2}$};
    \node (b2) at (2, -2) {$R(a, b_2)$};
    \node (c2) at (2, -3.5) {$S(b_2, c_2)$};
    \node[draw=none] (d2) at (2, -4.5) {$\cdots$};
    \node[draw=none] (bnn) at (4, -2) {$\cdots$};
    \node[draw=none] (bn2) at (8, -2) {$\cdots$};
    \node (bn) at (6, -2) {$R(a, b_n)$};
    \node (cn) at (6, -3.5) {$S(b_n, c_n)$};
    \node[draw=none] (dn) at (6, -4.5) {$\cdots$};
    \draw (a) -- (b1);
    \draw (a) -- (b2);
    \draw (a) -- (bn);
    \draw (b1) -- (c1);
    \draw (b2) -- (c2);
    \draw (bn) -- (cn);
    \draw[rounded corners=5pt] (-2,-1.5) rectangle ++(12,-1);
    \node[draw=none] (c) at (9.5,-2) {$N$};
    \node[draw=none] (c) at (9.5,-3.5) {$W$};
    \draw[rounded corners=5pt] (-2.5,-1) rectangle ++(13,-4);
  \end{tikzpicture}
  \caption{Illustration of the blow-up process of
  Theorem~\ref{thm:simplifychoiceuidfd}, where relation $R$ has an access on its
  first position, we blow up on the access to~$R$ with value~$a$, and there is (among
  other dependencies) an $\incd$ $R(x, y) \rightarrow \exists z ~ S(y, z)$}
  \label{fig:blowupp}
\end{figure}

\newcommand{\intdomain}{{\mathcal J}} %
\begin{clm} \label{clm:decomp}
Let $\ids$ be a set of~$\uincd$s
  and let $\fds$ be a set of FDs.
  Let $I$ and $N$ be instances, and let 
  $\intdomain \colonequals \adom(I) \cap \adom(N)$.
  Assume that~$I$ satisfies $\fds \cup \ids$, that
  $I \cup N$ satisfies $\fds$, and that whenever $a \in \intdomain$ occurs at a
  position~$(R, i)$ in~$N$, then it also occurs at~$(R,i)$ in~$I$.
  Let $W$ denote the restricted chase 
  of~$N$ by~$\ids$ 
  where we do not fire any triggers which map the
  exported variable to an element of  $\intdomain$.
  Then $I \cup W$ satisfies $\ids \cup \fds$.
\end{clm}

\begin{proof}

  Let us first notice that we have $\adom(W) \cap \adom(I) = \intdomain$.
  Indeed, it is a superset of $\adom(N) \cap \adom(I)$ so contains $\intdomain$,
  and all new domain elements in~$W$ are fresh by definition of the chase. What
  is more, we also notice that the facts of $W \setminus N$ never contain
  elements of~$\adom(I)$. This is because all triggers fired when constructing
  $W$ must have exported elements not in~$\intdomain$, hence not in~$\adom(I)$
  by what precedes.

  Now, we show that~$I \cup W$ satisfies $\ids$. Consider a 
  trigger $\trig$ for a $\uincd$ $\dep$ and let us show that it is not active. 
   The range of~$\trig$ is either in~$I$ or in~$W$. In the first case,
  as $I$ satisfies $\ids$, the trigger $\trig$ for~$\dep$ cannot be active.
  So consider the second case and assume $\trig$ were not active.
Then  it must map the exported variable
  to an element  of~$\intdomain$, i.e., it is a trigger which we did not fire in
  $W$. Let $R(\vec a)$ be the fact of~$W$ in the image of~$\trig$.
  This fact must be a fact of~$N$, because as we argued the facts of~$W \setminus N$
  do not contain elements of~$\intdomain$.
  Let $a_i$ be the image of the exported
  variable in~$\vec a$, with~$a_i \in \intdomain$. Hence, $a_i$ occurs at position~$(R, i)$ in
  $N$, so by our assumption on~$N$ it also occurs at position~$(R,i)$ in~$I$. Let $R(\vec b)$ be a
  fact of~$I$ such that~$b_i = a_i$. As $I$ satisfies $\ids$, for the match
of the body of~$\dep$ to~$R(\vec b)$ there is a corresponding
fact $F$ in~$I$ extending the match to the head of~$\dep$.
  But $F$ also serves as a witness in~$I \cup W$ for the match of the body of
  $\dep$,
  so  $\trig$ is not active, a contradiction.
  Hence, we have shown satisfaction of~$\ids$.
  
  We now show that~$I \cup W$ satisfies $\fds$.
  We begin by arguing that~$W$
  satisfies $\fds$. This is  because $N$ satisfies $\fds$.
  Now, every time the chase adds a fact $F$ to~$W$, all elements of~$F$ are
  fresh except one (the exported element), and that element did not occur at the
  position at which it occurs in~$F$, since otherwise the other fact where it did occur would
  witness that the trigger fired was not active. Thus, $F$ cannot be
  part of an FD violation. Hence, by induction, $W$
  satisfies~$\fds$.
  Now,
  assume by way of contradiction that there is an FD violation~$\{F, F'\}$ in~$I \cup
  W$. As $I$ and $W$ satisfy $\fds$ in isolation,
  it must be the case that one
  fact of the violation is in~$I$ and one is in~$W$: without loss of
  generality, assume that we have $F \in I$ and $F' \in W$.
  We cannot have $F'$ in $N$ because we know that $I \cup N$ satisfies $\fds$, so
  we must have $F' \in N \setminus W$. But as we argued, facts in $N \setminus
  W$ do not contain elements of~$\adom(I)$, so $F$ and $F'$ cannot constitute an FD
  violation as they are on disjoint elements.
  This establishes that $I \cup W$ satisfies $\fds$ and concludes the proof.
\end{proof}

We return to the proof of Theorem~\ref{thm:simplifychoiceuidfd}.
Recall that $W$ is the result of applying restricted chase steps to~$N$ with the $\uincd$s without firing
triggers whose exported element occurs in~$\vec t_1$.
Construct $I_1 \colonequals I_1^\dagger \cup W$, $I_2
\colonequals I_2^\dagger \cup W$, and 
$I_\acc \colonequals I_\acc^\dagger \cup W$.
By Claim~\ref{clm:decomp}, we know that~$I_1$
and $I_2$ satisfy the constraints.

Let
us then conclude our proof of
Theorem~\ref{thm:simplifychoiceuidfd} via the process
of Lemma~\ref{lem:enlargeprog}. We show the first conditions
on~$I_1$,
$I_2$, and $I_\acc^\dagger$ stated in Definition~\ref{def:sablow}:
\begin{itemize}
    \item We clearly have 
      $I_1^\dagger \subseteq I_1$.
    \item We have just shown that~$I_1$
      and $I_2$ satisfy the constraints.
    \item We now argue that~$I_2$ has a homomorphism to~$I_2^\dagger$.
  This argument is reminiscent of the proof of
    Theorem~\ref{thm:simplifyidsexistence}.
 We first define the homomorphism from~$I_2^\dagger \cup N$
to~$I_2^\dagger$ by mapping $I_2^\dagger$ to itself, and mapping the fresh elements of~$N$ so
    that the facts of~$N$ are mapped to~$R(\vec
t_1)$. 
This is possible because each fresh element in~$N$ occurs at only one
    position.
It is clear that this
is a homomorphism. We then extend this homomorphism inductively on each fact
created in~$W$ in the following way. Whenever a
fact $S(\vec b)$ is created by firing an active trigger $R(\vec a)$ 
    for a $\uincd$ $R(\vec x) \rightarrow S(\vec y)$ where $x_p = y_q$ is the
    exported variable
    (so we have $a_p = b_q$), consider the
fact $R(h(\vec a))$ of~$I_2^\dagger$ (with~$h$ defined on~$\vec a$ by induction
    hypothesis). As $I_2^\dagger$ satisfies
$\Sigma$, we can find a fact $S(\vec c)$ with~$c_q = h(a_p)$, so we can define
$h(\vec b)$ to be $\vec c$, and this is consistent with the existing image
    of~$a_p$.

\item 
  Clearly $I_\acc$ is a common subinstance of~$I_1$ and~$I_2$ by
    construction.
    We now show that~$I_\acc$
    is access-valid
    for~$I_1$ and $\aschema^\dagger$. Let  $(\mt', \accbind')$ be an access in
    $I_\acc$. 
We first consider the case when the range of the binding $\accbind'$ includes an element
    of~$\adom(I_\acc)
    \setminus \adom(I_\acc^\dagger)$, namely, an element of~$\adom(W) \setminus
    \adom(I_1^\dagger)$. In this case, all matching facts must be facts of~$W$. Thus, if
    $\mt'$ has no result bound then all matching facts to be returned are
    in~$I_\acc$, and 
    if $\mt'$ is result-bounded then any choice of a tuple from~$W \subseteq
    I_\acc$ (or no tuples, if this
    set is empty) is an
    output to the access which is valid for the choice
    simplification~$\aschema^\dagger$. 
The second case is when $\accbind'$
     only involves  elements of~$\adom(I_\acc^\dagger)$.  Then  $(\mt', \accbind')$ is actually an access
    on~$I_\acc^\dagger$. As $I_\acc^\dagger$ is access-valid in~$I_1^\dagger$, let 
    $U$ be a (possibly empty) output to the access from~$I_\acc^\dagger$ which is valid
    in~$I_1^\dagger$
    for~$\aschema^\dagger$. 
    Some tuples of~$W$, say $U'$, may also be matching tuples to the access $(\mt',
    \accbind')$ in~$I_1$. Now, if $\mt'$ has no result bound, then all
    matching facts to be returned are in $U \cup U'$ and hence
    in~$I_\acc^\dagger$. And
    if~$\mt'$ is result-bounded, then
    any choice of  a tuple in $U \cup U'$ (or no
    tuple, if $U \cup U'$ is empty) gives an output to the access which is
    in~$I_\acc$ and is
    valid for~$\aschema^\dagger$. 
    Hence, it is indeed the case that $I_\acc$ is access-valid
    for~$I_1$
    and~$\aschema^\dagger$.
\end{itemize}
We now show the four additional conditions of Definition~\ref{def:sablow}:
\begin{enumerate}
  \item It is clear by definition that~$I_\acc \supseteq I_\acc^\dagger$.
\item We must show that the access $(\mt, \accbind)$ is valid for~$\aschema$ in
  $I_\acc$. Indeed, there are now
    infinitely many matching tuples in~$I_\acc$, namely, those of~$N$. Thus
this access is valid for~$\aschema$ in~$I_1^\dagger$: we can choose as many tuples as
the value of the bound to obtain an output which is valid in~$I_1^\dagger$.

\item We must verify that, for any access $(\mt', \accbind')$ 
  of~$I_\acc^\dagger$ that has an output which is valid in~$I_1^\dagger$
    for~$\aschema$, we can construct such an output in~$I_\acc$ which is
    valid in~$I_1$
    for~$\aschema$.
    The argument is
    the same as in the second case of the fourth bullet point above: from the
    valid output to the access~$(\mt', \accbind')$ in~$I_1^\dagger$ for~$\aschema$, we construct
    a valid output to~$(\mt', \accbind')$ in~$I_1$ for~$\aschema$.
  \item  Let us consider any access in~$I_\acc$ 
    which is not an access in~$I_\acc^\dagger$.
The binding for this access must include
some element of~$\adom(W)$, so its matching tuples must be in~$W$, which are all
    in~$I_\acc$. Hence, 
    by construction any such accesses are valid for~$\aschema$. 
\end{enumerate}
This concludes the proof of Theorem~\ref{thm:simplifychoiceuidfd} using
Lemma~\ref{lem:enlargeprog}, correcting each access according to the above process.
\end{proof}

\section{Decidability using Choice Simplification}
\label{sec:complexitychoice}

In this section, we present the consequences of the choice simplifiability
results of the previous section,
in terms of decidability for expressive constraint languages.

\paragraph{Decidable equality-free constraints.}
Theorem~\ref{thm:simplifychoice} implies that monotone answerability is decidable 
for a wide variety of schemas. The approach applies to constraints that do not involve equality 
and have decidable query containment.
We state here one complexity result for the class of  \emph{frontier-guarded
TGDs} (FGTGDs): recall that these are TGDs whose body contains a single atom
including all exported variables.
The same approach  applies  to  
extensions of FGTGDs
with disjunction and negation \cite{bourhis2016guarded,gnfj}.

\begin{thm} \label{thm:decidegf}
  We can decide whether a CQ is 
monotonically answerable with respect to a schema with
  result bounds whose constraints are FGTGDs.
  The problem is $\twoexp$-complete.
\end{thm}
\begin{proof}
  Hardness holds because of a reduction from query containment with
  FGTGDs (see, e.g., Prop.~3.16 in~\cite{thebook}), already in
  the absence of result bounds, so we focus on $\twoexp$-membership.
By  Theorem~\ref{thm:simplifychoice} we can assume that all result bounds
are one, and by Proposition~\ref{prop:elimupper} we can replace
the schema with the relaxed version that contains
 only result lower bounds. Now, a result lower bound of~$1$ can be expressed
  as an $\incd$ as was illustrated in Example~\ref{ex:choice}. Thus, Proposition~\ref{prop:reduce} allows us to reduce
  monotone answerability to a query containment problem 
  with additional FGTGDs, and
this is decidable in~$\twoexp$ (see, e.g.,~\cite{bgo}).
\end{proof}

\paragraph{Complexity with $\uincd$s and FDs.}
We now turn to constraints that consist of $\uincd$s and FDs,
and use the choice simplifiability result of
Theorem~\ref{thm:simplifychoiceuidfd} to derive complexity results for monotone
answerability with result-bounded access methods:

\newcommand{\deciduidfd}{
  We can decide monotone answerability with respect to a schema
  with result bounds whose constraints are $\uincd$s and FDs. The problem is in $\twoexptime$.
}
\begin{thm} \label{thm:deciduidfd}
  \deciduidfd
\end{thm}

Compared to Theorem~\ref{thm:npidsbounds},
this result restricts to $\uincd$s rather than $\incd$s,
and has a higher complexity, but it allows FD constraints.
To the best of our knowledge, this result is new even in the setting without result bounds.

\newcommand{\separableconstraints}{
\item for each non-result-bounded method~$\mt$ accessing relation $R$ with
  input positions $\vec x$, the axiom $\left(\bigwedge_i \accessible(x_i)\right) \wedge
  R(\vec x, \vec y) \rightarrow R'(\vec x, \vec y)
\wedge \bigwedge_i \accessible(y_i)$
\item for each result-bounded method $\mt$ accessing relation $R$ with input
  positions $\vec x$, the axiom $\left(\bigwedge_i \accessible(x_i)\right) \wedge R(\vec x, \vec y)
  \rightarrow \exists \vec z ~ [R(\vec x, \vec z) \wedge R'(\vec x, \vec z)
    \wedge \bigwedge_i \accessible(z_i)]$
}
\begin{proof}
  By choice simplifiability (Theorem~\ref{thm:simplifychoiceuidfd}) we can assume that all result bounds are one.
  By Proposition~\ref{prop:reduce} we can reduce to a query
  containment problem 
  $Q \subseteq_\Gamma Q'$.
The constraints $\Gamma$ include $\Sigma$, its copy $\Sigma'$,
and accessibility axioms:
  
\begin{itemize}
\item $\left(\bigwedge_i \accessible(x_i)\right) \wedge
  R(\vec x, \vec y) \rightarrow R_\acc(\vec x, \vec y)$ for each
  non-result-bounded method $\mt$ accessing relation~$R$ and having input
  positions $\vec x$;
\item $\left(\bigwedge_i \accessible(x_i)\right) \wedge \exists \vec y ~
  R(\vec x, \vec y)
\rightarrow \exists \vec z ~ R_\acc(\vec x, \vec z)$ for each result-bounded
  method $\mt$ accessing relation~$R$ and having input positions $\vec x$;
\item $R_\acc(\vec w) \rightarrow R(\vec w) \wedge R'(\vec w) \wedge \bigwedge_i
  \accessible(w_i)$ for each relation~$R$.
\end{itemize}

Note that $\Gamma$ includes FDs and non-unary IDs; containment for these in general is undecidable \cite{mitchell1983implication}.
To show decidability, we will
explain how to rewrite these axioms in a way that makes $\Gamma$
  \emph{separable}~\cite{cali2003decidability}. That is, we will be able to
  drop the FDs of~$\Sigma$ and~$\Sigma'$ without impacting containment.
First, by inlining~$R_\acc$, we can rewrite the above axioms as follows:

\begin{itemize}
  \separableconstraints
\end{itemize}

We then modify the second type of axiom so that,
in addition to the variables $\vec x$ at input positions of~$\mt$ in~$R$,
  the axioms also export all variables at  positions $\detby(\mt)$ of~$R$ that are determined by the
  input positions. In other words, the second bullet point becomes:
  \begin{itemize}
\item for each result-bounded method $\mt$ accessing relation $R$ with $\vec
  x$ the variables at positions of~$\detby(\mt)$,
   $\left(\bigwedge_i \accessible(x_i)\right) \wedge R(\vec x, \vec y)
  \rightarrow \exists \vec z ~ [R(\vec x, \vec z) \wedge R'(\vec x, \vec z)
    \wedge \bigwedge_i \accessible(z_i)]$.
  \end{itemize}
This rewriting does not impact the soundness of the chase, as each chase step
  with a rewritten axiom can be
mimicked by a step with an original axiom followed by FD applications.

Let us now argue that we can indeed drop the FDs of $\Sigma$ and $\Sigma'$. 
That is, let
  us show that we never create a violation of an FD of~$\Sigma$ and~$\Sigma'$ in
  restricted chase proofs with the rewritten constraints.
To argue this, 
it suffices to consider restricted chase proofs where we first
fire the constraints in~$\Sigma$, then the accessibility axioms as rewritten
  above, and last the constraints in~$\Sigma'$. Indeed, at each of these three
  steps, we never create any new triggers for a preceding step. So let us show
  that these steps never introduce FD violations.

To show that the first steps do not introduce violations,
remember that the restricted chase with $\uincd$s can never introduce FD
violations, because we only fire active triggers --- this argument was spelled
  out at the end of the proof of Claim~\ref{clm:decomp}. 
  The same reasoning shows that the third step cannot create FD violations
  either.

For the second step, assume by contradiction that firing the rewritten axioms creates 
an FD violation, and
consider the first FD violation that is created. Either the violation is on a
primed relation, or it is on an unprimed relation. If it is on a primed
relation, it consists of 
a  first fact $F_1' = R'(\vec c, \vec d)$, and of a second fact $F_2' = R'(\vec f,
\vec g)$ which was just generated by firing an axiom
on some fact $F_2 = R(\vec f, \vec h)$. The accessibility axiom may be associated with an 
access method that is not result-bounded, in which case $\vec g$ and $\vec h$
are empty tuples; or it may relate to a result-bounded access method, in
which case all values in $\vec g$ are fresh. 
As we are only at the second step, we have not fired any dependencies
from~$\Sigma'$ yet,
so $F_1'$ must also
have been generated by firing an axiom on some fact
$F_1 = R(\vec c, \vec e)$, and again $\vec d$ is either empty or only consists of
fresh values. Now, we know that the determinant
of the violated FD 
must be within
the intersection of the positions of $\vec c$ and of $\vec f$, because it cannot
contain fresh values in any of the two facts $F_1'$ and $F_2'$. Hence, by the
modification that we did on the axioms, the determined position of the violated
FD must also be within the intersection of the positions of~$\vec c$ and
of~$\vec f$. This means that $F_1$ and $F_2$ are already a violation of the FD,
which means that $F_1', F_2'$ was not the first violation, a contradiction.

Now, if the violation is on an unprimed relation, it consists of a first fact $F_1' =
R(\vec c, \vec d)$, and of a second fact $F_2' = R(\vec f, \vec g)$ which was
just generated by an axiom for a result-bounded access method.
In this case, let
$F_2 = R(\vec f, \vec h)$ be the fact on which the axiom was fired. Because the elements of
$\vec g$ are fresh, the determinant of the violated FD must be within positions
of~$\vec f$. Now, the positions of~$\vec f$ are exactly those positions determined by the
input positions of the method, so the determined position of the violated FD must also be
within positions of~$\vec f$. This means that $F_1'$ and $F_2$ are already a
violation of the FD, again contradicting that $F_1$ and $F_2$ were the first
violation.

Thus, let $\gammasep$ denote the rewritten constraints without the FDs.
We have shown that monotone answerability is equivalent to \mbox{$Q
  \subseteq_{\gammasep} Q'$}.
As~$\gammasep$ contains only GTGDs, we can infer
decidability in~$\twoexp$ using~\cite{taming}, which concludes
the proof  of Theorem~\ref{thm:deciduidfd}.
\end{proof}

\section{General First-Order Constraints} \label{sec:general}

We have shown that, for many expressive
constraint classes, the value of result
bounds does not matter, and monotone answerability is decidable.
A natural question is then to understand what happens with schema simplification
and decidability for general FO constraints that may include the equality symbol.
In this case, we find that choice simplifiability no longer holds:

\begin{exa} \label{ex:complex}
Consider a schema $\aschema$ with two relations $P$ and~$U$ of arity~$1$.
There is an input-free method $\mt_P$ on~$P$ with result bound~$5$,
 and an input-free method $\mt_U$ on~$U$ with no result bound.
The first-order constraints $\Sigma$ say that~$P$ has exactly $7$ tuples,
and if one of the tuples is in~$U$,
then $4$ of these tuples must be in~$U$.
  Consider the query $Q: \exists x ~ P(x) \land U(x)$.
The query is monotonically answerable on~$\aschema$: the plan simply accesses $P$
  with~$\mt_P$, intersects the result with~$U$ using~$\mt_U$, and projects to
  return true or false depending on whether the intersection is empty or not.
  Thanks to~$\Sigma$, this will always return the correct result.

In the choice simplification $\aschema^\dagger$ of~$\aschema$, 
all we can do 
is access $\mt_U$, returning all of~$U$, and access $\mt_P$, returning a single
  tuple. If this tuple is not in~$U$, we have no information on  whether or not
$Q$ holds. Hence, we can easily see that $Q$ is not answerable
  on~$\aschema^\dagger$.
\end{exa}

The fact that simplification results fail does not immediately imply
that monotone answerability problems are undecidable.
However, we show that if we move to  constraints 
where containment is undecidable, 
then the monotone answerability problem is also undecidable, even in cases such as equality-free FO
which are choice simplifiable:

\newcommand{\foundecid}{It is undecidable to check if a conjunctive query $Q$ is
monotonically answerable
with respect to equality-free FO constraints.} 

\begin{prop} \label{prp:undec} 
\foundecid
\end{prop}

This result is true even without result bounds, and follows
from results in \cite{thebook}: we give a self-contained argument here.
Satisfiability for equality-free first-order constraints is undecidable  \cite{AHV}.
We will reduce from this to show undecidability of monotone answerability:

\begin{proof}
Assume that we are given an instance of a satisfiability problem 
consisting of  equality-free first-order constraints $\Sigma$.
We produce from this an instance of the monotone answerability problem where the schema has no access methods and has
constraints $\Sigma$, and we have a CQ~$Q$ consisting of a single $0$-ary relation~$A$ not mentioned
in~$\Sigma$.

We claim that this gives  a reduction from unsatisfiability to monotone answerability, and thus
shows that the latter problem is undecidable for equality-free first-order constraints.

If $\Sigma$ is unsatisfiable, then vacuously any plan answers $Q$: since answerability
is a condition where we quantify over all instances satisfying the constraints, this
is vacuously true when the constraints are unsatisfiable because we are
quantifying over the empty set.

Conversely, if there is some instance $I$ satisfying $\Sigma$, then we let $I_1$ be formed from~$I$
by setting $A$ to be true and $I_2$ be formed by setting $A$ to be false.
$I_1$ and $I_2$ both satisfy $\Sigma$ and have the same accessible part, so they form
a counterexample  to~$\amd$. Thus, there cannot be any  monotone plan  for~$Q$. This
establishes the correctness of our reduction, and concludes the proof of
Proposition~\ref{prp:undec}.
\end{proof}

The same undecidability result holds for other constraint languages where query containment is undecidable, such as general TGDs.

\section{Summary and Conclusion} \label{sec:conc}
We formalized the problem of answering queries in a complete way 
by accessing Web services that only return a bounded number of answers to each
access, assuming integrity constraints on the data.
We showed how to reduce this to a standard reasoning problem, query containment
with constraints. 
We have further shown simplification results for many classes of  constraints, 
limiting the ways in which  a query can be answered using result-bounded plans, 
thus simplifying the corresponding query containment problem.
By coupling these results with an analysis of query containment,
we have derived complexity bounds
for monotone answerability under several
classes of constraints.
 Table~\ref{tab:results} on p.~\pageref{tab:results} summarizes which simplifiability result
 holds for each constraint class, 
as well as the decidability and
complexity results.   

In our study of the answerability problem,
we have have also introduced refinements of technical tools which we hope could be useful in a wider
context. One example is the blowing-up method that we use in schema
simplification results. 
Our results on bounded-width
dependencies show that we can exploit the special form of query containments produced by
answerability problems with access method
-- namely, they are guarded TGDs where the ``side
atoms'' have a fixed signature. This
leads us to a finer-grained  analysis of the complexity of guarded TGDs, tracking
how a fixed side signature allows us to refine prior query answering
techniques~--- like the linearization approach
of \cite{gmp} and the tree-shrinking argument of \cite{johnsonklug}.
The paper demonstrates
how these model-theoretic and query-rewriting techniques can be applied to questions about
answerability with access methods, a setting quite different from prior motivations. We
believe the rewriting techniques in particular can be pushed to provide broader
results  on entailment with guarded TGDs, based on the distinction
between the guard signature and the ``side signature'':
for  some attempts in this direction, see Appendix G of \cite{oldarxiv}.

We now discuss limitations and open questions.

\paragraph{Complexity and expressiveness gaps.} 
Note that for the case of FDs and UIDs,
the complexity bounds are not tight.
The conference version sketches an approach to show that monotone
answerability for this class is in $\exptime$, with details
given in \cite{oldarxiv}. We do
not provide
a full presentation of this in this work.

We leave open the complexity of monotone answerability
with result bounds for some important
cases: full TGDs, and more generally weakly-acyclic TGDs.
Our choice approximation result applies here, but we do not know
how to analyze the chase even for the simplified containment problem.

On the expressiveness side, we also leave open the question of whether choice simplifiability holds
for general FDs and $\incd$s; that is, not restricting to $\uincd$s. We also
leave open the question of whether $\uincd$s and FDs, or even $\incd$s and FDs,
can be shown to be FD simplifiable

Note that all of our results forbid the use of constants in constraints. In particular,
our definition  of constraint classes like guarded TGDs forbids constants, differing in this respect
from some prior presentations of these classes.  We believe
all of the results in the paper still hold in the presence of constants with roughly
the same proofs,  but we have not verified this.

\paragraph{Monotone vs general plans.}
We have  restricted to \emph{monotone} plans 
throughout the paper. As explained in
Appendix~\ref{apx:ra},
the reduction to query containment still applies to plans  that can use
negation.
Our schema simplification results 
also extend easily to 
answerability with such plans, but lead to a more involved query containment
problem. Hence,
we do not know how to show decidability of the
answerability problem  for $\uincd$s and FDs with such plans.

In the case where constraints are dependencies, it is
 difficult to construct examples of CQs that require non-monotone plans. 
This suggests that the impact  of considering
richer plans is not large. But this is only anecdotal;
and in addition, the situation is completely different with more general
constraints --- e.g., with disjunction and negation --- where dealing
with general plans is obviously critical.

\paragraph{Finite vs unrestricted equivalence.}
We have defined 
answerability by requiring that the query and the plan agree on all instances,
finite and infinite. An alternative is to consider equivalence over finite instances only. We say that
a plan $\aplan$ \emph{finitely answers}~$Q$,
if for any finite instance $I$ satisfying the integrity constraints of~$\aplan$,
the only possible output of~$\aplan$s is~$Q(I)$.
Both finite and unrestricted answerability
have been studied
in past work on access methods in the absence of result bounds \cite{ustods,thebook}, just as finite and unrestricted variants
of other static analysis problems (e.g.,
query containment)
have long been investigated in database theory  (e.g., \cite{johnsonklug}).
The unrestricted variants usually
provide a cleaner theory and better algorithms, but the finite versions can be more precise.

In the presence of result bounds,  we know nothing about the finite variants. 
Our  analysis of the corresponding query containment problems can be extended
to  the finite variant of containment.
But a  major question is whether
the reductions to query containment from  Section~\ref{sec:reduce} still hold in the finite case.
The conference version of this paper \cite{confpaper} claimed that
these reductions could be extended, but the argument was found to be flawed in
the review process for the present paper. Thus the issue is left as an open
question for further work.

\paragraph{Practical impact.}
Our results provide a very comprehensive analysis of when there is an algorithm
for answering queries in the presence of result bounds. But there is a question of how
to interpret the ``bottom line'' of these results. For our expressiveness results, we feel it is reasonable to consider
them as negative. Example~\ref{ex:complex} shows that in the presence of complex constraints,
result-bounded methods can be useful for answering queries in extremely non-obvious
ways. In contrast, our simplification results show 
that for many   common constraint classes, result-bounded
methods can be useful only in limited ways. In particular, we show that this limitation holds
for many of the classes where we have decision procedures for the query answering
problem. This limitation is related to the fact that our notion of
answerability~--- the usual one
considered for access methods and for views~--- is difficult to achieve for queries that intrinsically rely on result
bounds. More relaxed notions have been explored in recent work \cite{romerosmartplans}, but only in very
restricted settings. In the setting of result bounds, weaker notions of
answerability are  an important topic for future investigation.

\paragraph{Answering vs answerability.}
In this paper we have focused only on the decision problem related to
answerability --- does there
exists a plan that answers the query. But we did not deal
with how to obtain the plans. Exactly the same complexity bounds apply to the plan-construction
problem as to the decision problem in each case we consider. Indeed, in this work
we have reduced
 the answerability question to a query containment question, and we then
analyzed  the complexity of determining whether there is a proof witnessing the containment resulting
from the reduction. In the case where the constraints are
dependencies,  the corresponding proofs are just chase sequences.  
The temporary
tables will store the state of the chase after each proof step. 
And there is a simple linear-time algorithm to extract a plan from a chase proof of the corresponding containment.
In the case where there are no result bounds, the method is given in \cite{ustods,thebook}.
In the plan we produce an access command for every firing of an accessibility axiom in the proof. When
our  existence-check or FD simplifiability results apply~--- for example the case of
$\incd$s given in Theorem~\ref{thm:simplifyidsexistence} and FDs in Theorem~\ref{thm:fdsimplify}
-- we can eliminate result bounds completely, and then use these algorithms out of the box.
But in the presence of results bounds, these algorithms generalize in the obvious way. When we fire
an accessibility axiom that corresponds to a result-limited method, we generate an access command
in the same way as in the absence of bounds.
Note that for languages with disjunction, constructing a plan is  more complex.
Instead of a chase proof, one needs a tableau proof, and instead of the straightforward algorithm given
in Chapter~4 of \cite{thebook}, one uses interpolation. Still the algorithms are linear in the size
of the proof, and extend to result-bounds. So again there is no distinction in the complexity between
answerability and plan generation.

\paragraph{Acknowledgments.}
The work  was funded by EPSRC grants
PDQ (EP/M005852/1), ED$^3$ (EP/N014359/1), and DBOnto (EP/L012138/1).
We are grateful to the journal reviewers for a multitude of helpful comments,
and in particularly for noticing some significant flaws in the arguments of
the conference version \cite{confpaper}.

\bibliographystyle{alphaurl}
\bibliography{algs}
\appendix
\section{Alternative Semantics for Plans} \label{app:idempotent}
In the body of the paper we defined a semantics for plans using valid access
selections, which assumed that 
multiple accesses with a result-bounded method always return the same output.
We also claimed that all our results held without this assumption.
We now formally define the alternative semantics where this assumption does not hold, and
show that indeed the choice of semantics makes no difference. In this appendix,
we will call
\emph{idempotent semantics} the one that we use in the main
body of the paper, and \emph{non-idempotent semantics} the one that we now
define.

Intuitively, the idempotent semantics
assumes that 
the access selection function is chosen for the entire plan,
so that all calls with the same input to the same
 access method return the same output. The non-idempotent
semantics makes no such assumption, and can choose a different valid access selection
for each access.
In both cases, the semantics is a function taking an instance $I$ for the input schema and
the input tables of the plan, and returning as output a set of \emph{possible outputs} for each output table of the plan.

Formally, given a schema $\aschema$ and instance $I$,
an \emph{access selection} is a function mapping each access on~$I$ to an
output of the
access, as defined in the body of the paper, and it is \emph{valid} if every output returned by
the access selection is a valid output for the corresponding access.
Given a plan,  a \emph{multi-selection assignment} associates a valid access selection
to each access command in the plan. An \emph{idempotent multi-selection assignment}  is one that always
assigns the same selection to a given method $\mt$, even if it occurs in multiple commands.
Given a multi-selection assignment $F$ for plan $\aplan$, we can associate to each instance $I$ 
an assignment mapping each variable in the plan to an instance of a relation,
by induction on the number of commands. 
For an access command $T \Leftarrow_\outmap \mt \Leftarrow_\inmap E$ 
the output is obtained by first evaluating $E$ to get a collection of tuples.
We then use the selection function that $F$ associates
with this command to get a set of results for each tuple, and put the union of the results
into~$T$.   The semantics of
middleware query commands is the usual semantics for relational algebra. The semantics
of concatenation of commands is now defined via induction. The output of the plan under the function $F$
is the value assigned to the output variable.

The difference between the output of a plan under the
idempotent semantics and the output under the non-idempotent semantics relates to
which assignments we consider.
For the
\emph{idempotent semantics}, given $I$, the possible outputs are those
that are returned by any idempotent multi-selection assignment.
For the \emph{non-idempotent semantics} the possible outputs are those
that are returned by any multi-selection assignment at all.

\begin{exa} \label{ex:twosem}
Consider a schema with a input-free access method $\mt$ with result bound~$5$ on relation~$R$.
Let $\aplan$ be the plan that accesses $\mt$ twice and then determines
  whether the intersection of the results is non-empty:\\[.3em]
$T_1 \Leftarrow \mt \Leftarrow \emptyset; \hfill
T_2 \Leftarrow \mt \Leftarrow \emptyset; \hfill
T_0 \colonequals \pi_{\emptyset} (T_1 \cap T_2); \hfill
  \return~T_0.$\\[.3em]
As $T_1$ and $T_2$ are identical under the idempotent semantics,
  $\aplan$ just tests if $R$ is non-empty.
  Under the non-idempotent semantics, $\aplan$ is non-deterministic,
since it can return empty or non-empty when $R$ contains at least $10$ tuples.
\end{exa}

Note that, in both semantics, when we use multiple access methods on the same relation,
there is no requirement that an access selection be ``consistent'':
if an instance $I$ includes a fact $R(a,b)$ and we have result-bounded access methods $\mt_1$ on the
first position of~$R$ and $\mt_2$ on the second position of~$R$, then an
access to~$\mt_1$ on~$a$ might return $(a,b)$ even if an access to~$\mt_2$ on
$b$ does not return
$(a,b)$. This captures the typical situation where distinct access methods use unrelated criteria to
determine which tuples to return.

It is clear that if a query that has a plan that answers it under the non-idempotent
semantics, then the same plan works under the idempotent semantics.
Conversely, Example~\ref{ex:twosem} shows that that a given plan may answer a query
under the idempotent semantics, while it does  not answer any query under the non-idempotent semantics.
However, if a query~$Q$ has  some plan that answers it under the idempotent
semantics, we can show that it also does under the non-idempotent semantics. We
formally state this as follows, recalling that an RA plan is a plan using the
full relational algebra (as introduced in the preliminaries):

\begin{prop} \label{prop:idempsuffices}
For any CQ  $Q$ over schema $\aschema$, there is a  monotone plan that
answers~$Q$ under  the idempotent semantics with respect to~$\aschema$ iff
there is a  monotone plan that answers~$Q$ under the non-idempotent semantics.
Likewise, there is an RA plan that answers~$Q$ under the idempotent semantics
  with respect to~$\aschema$ iff there is an RA plan that answers $Q$ under the
  non-idempotent semantics.
\end{prop}

We first give  the argument for \emph{RA plans} (i.e., non-monotone plans, which allow arbitrary relational algebra expressions).
If there is a plan $\aplan$ that answers $Q$ under the non-idempotent semantics, then clearly $\aplan$
also answers $Q$ under the idempotent semantics, because there are less
  possible outputs. 

In the other direction, suppose $\aplan$ answers $Q$ under the idempotent semantics.
Let $\kw{cached}(\aplan)$ be the function that executes $\aplan$, but whenever it encounters
an access $\mt$ on a binding $\accbind$ that has already been performed in a previous command, it uses
the values output by the prior command rather than making a new access, i.e., it
uses ``cached values''.
Executing $\kw{cached}(\aplan)$  under the non-idempotent semantics gives
  exactly the same outputs as executing $\aplan$ under the idempotent semantics,
  because $\kw{cached}(\aplan)$ never performs the same access twice.
  Further we can implement $\kw{cached}(\aplan)$ as an RA plan $\aplan'$: 
  for each access command
$T \Leftarrow \mt \Leftarrow E$  in~$\aplan$, we pre-process it in~$\aplan'$ by
removing from the output of~$E$ any tuples previously accessed in~$\mt$,
  using a middleware query command with the relational difference
  operator. We then perform an access to~$\mt$
with the remaining tuples, cache the output for further accesses, and
  post-process the output with a middleware query command to add back the
  output tuples cached from previous accesses.
Thus $\aplan'$ answers $Q$ under the idempotent semantics as required.

\medskip

Let us now give the argument for  \emph{monotone} plans (i.e., USPJ-plans), which are the plans
used throughout the body of the paper. Of course the forward direction
is proven in the same way, so we focus on the backward direction.
Contrary to plans that can use negation, we can no
longer avoid making accesses that were previously performed, because we can no
longer remove input tuples that we do not wish to query. However, we can still
cache the output of each access, and union it back when performing further
accesses.

Let $\aplan$ be a plan that answers $Q$ under the idempotent semantics.
We use Proposition~\ref{prop:elimupper} about the elimination of result upper
bounds to assume without loss of generality that~$\aplan$ answers the query~$Q$
on the schema $\relaxs(\aschema)$, where all result bounds of~$\aschema$ 
are replaced with result lower bounds only.

We define the plan $\aplan'$ from~$\aplan$, where access commands are modified
in the following way: whenever we perform an access for a method $\mt$ in an
access command $i$, we cache the input of access command $i$ in a special
intermediate table $\text{Inp}_{\mt,i}$ and its output in another table
$\text{Out}_{\mt,i}$, and then we add to the output of access command $i$ the
result of unioning, over all previously performed accesses with~$\mt$ for~$j <
i$, the set of tuples in  $\text{Out}_{\mt,j}$ whose restriction to the input positions lie
within
$\text{Inp}_{\mt,i} \cap \text{Inp}_{\mt,j}$. This can be implemented using the
relational join and project operators.
Informally, whenever we perform an access with a set
of input tuples, we add to its output the previous outputs of the accesses with
the same tuples on the same methods earlier in the plan. This can be implemented
using USPJ operators. For each table defined on the left-hand side of an access
or middleware command in~$\aplan$, we define its \emph{corresponding table} as
the table in~$\aplan'$ where the same result is defined: for middleware
commands, the correspondence is obvious because they are not changed from
$\aplan$ to~$\aplan'$; for access commands, the corresponding table is the one
where we have performed the postprocessing to incorporate the previous tuple
results.

We now make the following claim: 
\begin{clm} 
  \label{clm:sameres}
  Every possible output of~$\aplan'$ in the non-idempotent
semantics is a subset of a possible output of~$\aplan$ in the idempotent
  semantics, and is a superset of a possible output of~$\aplan$ in the
  idempotent semantics.
  \end{clm}
 
 This
suffices to establish that~$\aplan'$ answers the query $Q$ in the non-idempotent
semantics, because, as~$\aplan$ answers $Q$ in the idempotent semantics, its
only possible output on an instance $I$ in the idempotent semantics is $Q(I)$,
so Claim~\ref{clm:sameres} implies that the only possible output of~$\aplan'$
on~$I$ is also~$Q(I)$, so $\aplan'$ answers $Q$ under the non-idempotent
semantics, concluding the proof. So it suffices to prove
Claim~\ref{clm:sameres}. We now do so:

\begin{proof}
  Letting $O$ be a result of~$\aplan'$ under
  the non-idempotent semantics on an instance~$I$, and letting 
  $\aselect_1 \ldots \aselect_n$ be the choice of valid access selections used for each
  access command of~$\aplan'$ to obtain~$O$, we first show that $O$ is a
  superset of a possible output of~$\aplan$ in the idempotent semantics, and
  then show that $O$ is a subset of a possible output of~$\aplan$ in the
  idempotent semantics.

  To show the first inclusion, let us first consider the access selection
  $\aselect^-$ on~$I$ defined in
  the following way: for each access binding $\accbind$ on a method $\mt$,
  letting $\aselect_i$
  be the access selection for the first access command of~$\aplan$ where the
  access on~$\accbind$ is performed on~$\mt$, we define $\aselect^-(\mt,
  \accbind) \colonequals \aselect_i(\mt, \accbind)$; if the access is never
  performed, define $\aselect$ according to one of the~$\aselect_i$ (chosen arbitrarily). 
  We see that~$\aselect^-$ is a valid access selection for~$I$, because each
  $\aselect_i$
  is a valid access selection for~$i$, and for each access $\aselect^-$ returns the
  output of one of the~$\aselect_i$, which is valid.
  Now, by
  induction on the length of the plan, it is clear that for every table in the
  execution of~$\aplan$ on~$I$ with~$\aselect^-$, its contents are a \emph{subset}
  of the contents of the corresponding table in the execution of~$\aplan'$ on~$I$ with
  $\aselect_1 \ldots \aselect_n$. Indeed, the base case is trivial. The induction case
  for middleware commands follows from monotonicity of the USPJ operators. The
  induction case on access commands will follow because we perform an access with
  a subset of bindings. For each binding $\accbind$, if this is the first time we
  perform the access for this method on~$\accbind$, we obtain the same output in
  $\aplan$ as in~$\aplan'$. And if this is not the first time, in~$\aplan$ we
  obtain the output as we did the first time, and in~$\aplan'$ we still obtain it
  because we retrieve it from the cached copy. The conclusion of the induction
  is that the output of~$\aplan$ on~$I$ under $\aselect^-$ is a subset of the output
  $O$ of~$\aplan'$ on~$I$ under $\aselect_1 \ldots \aselect_n$.

  Let us now show the second inclusion by considering the access selection
  $\aselect^+$ on~$I$ defined in the following way: for each access binding
  $\accbind$ and method $\mt$, we define $\aselect^+(\mt, \accbind) \colonequals
  \bigcup_{1 \leq i \leq n} \aselect_i(\mt, \accbind)$. That is, $\aselect^+$ returns
  all outputs that are returned in the execution of~$\aplan'$ on~$I$ in the
  non-idempotent semantics with~$\aselect_1 \ldots \aselect_n$. This is a valid
  access selection, because for each access and binding it returns a superset
  of a valid output, so we are still obeying the result lower bounds, and there
  are no result upper bounds because we we are working with the schema
  $\relaxs(\aschema)$ where result upper bounds have been eliminated.
  Now, by induction
  on the length of the plan, analogously to the case above, we see that for
  every table in the execution of~$\aplan$ on~$I$ with~$\aselect^+$, its contents are
  a \emph{superset} of that of the corresponding table in the execution of
  $\aplan'$ on~$I$ with~$\aselect_1 \ldots \aselect_n$: the induction case is because
  each access on a binding in~$\aplan'$ cannot return more than the outputs  of
  this access in all the~$\aselect_i$, and this is the output obtained with
  $\aselect^+$. So we have shown that~$O$ is a subset of a possible output of
  $\aplan$, and that it is a superset of a possible output of~$\aplan$,
  concluding the proof of the claim.
\end{proof}

This concludes the proof of Proposition~\ref{prop:idempsuffices}.

\section{Proof of Theorem~\ref{thm:gtgd_one_pass}:\texorpdfstring{\\}{ }completeness of the downward-free chase} \label{sec:onepass}
Recall the statement:

\medskip

\thmgtgdonepass

\medskip

\begin{proof}
  We prove the result  by induction on $n$, calling $h_{n}$ the homomorphism
  produced for $n$.
We will  ensure inductively that our homomorphism $h_{n}$ preserves the tree
  structure of $T_{n}$.
That is,  
  there is additionally an injective homomorphism $h^{\mathrm{T}}_{n}$
  from the underlying tree of $T_{n}$ to 
  the underlying tree of $\overline{T_{m}}$ such that for each node $v$
  of~$T_{n}$ and each fact $G \in \factsof_{n}(v)$,
  the node $h^{\mathrm{T}}_{n}(v)$ contains the image fact
  of~$G$ obtained by mapping the elements of the
  fact following $h_{n}$.

For the base case $n=0$, we simply set $\overline{T_0}\coloneqq T_0$, take $m =
  n =0$, and let
  both $h_{n}$ and $h^{\mathrm{T}}_{n}$ be the identity.

  For the inductive case, there are two possibilities. The first possibility is
  that we performed a chase step when going from
  $T_{n-1}$ to~$T_n$ to fire a trigger $\rho$ at a node~$v$ to create a
  fact~$F$, then we simply take the image $h_{n-1}(\rho)$ of~$\rho$ by the
  homomorphism in the node $h^{\mathrm{T}}_{n-1}(v)$ and fire it there, creating
  the fact $h_{n-1}(F)$ that we can use to extend the homomorphism.

  The second possibility is the interesting one: we have performed a propagation
  step to go from~$T_{n-1}$ to~$T_n$, and we must
  explain how to ``mimic'' it in $\overline{T_m}$ while only performing upward
  propagation.
  To simplify the argument, letting $F$ be the fact that was just created
  in~$T_{n-1}$ in a node~$v$ and is propagated in~$T_n$, we assume that the
  propagation from~$T_{n-1}$ to~$T_n$ propagates $F$ to all nodes that have a
  guard for its elements: this is the most challenging case. 

  We first perform the upward
  propagation in~$\overline{T_m}$, that is, 
  we consider the ancestors of~$v$ having a guard of the elements of $F$ is propagated in~$T_n$, take their
  image by~$h^{\mathrm{T}}_{n-1}$, and propagate $\overline{F} \colonequals h_{n-1}(F)$ to these
  ancestors. Let $\overline{p}$ be the highest such ancestor
  in~$\overline{T_m}$.

  The key idea is now that we ``mimic'' downwards propagation by simply
  re-creating all the descendants of~$\overline{T_m}$, which will
  ``automatically'' propagate the fact $\overline{F}$ to them. More precisely,
  consider $U$ the set of all domain elements that occur in strict descendants
  of~$\overline{p}$ but do not occur in~$\overline{p}$, fix $U'$ a disjoint set of fresh element names
  of the same cardinality, and fix a bijection $h$ which maps~$U$ to~$U'$ and is
  the identity on the other elements of~$\overline{T_m}$. Now, consider the
  sequence $\overline{T_0}, \ldots, \overline{T_m}$ obtained thus far, which by
  induction does not contain any downwards propagation. Now, re-play that
  sequence but replacing the elements of~$U$ by~$U'$. More formally, all
  triggers and all created facts are mapped through~$h$. In particular, the
  non-full chase steps that created the children of~$\overline{p}$ will now create fresh
  child nodes, where the elements of~$U$ have been replaced by elements of~$U'$,
  further non-full chase steps on these children will continue creating a copy
  of their subtree, and full chase steps happening in their subtrees as well as
  upwards propagations are performed in the same way as in the original
  sequence. (Chase steps that create facts with no element of~$U$ are unchanged
  by the transformation, and doing them again recreates a fact that already
  exists, which has no effect.)

  After this process, we have extended $\overline{T_0}, \ldots, \overline{T_m}$
  by a sequence $\overline{T_{m+1}}, \ldots, \overline{T_{m'}}$, and 
  $\overline{T_{m'}}$ is a superinstance of~$\overline{T_m}$ which differs in
  that every child of~$\overline{p}$ now exists in two copies, one featuring elements
  of~$U$ and the other one featuring the corresponding elements of~$U'$, these
  two copies being the roots of subtrees between which $h$ is an isomorphism.
  Overall, the homomorphism $h$ maps $\overline{T_m}$ to~$\overline{T_{m'}}$ by mapping each
  original subtree to its copy. Let us call $h^{\mathrm{T}}$ the corresponding
  injective homomorphism at the level of tree nodes. We can compose $h_m$ and
  $h^{\mathrm{T}}_m$ with $h$ and $h^{\mathrm{T}}$ respectively to obtain
  well-defined homomorphisms $h_{m'}$ and $h^{\mathrm{T}}_{m'}$. Now, to show that they are suitable, the only point
  to verify is that we have correctly propagated the new fact, i.e., for all
  nodes $v'$ of~$T_n$ where $F$ is guarded, the node $h^{\mathrm{T}}_{m'}$ indeed contains $\overline{F}$.

  To understand why, notice that when we perform the sequence $\overline{T_{m+1}}, \ldots,
  \overline{T_{m'}}$, the node~$\overline{p}$ contains the new fact
  $\overline{F}$ that we wished to downwards propagate. Hence, while we created
  the new subtrees, $\overline{F}$ was added to every new child node whenever it
  was guarded by the elements of that node. Thus, $\overline{F}$ now exists in
  each node of the new subtrees where it is guarded.
  This establishes that the chase sequence $\overline{T_0}, \ldots,
  \overline{T_{m'}}$ and the homomorphisms $h_{m'}$ and $h^{\mathrm{T}}_{m'}$
  satisfy the conditions. This shows the inductive case, and concludes the
  inductive proof, establishing the result.
\end{proof}

\section{Proof of the Semi-Width Result
(Proposition~\ref{prop:semiwidthclassic})}
\label{apx:semiwidthproof}

In this appendix, we prove the $\np$ bound on containment for bounded semi-width
IDs, i.e., 
Proposition~\ref{prop:semiwidthclassic}. Recall its statement:

\begin{quote}
\semiwidthclassic
\end{quote}

To prove the result, let $\Sigma$ be the collection of~$\incd$s, and 
consider  a chase sequence based on the canonical database $I_0 \colonequals
\canondb(Q)$ of the conjunctive query
$Q$.
Recall the notion of a \emph{tree-like chase proof}
from the body of the paper (after
Proposition~\ref{prop:derivedaccessids}). 
This is a sequence of labelled trees --- the \emph{chase trees} of the proof ---
 one for each instance in the  chase sequence,
where the tree
associated with~$I_0$ consists of a single root node. In the case where
the constraints are all  $\incd$s, we will modify
the tree structure in this definition slightly, creating a new
node even when firing full rules. We do not perform propagation of facts, which will never
be needed for $\incd$s.
Thus the chase
will have non-root  nodes $n_F$ in one-to-one correspondence
with  generated facts $F$.
If performing a chase step on fact $F$ produces fact $F'$ in the sequence, then
the node~$n_{F'}$ is a child of the node~$n_F$.

Consider nodes $n$ and $n'$ in a chase tree within some tree-like chase proof,  with~$n$ a strict ancestor of~$n'$.
We say $n$ and $n'$ are \emph{far apart}
 if there are distinct generated facts $F_1$ and $F_2$ such that:
\begin{itemize}
\item  the node $n_1$
 corresponding to~$F_1$ and the node~$n_2$ corresponding to~$F_2$ are both ancestors
 of~$n'$ and descendants of~$n$,;
\item  $n_1$ is an ancestor of~$n_2$;
\item  $F_1$ and $F_2$ were generated by the same rule of~$\Sigma$; and
\item the equalities between values in positions within~$F_1$ are exactly the
  same as the equalities within~$F_2$,
and any values occurring in both $F_1$ and $F_2$ occur
in the same positions within  $F_2$ as they do in~$F_1$.
\end{itemize}
If such an $n$ and $n'$ are not far apart, we say
that they are \emph{near}.

A \emph{match} of~$Q$ in the chase tree is a mapping from the variables of~$Q$
to the elements of the chase tree which is a homomorphism, i.e., it also maps
every atom of~$Q$ to a fact in the chase tree.
Given a match $h$ of~$Q$ in the chase tree, its
\emph{augmented image} is the closure of its image
under least common ancestors, including by convention the root node. If $Q$ has size $k$ then
this has size $\leq 2k+1$.
For nodes $n$ and $n'$ in the augmented image,
we call $n$ the \emph{image parent of} $n'$ 
if $n$ is the lowest  ancestor
of~$n'$ in the augmented image.

\begin{lem} If $Q$ has a match $h$ in the final instance of a tree-like chase proof, where the final instance
has chase tree $T$, then there is 
another tree-like chase proof with final tree $T'$, and a match $h'$ with the property that if
$n$  is the image parent of~$n'$
then $n$ and $n'$ are near.
\end{lem}
\begin{proof}
We prove that given such an $h$ and $T$, we can
construct an $h'$ and $T'$ such that we decrease 
the sum of the depths of the violations.

If $n$ is far apart from~$n'$, then there are witnesses
$F_1$ and $F_2$ to this, corresponding to nodes $n_1$ and $n_2$ respectively.
   Informally, we will ``pull up'' the homomorphism
by replacing witnesses below $F_2$ with witnesses below $F_1$.
Formally, we create  $T'$ by first removing each
step of the chase proof  that generates a node that is a descendent
 of $n_1$.  Letting $T_1$ be the nodes in $T$ that do not lie below $n_1$, we 
will add nodes and the associated proof steps to $T'$.
Let $C_2$ be the chase steps in $T$ that generate a node below $n_2$, ordered 
as in~$T$, and let $T_2$ be the nodes produced by these steps.
We then add chase steps in $T'$ for each chase step in $C_2$. 
More precisely, 
we expand $T'$ by an induction on prefixes of $C_2$, building
$T'$ and a  partial function  $m$ from the domains
of facts in $\{n_2\} \cup T_2$ into the domain of facts  associated to
$n_1$ and its descendants in $T'$. The invariant is that $m$ preserves each fact of $T$ generated
by the chase steps in $C_2$ we have processed thus far in the induction, and that
$m$ is the identity on any values in $F_1$.
We initialize the induction by mapping the elements associated to $n_1$
to elements associated to $n_2$.
Our assumptions
on  $n_1$ and $n_2$ suffice to guarantee that we can perform such a mapping satisfying the invariant.
For the inductive case, suppose the next
chase step $s$ in $C_2$ uses  $\incd$ $\delta$, firing
on the fact associated to $v_i$ in $T$,  producing node $v_{i+1}$.
Then we perform a step $s'$ using
 $\delta$ and the fact associated to  $m(v_i)$ in $T'$.
If $\delta$ was a full $\incd$ we do not modify $m$, while if it is a non-full 
$\incd$
 we  extend $m$ to map the generated elements of $s$ to the corresponding
elements of $s'$.
 We can thus form $h'$ by revising $h(x)$ when $h(x)$ lies
below  $n_1$, setting $h'(x)$ to $m(h(x))$.
Note that there could not have been any elements in the augmented
image of $h$ in $T$ that hang off the path between $n_1$
and $n_2$, since $n$ and $n'$ were assumed to be adjacent in the augmented image.

In moving from $T$ and $h$ to $T'$ and $h'$ we reduce the sum of the depths of
nodes in the image, while no new violations 
are created, since the image-parent relationships  are preserved.
\end{proof}

Call a match $h$ of~$Q$ in the chase \emph{tight} if it
has the property given in the lemma above. The \emph{depth}
of the match is the depth of the lowest node in its image.
The next observation, also due to Johnson and Klug, is that when the width
is bounded, tight matches can not occur  far down in the tree:

\begin{lem} \label{lem:depthbound} If $\Sigma$ is a set of $\incd$s of width $w$
and the schema has arity bounded by $m$,
then any tight match  of size $k$  has all of its nodes at depth at most
  $k \cdot \card{\Sigma} \cdot m^{w+1} \cdot 2^w$.
\end{lem}

\begin{proof}
We claim that the length of the path  between a match element $h(x)$
and its image parent $h(x')$
 must be at most $\Delta \colonequals  \card{\Sigma} \cdot m^{w+1} \cdot 2^w$.
At most $w$ values from~$h(x')$ are present
in any fact on the path,  and thus the number of configurations
that can occur is at most $m^{w+1}$. Further, we multiply by a factor
  of~$\card{\Sigma}$ because we are accounting for the last rule used. We also
  multiply by a factor of $2^w$ to account for the possible equality patterns
  among the values in the positions of the fact that do not contain a null.
  Thus after
 $\Delta$ steps there will be two elements which
repeat both the rule and the configuration of the values, which
would contradict tightness.
Since the augmented image contains the root, this implies the bound above.
\end{proof}

Johnson and Klug's result follows from combining the previous two lemmas:
\begin{propC}[\cite{johnsonklug}]
  \label{prop:jkwidth}
For any fixed  $w \in \NN$, 
there is an $\np$ algorithm for query containment under  $\incd$s of
width at most $w$.
\end{propC}
\begin{proof}
We know it suffices to determine whether there  is a match in
a chase proof, and the previous lemmas tell us that 
the portion of a chase proof required to find a match is not large.
We thus guess a tree-like chase proof where the tree consists
of $k$ branches of depth at most
$k \cdot |\Sigma| \cdot m^{w+1} \cdot 2^w$, along with  
a match in them, verifying the validity of the branches according to the
rules of $\Sigma$.
\end{proof}

We now give the  extension of this argument for bounded semi-width.
Recall from the body that a collection of $\incd$s $\Sigma$ has \emph{semi-width} bounded
by $w$ if it can be decomposed as~$\Sigma = \Sigma_1 \cup \Sigma_2$
where $\Sigma_1$ has width bounded by $w$ and
the basic position graph of~$\Sigma_2$ is acyclic.
An easy modification of Proposition~\ref{prop:jkwidth}
now completes the proof of our semi-width result
(Proposition~\ref{prop:semiwidthclassic}):
\begin{proof}
We revisit the argument of Lemma~\ref{lem:depthbound}, claiming a bound
  with an extra factor of $\card{\Sigma}$ in it.
As in that argument, it suffices
  to show that, considering the extended image of a tight match of~$Q$ in
  a chase proof, then the distance between any node~$n'$ of the extended image
  and its closest ancestor~$n$ is bounded, i.e., it 
must be at most $|\Sigma|^2 \cdot m^{w+1} \cdot 2^w$.
Indeed, as soon as we apply a rule of~$\Sigma_1$ along
the path, at most $w$ values are exported, and so the
remaining path is bounded as before.
Since $\Sigma_2$ has an acyclic basic position graph, 
a value in~$n$ can propagate for at most $|\Sigma_2|$ steps 
when using rules of~$\Sigma_2$ only. Thus after
at most $|\Sigma_2|$  edges in a path
we will either have no values propagated (if we used only
rules from~$\Sigma_2$) or at most $w$ values (if we used
  a rule from~$\Sigma_1$). In particular, we cannot
have a gap of more than  $ | \Sigma_2|  \cdot 
|\Sigma| \cdot m^{w+1} \cdot 2^w$
in a tight match.
\end{proof}

\section{Generalization of Results to RA Plans}
\label{apx:ra}
In the body of the paper we dealt with monotone answerability. 
A \emph{relational algebra plan} or just \emph{RA plan} is defined as with a monotone
plan, but in addition to the  monotone relational algebra expressions we also
allow as an operator a relational \emph{difference operator} $\setminus$, which takes
as input instances of two relations with the same  arity.
It is known that the  queries defined by relational algebra are the same as those
defined by first-order logic with the active-domain semantics \cite{AHV}.
As we mentioned in the body of the paper, there are monotone  queries that
can be expressed using an RA plan, but not with a monotone plan. In the setting
of views, this surprising fact, which contradicted prior claims in the
literature (e.g., \cite{SVconf}),
was uncovered in \cite{NSV}. 
These counterexamples notwithstanding,  the bulk of the work in the literature on querying with access patterns
has focused
on monotone plans.

At the end of
Section~\ref{sec:prelims} and in Section~\ref{sec:conc}, we claimed that many of the  results in
the paper, including the reduction to query containment and the schema
simplification results, generalize in the ``obvious way''
to answerability where general relational algebra expressions are allowed. In addition, the results on complexity
for monotone answerability that are shown in the body extend to answerability
with RA plans, with one exception.
The exception is that
 we do not have a decidability result for $\uincd$s and FDs analogous to
 Theorem~\ref{thm:deciduidfd}, because
the containment problem is more complex.

We explain in the rest of the appendix how to adapt our results in the
unrestricted setting from
monotone answerability to RA answerability.
In the specific case of $\incd$s,
we will show (Proposition~\ref{prop:monotonecollapse}) that
RA answerability and monotone answerability coincide. This
generalizes a result known for views, and extends it to the setting with result
bounds.
\subsection{Variant of Reduction Results for RA Answerability} \label{subsec:detdeffull}

We first formally define the analog of~$\amd$ for the notion of RA answerability
that we study in this appendix. 
In the absence of result bounds, this corresponds to the notion of
\emph{access-determinacy}~\cite{thebook,ustods},
which states that two instances with the
same accessible part must agree on the query result. Here we
generalize this to the setting with result bounds, where the accessible instance
is not uniquely defined.

Given a schema $\aschema$ with constraints and methods which may have
result lower bounds as well as result upper bounds, a query $Q$ is said to be \emph{access-determined}
 if for any two instances $I_1$, $I_2$ satisfying the constraints of~$\aschema$,
 if there is a valid access selection $\sigma_1$ for $I_1$
 and a valid access selection $\sigma_2$ for $I_2$ such that
$\accpart(\sigma_1, I_1)=\accpart(\sigma_2, I_2)$,
then  $Q(I_1)=Q(I_2)$.

We will now show that access-determinacy is equivalent to query
containment, as with access monotonic determinacy.
We will stick to the setting where $Q$ is a CQ, for consistency with
the body of the paper. This restriction will also be essential
in the core results on expressiveness, decidability, and complexity to come.
However, \emph{the results in this subsection, concerning reduction to query
containment, hold also for a query $Q$ in  relational algebra}.

As we did with $\amd$,  it will be convenient to give an alternative definition
of access-determinacy that talks only about a subinstance of a single instance.

For a schema $\aschema$ a 
 common subinstance
$I_\acc$ of~$I_1$ and $I_2$ is \emph{jointly access-valid} if, for any access performed with a method of~$\aschema$ in
$I_\acc$, there is  a set of matching
tuples in~$I_\acc$ which is a valid output to the access in~$I_1$
and in~$I_2$. In other words, there is an access selection~$\aselect$
for~$I_\acc$
whose outputs are valid in~$I_1$ and in~$I_2$.

We now claim the analogue of
Proposition~\ref{prop:altdef}, namely:

\begin{prop}
  \label{prop:altdefra}
 For any schema $\aschema$ with arbitrary constraints $\Sigma$ and methods which may have result lower bounds
and result upper bounds,
  a CQ $Q$ is access-determined if and only if the following implication holds:
for any two instances $I_1, I_2$ satisfying $\Sigma$, if $I_1$ and $I_2$ have a
  common subinstance~$I_\acc$ that is jointly
  access-valid, then $Q(I_1) = Q(I_2)$.
\end{prop}

This result gives the alternative definition of access-determinacy that we will
use in our proofs.
Proposition~\ref{prop:altdefra} follows immediately from the following
proposition (the analogue of Proposition~\ref{prop:equivalence}):

\begin{prop}  \label{prop:commonaccessiblera}
Again assume a schema with arbitrary constraints  along with
methods that may have result lower and upper bounds.
The following are equivalent:
  \begin{enumerate}[(i)]
\item $I_1$ and $I_2$ have a common subinstance $I_\acc$ that is jointly access-valid.
\item There is a common accessible part $A$ 
of~$I_1$ and for~$I_2$.
\end{enumerate}
\end{prop}
\begin{proof}
Suppose  $I_1$ and $I_2$ have a common subinstance $I_\acc$ that is jointly access-valid. 
 This means that we can define an access selection
  $\aselect$ that takes any access performed with values of~$I_\acc$ and a
  method of~$\aschema$, and maps it to a set of matching tuples in~$I_\acc$ that
  is valid in~$I_1$ and in~$I_2$.
We can see that~$\aselect$ can be used as a valid access selection in~$I_1$ and $I_2$
  by extending it to return an arbitrary valid output to accesses in~$I_1$ that
  are not accesses in~$I_\acc$, and likewise to accesses in~$I_2$ that are not
  accesses in~$I_\acc$; we then have
  $\accpart(\aselect,I_1)=\accpart(\aselect,I_2)$ so we can define the
  accessible part $A$
  accordingly, noting that we have $A \subseteq I_\acc$.
Thus the first item implies the second.

Conversely, suppose that $I_1$ and $I_2$ have a common accessible part $A$,
  and let $\aselect_1$ and~$\aselect_2$ be the witnessing valid access selections
  for~$I_1$ and~$I_2$, i.e.,
$A = \accpart(\aselect_1, I_1)=\accpart(\aselect_2, I_2)$.
Let $I_\acc \colonequals A$, and let us show that $I_\acc$ is a common
  subinstance of~$I_1$ and~$I_2$ that is jointly access-valid. By definition we
  have $I_\acc \subseteq I_1$ and $I_\acc \subseteq I_2$. Now, to show that it
  is jointly access-valid in~$I_1$ and~$I_2$, consider any access
$\abind, \mt$ with values in~$I_\acc$. We know that there is $i$ such
that~$\abind$ is in~$\accpart_i(\aselect_1, I_1)$, therefore by definition of
  the fixpoint process and of the access selection~$\aselect_1$ there is a valid
  output to the access in~$\accpart_{i+1}(\aselect_1, I_1)$, hence in~$I_\acc$. 
  Thus we can choose an output in~$I_\acc$ which is valid
  in~$I_1$.
But this output must also be in~$\accpart(\aselect_2, I_2)$, and thus it is valid in
$I_2$ as well.
Thus, $I_\acc$ is jointly access-valid. This shows the converse implication and
concludes the proof.
\end{proof}

The following analogue of Proposition~\ref{prp:plantoproof} motivates these two
equivalent definitions of access-determinacy, showing that either
one is equivalent to the existence of an RA-plan that answers $Q$:
\begin{prop}
  \label{prop:plantoproofra}
Again assume a schema with arbitrary constraints, along with methods that
may have result upper bounds and result lower bounds.
If a CQ $Q$ has an RA plan $\aplan$ that answers it \wrt\ $\aschema$, then
$Q$ is access-determined over~$\aschema$.
\end{prop}

\begin{proof}
  Assume that $Q$ has an RA plan $\aplan$ that answers it. Using
  Proposition~\ref{prop:altdefra}, consider two instances $I_1$ and $I_2$
  satisfying the constraints of~$\aschema$, and having a common subinstance
  $I_\acc$ that is jointly access-valid. Let us show that $Q(I_1) = Q(I_2)$.
  Let $\sigma$ be a valid access selection for~$I_\acc$, and extend it to a
  valid access selection $\sigma_1$ for~$I_1$  and $\sigma_2$ for~$I_2$.
  Specifically, accesses with a binding in~$I_\acc$ on~$\sigma_1$ and~$\sigma_2$
  return the same result as~$\sigma$, which by definition is valid in~$I_1$
  and~$I_2$ in addition to being valid in~$I_\acc$. Further, accesses with a
  binding using values from~$\adom(I_1)\setminus \adom(I_\acc)$ for~$\sigma_1$
  return some valid response for~$\sigma_1$, and likewise for~$\sigma_2$.

Now, a simple induction shows that the intermediate tables produced by a plan
using $\sigma_1$ on $I_1$, using $\sigma_2$ on~$I_2$, and using $\sigma$
  on~$I_\acc$, must be the same, and must all consist of values
  from~$\adom(I_\acc)$.

Now, as $\aplan$ answers $Q$, we know that the output 
of $Q$ on $I_1$ is equal to that of $Q$ on $I_2$. This concludes the proof.
\end{proof}

Analogously to Theorem~\ref{thm:equiv},  we can show that access-determinacy is equivalent
to RA answerability.
The proof starts the same way as that of Theorem~\ref{thm:equiv}, noting
that in  the absence of  result bounds, this equivalence was shown in prior work:
\begin{thmC}[\cite{thebook,ustods}]
  \label{thm:determandplansclassic}
  For any CQ~$Q$ and schema $\aschema$ (with no result bounds)
  whose constraints $\Sigma$ are expressible
  in active-domain first-order logic, the following are equivalent:
\begin{enumerate}
\item $Q$ has an RA plan that answers it over~$\aschema$.
\item $Q$ is access-determined over~$\aschema$.
\end{enumerate}
\end{thmC}
The extension to result bounds is shown using the same reduction as
for Theorem~\ref{thm:equiv}, by just ``axiomatizing'' the result bounds as
additional constraints (by a direct analogue of Proposition~\ref{prop:elimresultboundplan}).
This gives the immediate generalization of
Theorem~\ref{thm:determandplansclassic} to schemas that may include result
bounds:

\begin{thm} \label{thm:equivra}
  For any CQ~$Q$ and schema $\aschema$ 
  whose constraints~$\Sigma$ are expressible
  in active-domain first-order logic, where methods may have result upper and result lower bounds, the following are equivalent:
\begin{enumerate}
\item $Q$ has an RA plan that answers it over~$\aschema$.
\item
$Q$ is access-determined over~$\aschema$.
\end{enumerate}
\end{thm}

Hence, we have shown the analogue of Theorem~\ref{thm:equiv} for the setting of
RA answerability and RA plans studied in this appendix.

\paragraph{Elimination of result upper bounds for RA plans.}
As with monotone answerability, it suffices to consider only result lower bounds.
Recall that $\relaxs(\aschema)$ is the schema obtained
from~$\aschema$ by removing result upper bounds and keeping only result lower
bounds. We have:

\begin{prop} \label{prop:elimupperra} Let $\aschema$ be a schema with
arbitrary constraints and access methods which may have result lower bounds and result
upper bounds.
A query $Q$ is RA answerable in~$\aschema$ if and only if it is RA answerable
in~$\relaxs(\aschema)$.
\end{prop}

\begin{proof}  
  The proof follows that of Proposition~\ref{prop:elimupper}. 
  We show the result for access-determinacy instead of RA answerability, thanks to
  Theorem~\ref{thm:equivra}, and we use Proposition~\ref{prop:altdefra}.
  Consider arbitrary instances $I_1$ and
  $I_2$ that satisfy the constraints, and let us show that any common subinstance
  $I_\acc$ of~$I_1$ and $I_2$  is jointly access-valid for~$\aschema$ iff
  it is jointly access-valid for~$\relaxs(\aschema)$: this implies
  the claimed result.

  In the forward direction, if $I_\acc$ is jointly access-valid for~$\aschema$,
  then clearly it is jointly access-valid for~$\relaxs(\aschema)$, as any output
  of an access on~$I_\acc$ which is valid in~$I_1$ and in~$I_2$ for~$\aschema$
  is also valid for~$\relaxs(\aschema)$.

  In the backward direction, assume $I_\acc$ is jointly access-valid
  for~$\relaxs(\aschema)$, and consider an access $(\mt, \abind)$ with values
  from~$I_\acc$. 
  If $\mt$ has no result lower bound, then there is only one possible output for the
  access, and it is valid also for~$\aschema$.
  Likewise, if $\mt$ has a result lower bound of~$k$ and there are $\leq k$ matching
  tuples for the access in~$I_1$ or in~$I_2$, then the definition of a result lower bound ensures
  that there is only one possible output which is valid for~$\relaxs(\aschema)$
  in~$I_1$ and~$I_2$, and it is again valid for~$\aschema$.
  Finally, if there are $>k$ matching tuples for the access,
  we let $J$ be a set of tuples in~$I_\acc$ which is is a valid output
  to the access in~$I_1$ and~$I_2$ for~$\relaxs(\aschema)$, and take any subset $J'$ of~$J$ with
  $k$ tuples; it is clearly a valid output to the access for~$\aschema$
  in~$I_1$ and~$I_2$. This
  establishes the backward direction, concluding the proof.
\end{proof}

Based on this, \emph{from now on we will assume only result lower bounds in our schema}.

\paragraph{Reduction to query containment.}
Since access-determinacy  can be expresses as a query containment,
in  Theorem~\ref{thm:equivra} we already established
an reduction of RA answerability to query containment. 
We will spell out what these axioms look like, focusing
 in the case where we have only result lower bounds. This is sufficient for our purposes
by Proposition~\ref{prop:elimupperra}.
The constraints will be a more ``more symmetrical'' version
of the axioms we saw in the case of access monotone determinacy.

Recall that $\accessible$ is a fresh unary relation, intuitively used to
describe which elements are accessible.
Given a schema $\aschema$ with constraints
and access methods with result bounds, the 
 \emph{access-determinacy containment} for~$Q$ and $\aschema$ is
the CQ containment $Q \subseteq_\Gamma Q'$ where
the constraints $\Gamma$ are defined as follows: they include the original
constraints $\Sigma$, the constraints $\Sigma'$ on the relations~$R'$, and the following
axioms (with implicit universal
quantification):
\begin{itemize}
  \item For each method $\mt$ that is not result-bounded, letting $R$ be the
    relation accessed by~$\mt$:
    \begin{align*}
    \Big(\bigwedge_i \accessible(x_i)\Big) \wedge
      \phantom{'}R(\vec x, \vec y) \rightarrow & R_\acc(\vec x, \vec y)\\
      \Big(\bigwedge_i \accessible(x_i)\Big) \wedge
      R'(\vec x, \vec y) \rightarrow & R_\acc(\vec x, \vec y)
    \end{align*}
    where $\vec x$ denotes the input positions of~$\mt$ in~$R$ and $i$ ranges
over these positions.
  \item For each method $\mt$ with a result lower bound of~$k$, letting $R$ be
    the relation accessed by~$\mt$, for all $j \leq k$:
    \begin{align*}
      \Big(\bigwedge_i \accessible(x_i)\Big) \wedge \exists^{\geq j} \vec y ~
      \phantom{'}R(\vec x, \vec y)
      \rightarrow & \exists^{\geq j} \vec z ~ R_\acc(\vec x, \vec z) \\
      \Big(\bigwedge_i \accessible(x_i)\Big) \wedge \exists^{\geq j} \vec y ~
  R'(\vec x, \vec y)
      \rightarrow & \exists^{\geq j} \vec z ~ R_\acc(\vec x, \vec z)
    \end{align*}
    where $\vec x$ denotes the input positions of~$\mt$ in~$R$.
\item For every relation $R$ of the original signature:
  \[R_\acc(\vec w) \rightarrow R(\vec w) \wedge R'(\vec w) \wedge \bigwedge_i
    \accessible(w_i).\]
\end{itemize}

The intuition, like for~$\amd$ containment, is that the constraints $\Gamma$ are
axiomatizing the definition of access-determinacy, i.e., enforcing that $I_\acc$
is jointly access-valid.
The only difference from the~$\amd$ containment is that the additional
constraints are now symmetric in the two signatures, primed and unprimed.
The analogue of Proposition~\ref{prop:reduce} then follows immediately from Theorem~\ref{thm:equivra} and the definition of access-determinacy:

\begin{prop} \label{prop:reducera}
Let $Q$ be a CQ, and let $\aschema$ be a schema with constraints expressible
  in active-domain first-order logic and with access methods that may have
  result upper and lower bounds. Then the following are equivalent:
\begin{itemize}
\item $Q$ has an RA plan that answers it over~$\aschema$.
\item $Q$ is access-determined over~$\aschema$.
\item The containment corresponding to access-determinacy holds.
\end{itemize}
\end{prop}

Based on this reduction, all of our  arguments about answerability
with RA-plans can deal with the semantic notion of
access-determinacy and the corresponding entailments, and we will always
make use of this in what follows.
Also following the convention in the body of the paper,
in reasoning about access-determinacy and these
entailments,  \emph{we can
restrict to the case of Boolean CQs}, since non-Boolean CQs can be considered
Boolean CQs with additional constants. We perform this restriction
in proofs by default in the remainder of this section.

\subsection{Full Answerability and Monotone Answerability}
We show that there is no difference between full answerability and monotone answerability
when constraints consist of $\incd$s only.
This is a generalization of an observation that is known for views
(see, e.g., Proposition~2.15 
in~\cite{thebook}):

\begin{prop} \label{prop:monotonecollapse}
Let $\aschema$ be a schema with access methods and constraints $\Sigma$ consisting
of inclusion dependencies, and $Q$ be a CQ that is access-determined.
Then $Q$ is $\amd$.
\end{prop}
\begin{proof}
  We know by Propositions~\ref{prop:elimupper} and~\ref{prop:elimupperra} that
  we can work with $\relaxs(\aschema)$ which has only 
  result lower bounds, so we do so throughout this proof.
Towards proving $\amd$, assume by way of contradiction that
  we have:
\begin{itemize}
\item  instances $I_1$ and $I_2$ satisfying
$\Sigma$;
\item    an accessible part $A_1$ of~$I_1$ with valid access selection $\aselect_1$,
and an accessible part $A_2$ of~$I_2$ with valid access selection
  $\aselect_2$;
\item $A_1 \subseteq A_2$;
\item $Q$ holding in~$I_1$
but not in~$I_2$.
\end{itemize}
 We first modify $I_2$ and $A_2$ to~$I_2^+$ and $A_2^+$ by replacing each element that is in~$I_1$ but not in~$A_1$ by
a copy that is not in~$I_1$; we modify the access selection from $\aselect_2$ to
  $\aselect_2^+$ accordingly. Since $I_2^+$
is isomorphic to~$I_2$,
it is clearly true that the access selection $\aselect_2^+$ is valid in $I_2^+$,
  that  $A_2^+$ is the
  accessible part of~$I_2^+$ corresponding to~$\aselect_2^+$, that $I_2^+$ satisfies $\Sigma$
and that~$Q$ fails in~$I_2^+$. Further we still have $A_1 \subseteq A_2^+$ by
  construction. What we have ensured at this step is that values of $I_2^+$ that
  are in~$I_1$ must be in~$A_1$.

Consider now $I_1^+\colonequals I_1 \cup I_2^+$. It is clear that~$Q$ holds
  in~$I_1^+$, and $I_1^+$ also
  satisfies~$\Sigma$ because $\incd$s are preserved under taking unions.  We will show
that $I_1^+$ and $I_2^+$ have a common
accessible part $A_2^+$, which will contradict
the assumption that $Q$ is access-determined.

  Towards this goal, define an access selection $\aselect_1^+$ on~$I_1^+$ as follows:
  \begin{itemize}
    \item For any access $(\mt, \accbind)$ made with a binding where all values are
  in~$A_1$, we let $\aselect_1^+(\mt, \accbind) \colonequals \aselect_1(\mt,
  \accbind) \cup \aselect_2^+(\mt, \accbind)$: note that all returned tuples are
      in $A_2^+$, because
  the second member of the union is contained in~$A_2^+$, while the first is contained in $A_1$ which is
  a subset of~$A_2^+$.
    \item For any access $(\mt, \accbind)$ made with a binding where all values are
  in~$A_2^+$ and some value is not in~$A_1$, we let $\aselect_1^+(\mt, \accbind)
      \colonequals \aselect_2^+(\mt, \accbind)$: again all tuples returned here
      are in $A_2^+$.
\item For any access $(\mt, \accbind)$ made with a binding where
  some value is not in~$A_2^+$, we choose an arbitrary set of tuples of~$I_1^+$ to
      form a valid output.
  \end{itemize}
  We claim that
  $\aselect_1^+$ is a valid access selection and that performing the fixpoint
  process with this access selection yields $A_2^+$ as an accessible part
  of~$I_1^+$. To show this, first notice that performing the fixpoint
  process with $\aselect_2^+$ indeed returns~$A_2^+$: all facts of~$A_2^+$ are
  returned because this was already the
  case in~$I_2^+$, and no other facts are returned because it is clear by induction that the fixpoint will
  only consider bindings in~$A_2^+$, so that the choices made in the third point of
  the list above have no impact on the accessible part that we obtain.

  So it suffices to show that $\aselect_1^+$ is valid,
  i.e., that for any access $(\mt, \abind)$ with a binding $\abind$
  in~$A_2^+$, the access selection $\aselect_1^+$ returns a set of tuples which
  is a valid output to the access. For the first point in the list, we know that
  the selected tuples are the union of a valid result to the access in~$I_1$
  and of a valid result to the access in~$I_2^+$, so it is clear that it
  consists only of matching tuples in~$I_1^+$. We then argue that it is valid by
  distinguishing two cases. If~$\mt$ is not result-bounded, then the output is
  clearly valid, because it contains all matching tuples of~$I_1$ and all
  matching tuples of~$I_2^+$, hence all matching tuples of~$I_1^+$. Now suppose $\mt$ has a
  result lower bound of~$k$. Suppose that for  $j \leq k$
there are $\geq j$ matching tuples in~$I_1^+$. We will show that the output of the access contains 
  $\geq j$ tuples. There are two sub-cases. The first sub-case is when there are $\geq j$ matching tuples in~$I_1$.
In this sub-case we can 
  conclude because $\aselect_1(\mt, \accbind)$ must return $\geq j$
  tuples. The second sub-case is when there are $<j$ matching tuples in~$I_1$.
  In this sub-case, 
  $\aselect_1(\mt, \accbind)$ must return all of them, so these matching tuples
  are all in~$A_1$.  Hence they are all in $A_2^+$ because $A_1 \subseteq A_2^+$.
  Thus the returned tuples are in $I_2^+$. Thus, in the second sub-case, all
  matching tuples in~$I_1^+$ for the access are actually in~$I_2^+$, so we
  conclude because $\aselect_2^+(\mt, \accbind)$ must return $\geq j$ tuples. This 
  shows that the outputs of accesses defined in the first point are valid.
 
  For accesses corresponding to the second point in the list, by the
  construction used to create $I_2^+$
  from~$I_2$, we know that the value in~$\accbind$ which is not in~$A_1$ cannot
  be in~$I_1$ either. Thus  all matching tuples of the access are in~$I_2^+$.
  So we conclude because $\aselect_2^+$ is a valid access selection of~$I_2^+$.
  For accesses corresponding to the third point, the output is always valid by definition.
  Hence, we have
  established that $\aselect_1^+$ is valid, and that it
  yields $A_2^+$ as an accessible part of~$I_1^+$.

  We have thus shown that~$I_1^+$ and $I_2^+$ both have $A_2^+$ as an accessible
  part.
  Since $Q$ holds in~$I_1^+$, by access-determinacy
$Q$ holds in~$I_2$, and this contradicts our initial assumption, concluding the
  proof.
\end{proof}

From Proposition~\ref{prop:monotonecollapse} we immediately
see that \emph{in the case where the constraints
consist of $\incd$s only, 
all the results about monotone answerability with result bounds 
transfer to answerability}. This includes simplification results and
complexity bounds.

\subsection{Blowup for RA Answerability}
We now explain how the method of ``blowing up counterexamples'' introduced
in the body extends to work with access-determinacy.  We consider a
counterexample to access-determinacy in the simplification (intuitively a pair of
instances that satisfy the constraints and have a common subinstance that is jointly access-valid but one
satisfy the query and one does not), and we show that we can blow it up to a
counterexample to access-determinacy in the original schema. 
As we did in the body, we stick to Boolean CQs in all of our arguments~--- the
results extend 
to the non-Boolean case by the usual method of changing free variables to constants.
Formally,
  a \emph{counterexample to access-determinacy} for a Boolean CQ~$Q$ and a
  schema $\aschema^\dagger$ is a pair of
  instances $I_1^\dagger, I_2^\dagger$ both satisfying the schema constraints,
  such that~$I_1^\dagger$ satisfies $Q$ while $I_2^\dagger$
  satisfies $\neg Q$, and $I_1^\dagger$ and $I_2^\dagger$ have a common  subinstance
  $I_\acc^\dagger$ that is jointly access-valid.

It is clear that, whenever there is a counterexample to access-determinacy for schema
$\aschema$ and query~$Q$, then $Q$ is \emph{not} access-determined \wrt\
$\aschema$.
We now state the 
blowup lemma that we use. It is the direct analogue of
Lemma~\ref{lem:enlarge}, and the intuition is similar: we will obtain our
counterexample for~$\aschema$ by ``blowing up'' a counterexample to
access-determinacy for~$\aschema^\dagger$.
Here is the formal statement:

\begin{lem}
  \label{lem:enlargera}
  Let $\aschema$ and $\aschema^\dagger$ be schemas and $Q$ a Boolean CQ on the common relations
  of $\aschema$ and~$\aschema^\dagger$ such that $Q$ is not
  access-determined in~$\aschema^\dagger$. Suppose that for some counterexample
  $I_1^\dagger,
  I_2^\dagger$ to access-determinacy for~$Q$ in~$\aschema^\dagger$ we can construct instances
  $I_1$ and $I_2$ that satisfy the constraints of
  $\aschema$, that have a common subinstance $I_\acc^\dagger$ that is jointly access-valid for~$\aschema$,
  such that $I_2$ has a homomorphism to~$I_2^\dagger$, and
  the restriction of $I_1^\dagger$ to the relations of~$\aschema$ is a subinstance
  of~$I_1$. Then $Q$ is not access-determined in~$\aschema$.
\end{lem}

\begin{proof}
  We prove the contrapositive of the claim. Let $Q$ be a query which is not
  access-determined in~$\aschema^\dagger$, and let $\{I_1^\dagger, I_2^\dagger\}$ be a counterexample.
  Using the hypothesis, we construct $I_1$
  and $I_2$. It suffices to observe that they are a counterexample to
  access-determinacy for~$Q$ and $\aschema$, which we show. First, they satisfy the
  constraints of~$\aschema$ and have a common subinstance which is jointly
  access-valid. Second, as~$I_1^\dagger$ satisfies $Q$, as all relations used in~$Q$ are
  on~$\aschema$, and as the restriction of $I_1^\dagger$ is a subset of~$I_1$, we know
  that~$I_1$ satisfies $Q$. Finally, since~$I_2^\dagger$ does not satisfy $Q$
  and $I_2$ has
  a homomorphism to~$I_2^\dagger$, we know that~$I_2$ does not satisfy~$Q$. Hence,
  $I_1, I_2$ is a counterexample to access-determinacy of~$Q$ in~$\aschema$,
  which concludes the proof.
\end{proof}

\subsection{Choice Simplifiability for RA answerability} 
Recall that the choice simplification of a result-bounded schema is obtained by
changing every result-bounded method to have bound~$1$.
We say that a schema $\aschema$ is \emph{choice simplifiable for RA plans} 
if any CQ that has an RA plan over~$\aschema$ has one over its choice simplification.
The following result is the counterpart to Theorem~\ref{thm:simplifychoice}:

\begin{thm} \label{thm:simplifychoicera}
 Let $\aschema$ be a schema with constraints in
  equality-free first-order logic (e.g., TGDs), and let $Q$ be a CQ that is access-determined \wrt\ $\aschema$.
Then $Q$ is also access-determined in the choice simplification
  $\aschema^\dagger$ of~$\aschema$.
\end{thm}

The proof follows that of Theorem~\ref{thm:simplifychoice} with no surprises,
using Lemma~\ref{lem:enlargera}.

\begin{proof}
 We fix a counterexample $I_1^\dagger, I_2^\dagger$ 
 to access-determinacy in $\aschema^\dagger$:
  we know that $I_1^\dagger$ satisfies the query, $I_2^\dagger$ violates the query,
  $I_1^\dagger$ and $I_2^\dagger$ satisfy the equality-free first order constraints
  of~$\aschema$,
  and $I_1^\dagger$ and $I_2^\dagger$ have a common subinstance $I_\acc^\dagger$ which is jointly
  access-valid for~$\aschema^\dagger$.
  We  expand $I_1^\dagger$ and~$I_2^\dagger$ to~$I_1$ and $I_2$ that have a common subinstance that is jointly access-valid
  for~$\aschema$, to conclude using Lemma~\ref{lem:enlargera}.
  Our construction is identical to the blow-up used in Theorem~\ref{thm:simplifychoice}:
  for each element $a$ in the domain of~$I_1^\dagger$, introduce infinitely many
  fresh elements $a^j$ for~$j \in \NN_{>0}$, and identify $a^0 \colonequals a$.
  Now, define $I_1 \colonequals \mathrm{Blowup}(I_1^\dagger)$, where 
  $\mathrm{Blowup}(I_1^\dagger)$ is the instance with facts $\{R(a_1^{i_1} \ldots a_n^{i_n}) \mid R(\vec
  a) \in I_1^\dagger, \vec i \in \NN^n\}$. Define $I_2$
  from~$I_2^\dagger$ in the same way.

The proof of Theorem~\ref{thm:simplifychoice} already showed that~$I_1^\dagger$ and
  $I_1$ agree on all equality-free first-order
  constraints, that
$I_1^\dagger$ still satisfies the query, and $I_2^\dagger$ still
violates the query.
All that remains is to construct a common subinstance 
that is jointly access-valid for~$\aschema$. We do this
as in the proof of Theorem~\ref{thm:simplifychoice}, 
  setting  $I_\acc \colonequals \mathrm{Blowup}(I_\acc^\dagger)$. 
  To show that $I_\acc$ is jointly access-valid, consider any access $(\mt,
  \accbind)$ with values from~$I_\acc$. If there are no matching tuples
  in~$I_1^\dagger$ and in~$I_2^\dagger$,
  then there are no matching tuples in~$I_1$ and~$I_2$ either. Otherwise,
  there must be some matching tuple in $I_\acc^\dagger$ because it is
jointly access-valid in~$I_1^\dagger$ and~$I_2^\dagger$ for~$\aschema^\dagger$. Hence,
  sufficiently many copies exist in~$I_\acc$ to satisfy
the original result bounds, so  we can find a valid response to the access
  in~$I_\acc$. Hence, $I_\acc$ is indeed jointly access-valid, which completes
the proof.
\end{proof}

As with choice simplification for~$\amd$, this result can be applied immediately
to TGDs. In particular,
if we consider frontier-guarded TGDs, the above result says that we can assume any result bounds
are $1$, and thus the  query containment problem produced
by Proposition~\ref{prop:reducera} will  involve only frontier-guarded TGDs.
We thus get the following analog of Theorem~\ref{thm:decidegf}:

\begin{thm} \label{thm:decidegfra}
  We can decide whether a CQ is RA answerable with respect to a schema with
  result bounds whose constraints are frontier-guarded TGDs.
  The problem is $\twoexp$-complete.
\end{thm}

\subsection{FD Simplifiability for RA plans} 

We now turn to FD simplification. Recall that the FD simplification of a
result-bounded schema is intuitively defined by adding an auxiliary relation $R_\mt$ for
every result-bounded method~$\mt$, relating it with inclusion dependencies to
the relation~$R$ accessed by~$\mt$, and replacing $\mt$ by a non-result-bounded
method on~$R_\mt$. The new method makes it possible to retrieve the values for
the output positions that are determined by the input positions of~$\mt$
according to the FDs. The formal definition is given in Section~\ref{sec:simplify}.
A schema is \emph{FD simplifiable for RA plans} if every CQ having a plan over the schema
has an RA plan in its FD simplification. 

We now show that schemas whose constraints consist  only of FDs are FD
simplifiable, which is the analogue of Theorem~\ref{thm:fdsimplify}:

\begin{thm} \label{thm:fdsimplifyra} 
  Let $\aschema$ be a schema whose constraints are FDs, and let
$Q$ be a CQ that is RA answerable in~$\aschema$. Then $Q$ is RA answerable in
  the FD simplification $\aschema^\dagger$ of~$\aschema$.
\end{thm}
\begin{proof}
  We use Lemma~\ref{lem:enlargera} and assume that we have a counterexample
  $I_1^\dagger, I_2^\dagger$ to determinacy for~$\aschema^\dagger$, with $Q$
  holding in~$I_1^\dagger$, with $Q$ not holding in~$I_2^\dagger$, and with
  $I_1^\dagger$ and $I_2^\dagger$
  having a common subinstance $I_\acc^\dagger$ which is jointly access-valid
  in~$I_1^\dagger$
  and~$I_2^\dagger$ for~$\aschema^\dagger$. We will upgrade these
  to~$I_1$, $I_2$, $I_\acc$ having the same property for~$\aschema$, by
  blowing up accesses one after the other. We fix a valid access selection
  $\sigma$ for~$I_\acc^\dagger$ which returns outputs for accesses that are valid
  in~$I_1^\dagger$ and in~$I_2^\dagger$.

  We construct $I_1$, $I_2$ and $I_\acc$ exactly as in 
  the proof of Theorem~\ref{thm:fdsimplify} with the access selection~$\sigma$,
  performing the same blowing up process (recall that it distinguished between the
  ``dangerous'' and the ``non-dangerous'' methods).
The only point to show is that
  $I_\acc$ is jointly access-valid in~$I_1$. To do so, we distinguish 
  several possibilities,  again following the distinction in that proof. If
  $\abind$ contains values from $\adom(I_\acc)
  \setminus \adom(I_\acc^\dagger)$, then all matching tuples in either $I_1$
  or~$I_2$ are
  in $I_\acc$ and there is nothing to show. If the method used is not
  result-bounded, then the matching tuples in $I_1$ and $I_2$
  are either tuples of~$I_1^\dagger$ and~$I_2^\dagger$ that were already
  in~$I_\acc^\dagger$, or tuples
  added to $I_1$ and $I_2$ that were added to $I_\acc$ as well. If the
  method is dangerous, 
  then one possibility
is that the corresponding access on $I_\acc^\dagger$ according
  to~$\sigma$ either returned no tuples, in which case there are no matching
  tuples in~$I_1$ and~$I_2$ and all potential matching tuples are in
  $I_\acc\setminus I_\acc^\dagger$. Or the access returned precisely one tuple, which was added
  to $I_\acc$.

  For the case of an access with a result-bounded and dangerous method, there
  are again two cases. If we did not blow up the access, then $\sigma$ did not
  return a result, and there were no matching tuples in~$I_1^\dagger$
  or~$I_2^\dagger$ for the
  access, so all potential matching tuples in $I_1$ or~$I_2$ are in
  $I_\acc$. If we did blow up the access, then we know by construction
  that $I_\acc$ contains
  infinitely many tuples, from which we can construct a response which is valid
  both in~$I_1$ and~$I_2$. This concludes the proof using
  Lemma~\ref{lem:enlargera}.
\end{proof}

\subsection{Complexity of RA answerability for FDs}
\label{apx:complexityfdspmr}
In Theorem~\ref{thm:decidfd} we showed that monotone answerability with FDs was decidable in the lowest possible
complexity, i.e., $\np$. 

The argument involved first showing
\emph{FD simplifiability}, which
allowed us to eliminate result bounds at the cost of adding additional $\incd$s.
We then simplified the resulting rules to
ensure that the chase would terminate.  This relied on the fact that the axioms for~$\amd$ would include
rules going from~$R$ to~$R'$, but not vice versa. Hence, the argument
does not generalize for the rules that axiomatize RA plans.

However,  we can repair the argument at the cost of adding an additional assumption.
A schema $\aschema$ with access methods is \emph{single method per relation}, abbreviated $\smpr$, if for
every relation there is at most one access method. This assumption was made
 in many works on access methods \cite{access1,access2},
 although we do not make it by default
elsewhere in this work. We can then show the following analogue of
Theorem~\ref{thm:decidfd} with this additional assumption:

\begin{thm} \label{thm:decidfdrasmpr}
  We can decide whether a CQ~$Q$ is RA answerable with respect to an $\smpr$ schema with
  result bounds whose constraints are FDs. The problem is $\np$-complete.
\end{thm}

We will actually show something stronger:
for~$\smpr$ schemas with constraints consisting of FDs only,
 there is no difference between full answerability and monotone answerability.
Given Theorem~\ref{thm:decidfd}, this immediately implies
Theorem~\ref{thm:decidfdrasmpr}.

\begin{prop} \label{prop:monotonecollapsefdsmpr}
Let $\aschema$ be a  schema with access methods satisfying $\smpr$ and constraints $\Sigma$ consisting
of functional dependencies, and $Q$ be a CQ that is access-determined.
Then $Q$ is $\amd$.
\end{prop}
\begin{proof}
We know from Theorem~\ref{thm:fdsimplifyra} that the schema is FD simplifiable,
so we can eliminate result bounds as follows. Recall that $\detby(\mt)$ denotes
  the positions of the relation of~$\mt$ that are determined by the input
  positions of~$\mt$ according to the FDs. Recall the form of the FD
  simplification:

\begin{itemize}
  \simplifyfddef
\end{itemize}

By Proposition~\ref{prop:reducera} we know that~$Q$ is access-determined exactly
when $Q \subseteq_\Gamma Q'$, where $\Gamma$ contains two copies of the above schema
and also axioms of the following form for each access method~$\mt$:
\begin{itemize}
  \item (Forward):
    \[\Big(\bigwedge_i \accessible(x_i)\Big) \wedge S(\vec x, \vec y) \rightarrow
    \Big(\bigwedge_i \accessible(y_i)\Big)
    \wedge S'(\vec x, \vec y).\]
\item (Backward): 
  \[\Big(\bigwedge_i \accessible(x_i)\Big) \wedge S'(\vec x, \vec y) \rightarrow
    \Big(\bigwedge_i \accessible(y_i)\Big)
    \wedge S(\vec x, \vec y).\]
\end{itemize}
where $\vec x$ denotes the input positions of~$\mt$.
Note that~$S$ may be one of the original relations, or one of the relations~$R_\mt$ produced by the transformation
above, depending on whether $\mt$ originally had result bounds or not.

We now show that  chase proofs with~$\Gamma$ must in fact be very simple under
the 
$\smpr$ assumption:
\begin{clm} Assuming our schema is $\smpr$, consider any chase
sequence for~$\Gamma$. Then:
\begin{itemize}
\item Rules of the form
$R_\mt(\vec x, \vec y)
\rightarrow \exists \vec z ~
  R(\vec x, \vec y, \vec z)$
will never fire.
\item  Rules of the form
$R'(\vec x, \vec y, \vec z) \rightarrow R'_\mt(\vec x, \vec y)$
will never fire.
\item FDs will never fire (assuming they were satisfied on the initial
instance).
\item (Backward) axioms will never fire.
\end{itemize}
\end{clm}
  Note that the last item suffices to conclude that
  Proposition~\ref{prop:monotonecollapsefdsmpr} holds, since a proof of access-determinacy
in which  (Backward) axioms never fire is
 a proof of $\amd$.
So it suffices to prove the claim.
We do so by induction on the length of a chase proof.
We consider the first item.
Consider a fact $R_\mt(\vec c, \vec d)$. Since the (Backward) axioms never
  fire (fourth point of the induction), the fact must have been produced from a fact $R(\vec c, \vec d, \vec e)$. Hence
the axiom can not fire on this fact, because we only fire
active triggers.

We move to the second item, considering a fact $R'(\vec c, \vec d, \vec e)$.
By $\smpr$ and the inductive assumption that FDs do not fire, this fact can only have been produced 
via applying a (Backward) axiom to a fact of the form  $R'_\mt(\vec c, \vec d)$.
Since inductively we know that such rules doe not fire, this completes the inductive step.

Turning to the third item, we first consider a potential violation of an FD $D
\determines r$ on an unprimed relation~$R$. This consists of facts $R(\vec c)$ and $R(\vec d)$ agreeing
on positions in~$D$ and disagreeing on position~$r$. As the initial instance is
always assumed to satisfy the FDs, these facts are not in the initial instance.
But they could not have been otherwise produced, as we know by induction (first
and fourth points) that
none of the rules with an unprimed relation $R$ in their head will fire.
Now let us turn to facts that are potential violations of the primed copies
of the FDs, for some relation~$R'$. The existence of the violation implies that
there is an access method on the corresponding relation~$R$ in the original
schema, since otherwise there could be no relation $R'_{\mt}$, and such a violation
could not have occurred.  By the $\smpr$ assumption there is exactly one such method.

We first consider the case where this access method has result bounds.
We know that the facts in the violation must have been produced by
the rule going from~$R'_\mt$ to~$R'$ (noting that in this case the Forward rule
creates $R'_\mt$-facts, not $R'$-facts). Let us write the facts of the violation
as $R'(\vec c_1, \vec d_1, \vec e_1)$ and $R'(\vec c_2, \vec d_2, \vec e_2)$. Assume
that  $R'(\vec c_2, \vec d_2, \vec e_2)$ was the latter of the two facts to be created,
then $\vec e_2$ would have been chosen fresh. Hence the violation must occur within
the positions corresponding to~$\vec c_1, \vec d_1$ and $\vec c_2, \vec d_2$.
But by induction (third point), and by the~$\smpr$ assumption, these facts must have been created from
facts $R'_\mt(\vec c_1, \vec d_1)$ and $R'_\mt(\vec c_2, \vec d_2)$ where $\mt$
is the only access method on~$R$, and in turn these must
have been created from facts $R_\mt(\vec c_1, \vec d_1)$ and $R_\mt(\vec c_2, \vec d_2)$. These last
must (again, by induction, using the third and fourth points) have been created from facts
$R(\vec c_1, \vec d_1, \vec f_1)$ and $R(\vec c_1, \vec d_1, \vec g_1)$. But then we have
an earlier violation of the FDs on these two facts, which is a contradiction.

We now consider the second case, where the access method on~$R$ has no result bounds in the
original schema. In this case there
is no relation~$R_\mt'$ and the facts of the violation must have been produced
by applying the Forward rule, which can only apply to the
relation $R$.  But then the $R$-facts used to create them must
themselves be an earlier violation of the corresponding FD on~$R$, which is
again a contradiction. Hence, we have shown the third item.

Turning to the last item, there are two kinds of Backward rules to consider.
First, the ones involving a primed relation $R'$ and the original relation $R$,
where there is an access method without result bounds on~$R$ in the original
schema. Secondly, the ones
involving a primed relation $R'_\mt$ and the unprimed relation $R_\mt$ where
there is an access method with result bounds on~$R$ in the original schema. 
For the first kind of axiom, any $R'$-fact can only have been created from an
$R$-fact using the Forward axioms, and so the Backward axiom cannot fire.
For the second kind of axiom, we show the claim 
by considering a fact $R'_\mt(\vec c, \vec d)$. Using the second point of the
induction, it can
only have been generated by a fact $R_\mt(\vec c, \vec d)$, and thus (Backward)
could not fire, which establishes the desired result.
\end{proof}

Without $\smpr$, we can still argue that RA answerability is decidable, and show
a singly exponential complexity upper bound:
\begin{thm} \label{thm:decidfdraexp}
For general schemas with access methods and constraints $\Sigma$
consisting of FDs, RA answerability is decidable in~$\exptime$.
\end{thm}
\begin{proof}
We consider the query containment problem for RA answerability obtained after
  eliminating result bounds, and let
$\Gamma$ be the corresponding constraints as in
  Proposition~\ref{prop:monotonecollapsefdsmpr}.

Instead of claiming that neither the FDs nor the backward axioms will fire, as in the
case of~$\smpr$, we argue only that  the FDs will not fire.
From this it follows that the constraints consist only of $\incd$s
and accessibility axioms, leading to an $\exptime$ complexity upper bound:
one can apply 
the~$\exptime$
complexity result without result bounds from~\cite{bbbicdt}.

We consider a chase proof with~$\Gamma$, and claim, for each
relation~$R$ and each result-bounded method $\mt$ on~$R$, the following
  invariant:
\begin{itemize}
\item Every $R_\mt$-fact and every $R'_\mt$-fact
 is a projection of some $R$-fact or some $R'$-fact.
\item All the FDs are satisfied in the chase instance,
and further  for any relation~$R$, the relation $R \cup R'$  satisfies any FDs  on relation $R$,
That is: for any FD $D \determines r$ on relation $R$,
we cannot have an $R$-fact and an $R'$-fact that agree on positions in~$D$ and disagree
on some position in $r$.
\end{itemize}
The second item of the invariant implies that the FDs do not fire, which as we
  have argued is sufficient to conclude our complexity bound.

 The invariant is initially true, by assumption that
  FDs are satisfied
  on the initial instance.
When firing an $R$-to-$R_\mt$ axiom or an $R'$-to-$R'_\mt$ axiom, the first
item is preserved by definition,  and the second is trivially preserved since
there are no FDs on~$R_\mt$ or $R'_\mt$.

When firing an accessibility axiom, either forward or backward, again the first  and the
  second  item are clearly preserved.

Now, consider the firing of an $R_\mt$-to-$R$ axiom.
 The first item is trivially  preserved, so we must only show the second.

 Consider the fact
$R_\mt(a_1 \ldots a_m)$ and the generated 
fact $F = R(a_1 \ldots a_m, b_1 \ldots b_n)$
   created by the rule firing. Assume that~$F$ is part of an FD violation with some
  other fact $F'$ which is of the form $R(a'_1 \ldots a'_m, b'_1 \ldots b'_m)$ or
$R'(a'_1 \ldots a'_m, b'_1 \ldots b'_m)$.

We know that the left-hand-side of the FD cannot contain any
 of the positions of the~$b_i$, because they are fresh nulls. Hence, the
  left-hand-side of the FD is
 included in the positions of~$a_1 \ldots a_m$. But now, by definition of the FD
  simplification, the right-hand-side of the FD cannot correspond
to one of  the~$b_1 \ldots b_n$, since otherwise that position
would have been  included in~$R_\mt$.
  So the right-hand-side is also one of the positions of~$a_1 \ldots a_m$, and
  in particular we must have $a_i \neq a_i'$ for some $1 \leq i \leq m$ in the
  right-hand-side of the FD\@.

 Now we use the first item of the inductive
  invariant on the fact $R_\mt(a_1 \ldots a_m)$: there was already a fact $F''$, either an $R$ or
$R'$-fact, with tuple of values $(a_1 \ldots a_m, b''_1 \ldots b''_m)$.
As there is $1 \leq i \leq m$ such that~$a'_i \neq a_i$, the tuples of values
of~$F'$ and $F''$ must be different. 
But now, as~$F$ and $F'$ are an FD violation on the positions
  $a_1 \ldots a_m$, then $F'$ and $F''$ are seen to also witness an FD
   violation in~$R \cup R'$ that existed before the firing. This
  contradicts the first point of the invariant, so we conclude that the second
  item is preserved when firing an $R_\mt$-to-$R$ axiom.

When firing $R'_\mt$-to-$R'$ rules, the symmetric argument applies.

This completes the proof of the invariant, and concludes the proof of
Theorem~\ref{thm:decidfdraexp}.
\end{proof}

\subsection{Choice Simplifiability for RA plans with $\uincd$s and FDs}

We last turn to the adaptation of our choice simplifiability result for
$\uincd$s and FDs (Theorem~\ref{thm:simplifychoiceuidfd}). Here is the statement
for the case of RA plans:

\begin{thm} \label{thm:simplifychoiceuidfdra}
Let schema $\aschema$ have  constraints  given by
 $\uincd$s and arbitrary FDs,  and $Q$ be a CQ that is access-determined \wrt\ $\aschema$.
Then $Q$ is also access-determined in the choice simplification of~$\aschema$.
\end{thm}

We will proceed in a similar fashion to Theorem~\ref{thm:simplifychoiceuidfd},
i.e., fixing one access at a time. Here is the analogue of the single-access
blowup (Definition~\ref{def:sablow}), where we simply replace the access-valid
subinstance by a jointly access-valid subinstance:

\begin{defi}
  Let $\aschema$ be a schema and $\aschema^\dagger$ be its choice simplification,
  and let $\Sigma$ be a set of constraints.

  Consider two instances $I_1^\dagger, I_2^\dagger$ that satisfy~$\Sigma$, and a common subinstance
  $I_\acc^\dagger$ which is jointly access-valid in~$I_1^\dagger$ and~$I_2^\dagger$
  for~$\aschema^\dagger$.
  Let $(\mt, \accbind)$ be an access in~$I_\acc^\dagger$

  A \emph{single-access RA blowup} of $I_1^\dagger, I_2^\dagger$
  and~$I_\acc^\dagger$ for $(\mt, \accbind)$ is a pair of
  instances $I_1, I_2$ that 
  satisfy $\Sigma$,
  such that $I_1$ is a superinstance of~$I_1^\dagger$,
  $I_2$ has a homomorphism to~$I_2^\dagger$,
  $I_1$ and $I_2$ have a common subinstance $I_\acc$ which is 
  jointly access-valid in~$I_1$ and~$I_2$ for~$\aschema^\dagger$, 
  and the following hold:
  \begin{enumerate}
    \item $I_\acc$ is a superinstance of~$I_\acc^\dagger$;
    \item there is an output to the access $\mt, \accbind$ in~$I_\acc$ which is valid
      in~$I_1$
  for~$\aschema$;
    \item for any access in~$I_\acc^\dagger$ having an output
      in~$I_\acc^\dagger$ which is valid
      for~$\aschema$ in~$I_1^\dagger$, there is an output to this access
      in~$I_\acc$ 
      which is valid for~$\aschema$ in~$I_1$;
  \item for any
    access in~$I_\acc$ which is not an access in~$I_\acc^\dagger$, there is an
      output in~$I_\acc$ which is valid for~$\aschema$ in~$I_1$;
  \end{enumerate}
\end{defi}

We use the following blowup lemma as an analogue of
Lemma~\ref{lem:enlargeprog}:

\begin{lem}
  \label{lem:enlargeprogra}
  Let $\aschema$ be a schema and $\aschema^\dagger$ be its choice simplification,
  and let $\Sigma$ be the set of constraints.

  Assume that, for any CQ~$Q$ not access-determined in~$\aschema^\dagger$, for any
  counterexample $I_1^\dagger, I_2^\dagger$ of access-determinacy for~$Q$ and
  $\aschema^\dagger$ with a common subinstance $I_\acc^\dagger$ 
  jointly access-valid in~$I_1^\dagger$ and~$I_2^\dagger$ for~$\aschema^\dagger$,
  for any access $\mt, \accbind$ in~$I_\acc^\dagger$,,
  we can construct a single-access RA blowup of $I_1^\dagger, I_2^\dagger$
  and~$I_\acc^\dagger$
  for~$(\mt, \accbind)$.

  Then any CQ which is access-determined in~$\aschema$ is also
  access-determined in~$\aschema^\dagger$.
\end{lem}

The proof of this lemma is exactly like that of Lemma~\ref{lem:enlargeprog}.

We are now ready to prove Theorem~\ref{thm:simplifychoiceuidfdra} using the process
of Lemma~\ref{lem:enlargeprogra}. We proceed similarly to the proof of
Theorem~\ref{thm:simplifychoiceuidfd}.

\begin{proof}
Let $\aschema$ be the schema, let $\aschema^\dagger$ be its choice
simplification, and let $\Sigma$ be the set of constraints.
Let $Q$ be a CQ which is not access-determined in~$\aschema^\dagger$,
let $I_1^\dagger, I_2^\dagger$ be a counterexample to access-determinacy,
and let $I_\acc^\dagger$ be a common subinstance of~$I_1^\dagger$ and~$I_2^\dagger$
for~$\aschema^\dagger$
which is jointly access-valid in~$I_1^\dagger$ and~$I_2^\dagger$ for~$\aschema^\dagger$.
Let $(\mt, \accbind)$ be
an access on relation~$R$ in~$I_\acc^\dagger$:
we know that this access has an output which is valid for~$\aschema^\dagger$, but it
does not necessarily have one
which is valid for~$\aschema$. 
Our proof is to follow the single-access RA blowup process and build 
superinstances $I_1$, $I_2$, and $I_\acc$ of $I_1^\dagger$, $I_2^\dagger$, and
$I_\acc^\dagger$ respectively, which satisfy the conditions.

As in the proof of
Theorem~\ref{thm:simplifychoiceuidfdra}, if there are no matching tuples
in~$I_1^\dagger$
for the access~$(\mt, \accbind)$, then there are no matching tuples in~$I_\acc^\dagger$ either,
so the access $(\mt, \accbind)$ already has a valid output for~$\aschema$ and there
is nothing to do. The same holds if there are no matching tuples in~$I_2^\dagger$.
Now, if there is exactly one matching tuple in~$I_1^\dagger$ and exactly one matching
tuple in~$I_2^\dagger$, as~$I_\acc^\dagger$ is jointly access-valid for~$\aschema^\dagger$, it
necessarily contains these matching tuples, so that, as~$I_\acc^\dagger \subseteq
I_1^\dagger$
and $I_\acc^\dagger \subseteq I_2^\dagger$, the matching tuple in~$I_1^\dagger$
and~$I_2^\dagger$ is the same,
and again there is nothing to do: the access $(\mt, \accbind)$ already has a
valid output for~$\aschema$.

Hence, the only interesting case is when there is a matching tuple to the access
in~$I_1^\dagger$ and in~$I_2^\dagger$, and there is more than one matching tuple in one of the
two. As $I_1^\dagger$ and $I_2^\dagger$ play a symmetric role in the hypotheses of
Lemma~\ref{lem:enlargeprogra}, we assume without loss of generality that it is
$I_1^\dagger$ which has multiple matching tuples for the access.

As $I_\acc^\dagger$ is access-valid in~$I_1^\dagger$ for~$\aschema^\dagger$, we know
that~$I_\acc^\dagger$ contains at least one of these tuples, say $\vec t_1$. As $I_\acc
\subseteq I_2^\dagger$, then $I_2^\dagger$ also contains $\vec t_1$. As in the
proof of Theorem~\ref{thm:simplifychoiceuidfd},
we take $\vec t_2$ a different matching tuple in~$I_1^\dagger$, let $C$ be the non-empty
set of positions where $\vec t_1$ and $\vec t_2$ disagree, and observe that
there is no FD implied from the complement of~$C$ to a position of~$C$.

We define $W$ as in the proof of
  Theorem~\ref{thm:simplifychoiceuidfd},
  and construct $I_1 \colonequals I_1^\dagger \cup W$ and $I_2
  \colonequals I_2^\dagger \cup W$ as in that proof. We
show that $(I_1, I_2)$ is a counterexample to determinacy for~$Q$
and~$\aschema^\dagger$:

\begin{itemize}
  \item We know by Claim~\ref{clm:decomp} that $I_1$ and $I_2$ satisfy the $\uincd$s and the
FDs of~$\Sigma$.
    \item We clearly have $I_1^\dagger \subseteq I_1$.
    \item The homomorphism
      from~$I_2$ to~$I_2^\dagger$ is
    defined as in the proof of Theorem~\ref{thm:simplifychoiceuidfd}.
    \item We define $I_\acc \colonequals I_\acc^\dagger \cup W$
      a common subinstance of~$I_1$ and~$I_2$ and we 
      must show that $I_\acc$ is 
    jointly access-valid in~$I_1$ and~$I_2$ for~$\aschema^\dagger$.
    We do this as in the
    proof of Theorem~\ref{thm:simplifychoiceuidfd}.
    First, for accesses that include an element of~$\adom(I_\acc) \setminus
    \adom(I_\acc^\dagger)$, the matching tuples are all in~$W$ so they are
    in~$I_\acc$.
    Second,
    for accesses on~$\adom(I_\acc^\dagger)$, the matching tuples include the
    result $U$ of this access in $I_\acc^\dagger$, which was valid in~$I_1^\dagger$ and
    $I_2^\dagger$, and possible additional matching tuples $U'$ from~$W$ which are
    in~$I_\acc$, and these are the only possible matching tuples.
    Thus, we can construct a valid output
    to this access for~$\aschema^\dagger$ from $U$ and~$U'$.
\end{itemize}
What remains to be able to use Lemma~\ref{lem:enlargeprogra} is to show the
four additional conditions:

\begin{enumerate}
  \item It is immediate that~$I_\acc \supseteq I_\acc^\dagger$.
  \item The access $(\mt, \accbind)$ has an output in~$I_\acc$ which is valid for~$\aschema$ 
    in~$I_1$ and~$I_2$. This is established as in the proof of
    Theorem~\ref{thm:simplifychoiceuidfd}: there are now
    infinitely many matching tuples for the access in~$I_1$
    and~$I_2$, so we can choose
    as many as we want in~$W$ to obtain an output in~$I_\acc$ which is valid
    for~$\aschema$ in~$I_1$ and~$I_2$.
  \item For every access of~$I_\acc^\dagger$ that has an output which is valid
    for~$\aschema$ in~$I_1^\dagger$
    in~$I_2^\dagger$, then we can construct such an output in~$I_\acc$ which is 
    valid for~$\aschema$ in~$I_1$ and~$I_2$.
    This is similar to the fourth bullet point above. From the 
    output $U$ to the access in~$I_\acc^\dagger$ which is valid for~$I_1^\dagger$ and
    $I_2^\dagger$,
    we construct an 
    output to the access in~$I_\acc$ which is valid for~$I_1$
    and~$I_2$, using the tuples
    of~$U$ and the matching tuples in~$W$.
  \item All accesses of~$I_\acc$ which are not accesses
    of~$I_\acc^\dagger$ have an
    output which is valid
    for~$\aschema$ in~$I_1$ and~$I_2$. As before, such accesses must include
    an element of~$W$, so by the fourth bullet point all matching tuples are
    in~$W$, so they are all in~$I_\acc$.
\end{enumerate}
Hence, we have explained how to fix the access $(\mt, \accbind)$, so we can
conclude using Lemma~\ref{lem:enlargeprogra} that we obtain a counterexample to
access-determinacy of~$Q$ in~$\aschema$ by fixing all accesses. This concludes
the proof.
\end{proof}

{\footnotesize
\begin{table}[t]
\caption{Summary of results on simplifiability and complexity of RA answerability}
\label{tab:resultsfull}
  {
  \renewcommand{\tabcolsep}{3pt}
  \begin{tabular}{lll}
\toprule
{\bfseries Fragment} &
{\bfseries Simplification} & 
{\bfseries Complexity} \\
\midrule
  IDs & Existence-check  (Thm~\ref{thm:simplifyidsexistence}, Prop.~\ref{prop:monotonecollapse}) & $\exptime$-complete  
(Thm~\ref{thm:decidids}, Prop.~\ref{prop:monotonecollapse})\\
  Bounded-width IDs & Existence-check (see above) & $\np$-complete
    (Thm~\ref{thm:npidsbounds}, Prop \ref{prop:monotonecollapse}) \\
  FDs & FD (Thm~\ref{thm:fdsimplifyra}) & 
    In $\exptime$ (Thm~\ref{thm:decidfdraexp}) \\
  FDs under $\smpr$ & FD (see above) &
    $\np$-complete (Thm~\ref{thm:decidfdrasmpr})\\
  FDs and UIDs & Choice  (Thm~\ref{thm:simplifychoiceuidfdra})
    &  \emph{Open}   \\
  Equality-free FO & Choice  (Thm~\ref{thm:simplifychoicera})
    & Undecidable (same proof as Prop~\ref{prp:undec}) \\
  Frontier-guarded TGDs & Choice  (see above) & $\twoexp$-complete  (Thm~\ref{thm:decidegfra}) \\
\bottomrule
\end{tabular}
  }
\end{table}
}

\subsection{Summary of Extensions to Answerability with RA plans}
Table \ref{tab:resultsfull} summarizes the expressiveness and complexity results
for RA plans. There are two differences with the corresponding table for
monotone answerability (Table~\ref{tab:results} in the body):
\begin{itemize}
\item For RA plans, while we know that choice simplifiability holds with FDs and
  UIDs, we do not know whether answerability is decidable. Indeed, in the
    monotone case, when proving Theorem~\ref{thm:deciduidfd}, we had used a separability argument
    to show that FDs could be ignored for FDs and UIDs (see the 
    proof of Theorem~\ref{thm:deciduidfd} in Section~\ref{sec:complexitychoice}).
We do not have such an
    argument for answerability with RA plans.
\item For RA plans, our tight complexity bound for answerability with FDs in
  isolation holds only under the $\smpr$ assumption; see
    Appendix~\ref{apx:complexityfdspmr} for
    details.
\end{itemize}

\end{document}